\documentclass[%
 reprint,
 superscriptaddress,
 amsmath,amssymb,
 aps, 
]{revtex4-2}

\usepackage{graphicx}
\usepackage{dcolumn}
\usepackage{bm}

\usepackage{multirow}
\usepackage{bm}
\usepackage{textcomp}
\usepackage{longtable}
\usepackage{booktabs}
\usepackage{epsfig}
\usepackage{rotating}
\usepackage{xfrac}
\usepackage{threeparttable}
\usepackage{lipsum} 
\usepackage{float}
\usepackage{dcolumn}
\usepackage{multirow}

\usepackage{rotating}
\RequirePackage{hyperref}
\hypersetup
{
unicode=false,%
pdftoolbar=true,
pdfmenubar=true,
pdffitwindow=false,
pdfstartview=FitH,
pdftitle={PRC},
pdfauthor={LAM Yi Hua},
pdfsubject={Nuclear Physics},
pdfcreator={LAM Yi Hua},
pdfproducer={LAM Yi Hua},
pdfkeywords={NucPhys},
bookmarksnumbered,%
colorlinks=true,%
linkcolor=magenta,%
citecolor=blue,
filecolor=cyan,%
urlcolor=blue,
breaklinks, 
plainpages=false}
\usepackage{bibunits}
\defaultbibliographystyle{apsrev4-1_titled}
\defaultbibliography{my-final-bib-from-jabref_titles,references}
\usepackage{CJKutf8}
\newcommand{\cntextSimKai}[1]{\begin{CJK*}{UTF8}{gkai}#1\end{CJK*}}
\newcommand{\cntextTraKai}[1]{\begin{CJK*}{UTF8}{bkai}#1\end{CJK*}}

\usepackage{xcolor}
\definecolor{Mycolor}{HTML}{C5E0B3}
\definecolor{red}{HTML}{ee6677}
\definecolor{blue}{HTML}{4477aa}
\definecolor{green}{HTML}{228833}
\definecolor{magenta}{HTML}{ee3377}
\definecolor{cyan}{HTML}{66ccee}
\definecolor{yellow}{HTML}{ccbb44}
\definecolor{grey}{HTML}{bbbbbb}
\begin{document}

\preprint{APS/123-QED}

\title{\textbf{Shape evolution of krypton isotopes calculated with\\axially deformed relativistic Hartree-Bogoliubov approach}}%

\author{Zi~Xin~Liu~(\cntextSimKai{刘子鑫})}
\affiliation{Advanced Energy Science and Technology, Guangdong Laboratory, Huizhou 516000, China}
\affiliation{CAS Key Laboratory of High Precision Nuclear Spectroscopy, \href{https://ror.org/03x8rhq63}{Institute of Modern Physics}, \href{https://ror.org/034t30j35}{Chinese Academy of Sciences}, Lanzhou 730000, China}

\author{Yi~Hua~Lam~(\cntextTraKai{藍乙華})}%
\email{lamyihua@zstu.edu.cn}
\homepage{https://lamyihua.github.io}
\affiliation{Zhejiang Key Laboratory of Quantum State Control and Optical Field Manipulation, Department of Physics, \href{https://ror.org/03893we55}{Zhejiang Sci-Tech University}, 310018 Hangzhou, China}
\affiliation{Astrophysical Big Bang Laboratory, Pioneering Research Institute, \href{https://ror.org/01sjwvz98}{RIKEN}, Wako, Saitama 351-0198, Japan}

\author{Peter~Ring}
\affiliation{Physik-Department der \href{https://ror.org/02kkvpp62}{Technische Universit{\"a}t München}, Garching D-85748, Germany} 
\affiliation{State Key Laboratory of Nuclear Physics and Technology, School of Physics, \href{https://ror.org/02v51f717}{Peking University}, Beijing 100871, China}

\date{\today}

\begin{abstract}
We perform a systematic study of the structure and properties of the krypton isotopic chain including both even-even and odd-$A$ nuclei based on the axially deformed relativistic Hartree-Bogoliubov approach. Five effective interactions of three families of covariant density functionals, i.e., PC-L3R, DD-PCX, DD-PC1, DD-MEX, and DD-ME2, are employed to calculate potential energy surfaces of krypton isotopes. $^{74,75}$Kr and $^{90,91,92}$Kr are determined as typical candidates of shape coexistence. The potential surfaces originating from the PC-L3R, DD-PCX, and DD-MEX interactions exhibit an abrupt shape transition from oblate to prolate for $^{73\text{-}74}$Kr, whereas DD-PC1 and DD-ME2 preserve an oblate ground-state shape. Such discrepancies are attributed to the occupations of single-particle levels at the vicinity of the Fermi surface described by these functionals. Moreover, the comparison between spherical and deformed calculations verifies the indispensability of deformation degrees of freedom in this region. The consideration of deformation effects improves the description of two-neutron separation energies, of which its evolution clearly demonstrates the $N=50$ and $82$ shell closures. Interestingly, PC-L3R predicts a more extended two-neutron drip line up to $^{132}$Kr, in agreement with the NL3$^*$ and PC-PK1 nonlinear effective interactions, whereas other functionals estimate a rather short isotopic chain up to $^{119}$Kr. This anomalous extension implies a significant softening or even collapse of the traditional $N=82$ shell closure near the neutron-rich drip line, highlighting the need for future studies based on triaxial deformation and beyond-mean-field correlations in this nuclear region.
\end{abstract}

\maketitle

\section{Introduction}

In recent years, remarkable progress in experimental techniques and the advent of next-generation radioactive isotope beam facilities have allowed experimental nuclear physicists to systematically investigate the properties of exotic nuclei far from the valley of stability~\cite{PPNP2021Yamaguchi,PPNP2025Nan}, for instance, the Second Generation System On-Line Production of Radioactive Ions (SPIRAL2) at GANIL in France \cite{SPIRAL2}, the Facility for Rare Isotope Beams (FRIB) in the US \cite{FRIB}, the Radioactive Ion Beam Factory (RIBF) at RIKEN in Japan \cite{RIBF}, the Facility for Antiproton and Ion Research (FAIR) in Germany \cite{FAIR}, the Rare Isotope Accelerator complex for Online Experiments (RAON) in Korea \cite{RAON}, and in particular, the recently commissioning High-Intensity Heavy-Ion Accelerator Facility (HIAF) in China is expected to provide groundbreaking discoveries on the limits of nuclear existence, exotic nuclear structures, and nuclear processes in stellar environments \cite{HIAF}. New operations of these state-of-the-art facilities extend experimental explorations of neutron-rich nuclei from the medium-heavy mass region toward the neutron drip line, thus offering unprecedented opportunities to test and constrain modern nuclear structure theories.

The shape of an atomic nucleus is determined by the number of nucleons and nucleon-nucleon interactions. Nuclear shape evolution has long been concerned with understanding the occupations of nucleons in orbital shells described by the effective interactions. In recent years, increasing attention has been devoted to the phenomenon of shape coexistence~\cite{PPNP2024Leoni,ADNDT2024Guo}, e.g., studies focusing on light nuclei~\cite{PRL2000Sarazin,PRC2018Hadyifmmode}, medium-mass nuclei~\cite{PPNP1999Petrovici,PLB2011Tomas,PRL2018Togashi,PRC2022Maya}, and heavy and superheavy nuclei~\cite{PRL1973Hendrie,PRC2002Niksic65,RMP2011Heyde}. This phenomenon referring to the coexistence of quantum states with distinct intrinsic shapes within a single atomic nucleus challenges the conventional shell-model picture and provides a valuable window into unraveling the microscopic mechanisms underlying nucleon-nucleon interactions~\cite{PPNP2022Paul}.

Rich structural features and shape coexistence phenomena have been extensively observed in Kr, Sr and Zr isotopes. Experimentally, accumulated evidence of shape coexistence has been reported for $^{72,74}\text{Kr}$ \cite{PRC1997Chandler,PRL2003Bouchez,PLB2004Daniel}, $^{98}\text{Sr}$~\cite{PRC2016Park}, and $^{98}\text{Zr}$\cite{PRL2018Singh}. Theoretically, a variety of nuclear structure approaches, including macroscopic-microscopic models~\cite{PRC2011Randrup,PRC2017Ward,NPA2026Aggarwal}, self-consistent mean-field methods~\cite{RMP2003Bender,PRC2017Abusara}, and beyond-mean-field theories~\cite{PLB2003Duguet,PRC2004Rodriguez}, have been employed to interpret and predict shape coexistence phenomena. In particular, the krypton isotopic chain, with its proton number $Z=36$ close to the $Z=40$ subshell closure and neutron numbers spanning the region where the $N=40$ subshell gap weakens, exhibits rich manifestations of shape coexistence and evolution~\cite{PRC1997Chandler,PRL2003Bouchez,PLB2004Daniel,PRC2017Abusara,PRC2018Bhuyan,PRC2023Zhang}.
This makes the krypton isotopic chain an ideal probe for advancing nuclear structure theory.

Covariant density functional theory (CDFT) has achieved remarkable success in describing the ground-state properties of both stable and exotic nuclei~\cite{ZPA1991Kucharek_339_23,PPNP1996Ring_37_193,PRL1996Meng_77_3963,PRL1997Poschl_79_3841,JPG2017Lv,PRC2018Liu,ADNDT2018Xia,PLB2018Afanasjev,PRC2021Fan,PRC2021Vale,PRC2021Perera,ADNDT2022Zhang,PLB2023Liu,ADNDT2024Liu,ADNDT2024Guo,PRC2025Osei}. In particular, the relativistic Hartree-Bogoliubov (RHB) approach provides a self-consistent and microscopic description of the properties of nuclear ground states by unifying the mean field and pairing correlations \cite{PR2005Vretenar_409_101}. This theoretical framework naturally incorporates the spin-orbit potential arising from meson-exchange interactions and correctly reproduces relativistic effects such as pseudospin symmetry~\cite{PPNP2006MengJ_57_470}. Currently, two families of commonly used effective interactions for the CDFT framework are meson-exchange \cite{NPA1994Sugahara,PRC1997Lalazissis,PRC2004Long,PRC2005Lalazissis_71_024312,PLB2013Afanasjev} and point-coupling functionals~\cite{PRC1992Nikolaus,PRC2008Niksic,PRC2010Zhao,PRC2019Yuksel,PLB2020Taninah,PLB2023Liu}. In recent years, the RHB approach received wide attentions for its successful description of multiple nuclear phenomena~\cite{ADNDT2018Xia,ADNDT2022Zhang,ADNDT2024Liu,ADNDT2024Guo}, precise predictions of proton separation energy \cite{Lam2022a}, and has been widely applied to investigate nuclear shape evolution and shape coexistence phenomena. For instance, \citet{PRC2017Abusara}, and \citet{PRC2023Zhang} investigated the shape evolution and predicted shape coexistence in the Kr isotopic chain. Nevertheless, a comprehensive understanding of shape coexistence here remains lacking, especially systematic studies on odd-$A$ nuclei.

In this work, we systematically investigate the ground-state properties of even-even and odd-$A$ nuclei in the krypton isotopic chain based on the axially symmetric RHB approach with the PC-L3R~\cite{PLB2023Liu}, DD-PCX~\cite{PRC2019Yuksel}, DD-PC1~\cite{PRC2008Niksic}, DD-MEX~\cite{PLB2020Taninah}, and DD-ME2~\cite{PRC2005Lalazissis_71_024312} effective interactions and the respective finite-range separable pairing force. By analyzing the evolution of potential energy curves, separation energies, and deformation parameters as function of neutron numbers, we explore the microscopic mechanisms of shape coexistence in the krypton isotopic chain. This study not only contributes to understanding the physical laws governing shape evolution in this isotopic chain but also provides theoretical references for future experimental investigations. The paper is organized as follows. We briefly introduce the axially symmetric RHB approach in Sec.~\ref{Sec:Theory}, then present numerical results and discussions in Sec.~\ref{Sec:Results}. Summary of the present findings is shown in Sec.~\ref{Sec:summary}.

\section{Relativistic Hartree-Bogoliubov Approach}
\label{Sec:Theory}

The RHB model can be derived within the framework of covariant density functional theory, which unifies the self-consistent mean field and a pairing field. 
The covariant density functional is based on the effective Lagrangian density described by either effective point-coupling or meson-exchange interactions (see Refs.~\cite{PRC1998Lalazissis,PRC2005Lalazissis_71_024312,PRC2008Niksic,PRC2010Zhao,PRC2019Yuksel,PLB2020Taninah,PLB2023Liu} for details). The RHB equation for nucleons is derived by the variational procedure, 
\begin{equation}\label{eq:RHB}
  \int d^3 \bm{r}' \left(
   \begin{array}{cc}
    h_\mathrm{D}-\lambda_{\tau} &  \Delta \\
     -\Delta^{\ast} &  -h^{\ast}_\mathrm{D}+\lambda_{\tau} \\
  \end{array}
\right)\left(
   \begin{array}{c}
    U_{k}\\
    V_{k}\\
  \end{array}
\right)=E_{k}\left(
   \begin{array}{c}
    U_{k}\\
    V_{k}\\
  \end{array}
\right),
\end{equation}
where $\lambda_{\tau}$ ($\tau\!=\!\mathrm{n}, \mathrm{p}$) is the chemical potential, $U_{k}$ and $V_{k}$ are quasiparticle wave functions, $h_\mathrm{D}$ is the Dirac Hamiltonian and $E_{k}$ is the quasiparticle energy,
\begin{equation}\label{Dirac-H}
  h_\mathrm{D}(\bm r)=\bm{\alpha}\cdot\bm{p}+V(\bm r)+ \beta(M+S(\bm r)) \, ,
\end{equation}
where $\bm{\alpha}$ and $\beta$ are the Dirac matrices, $\bm{p}$ is the momentum operator, and $S(\bm r)$ and $V(\bm r)$ are the scalar and vector potentials, respectively. The local densities are written as
\begin{eqnarray}
  \rho_{S}(\bm r) &=& \sum_{k>0}\bar{V}_{k}(\bm r)V_{k}(\bm r) \,, \nonumber\\
  \rho_{V} (\bm r)&=&  \sum_{k>0} V^{\dag}_{k}(\bm r)V_{k}(\bm r)\,,~{\rm and} \\
  \rho_{TV}(\bm r) &=& \sum_{k>0} V^{\dag}_{k}(\bm r)\tau_{3}V_{k}(\bm r) \,. \nonumber
\end{eqnarray}
The pairing field $\Delta$ in Eq.~(\ref{eq:RHB}) reads
\begin{equation}
\label{eq:Delta}
\Delta_{n_{1}n'_{1}}=\frac{1}{2}\sum_{n_{2}n'_{2}}\langle n_{1}n'_{1}|V^{pp}|n_{2}n'_{2} \rangle \kappa_{n_{2}n'_{2}} \, .
\end{equation}
The separable form of the pairing force \cite{PLB2009Tian_676_44} applied for the study of pairing properties in nuclei close to and far from stability is used in Eq.~(\ref{eq:Delta}). Compared to using the finite range Gogny force, implementing the separable pairing force is capable of significantly reducing computational costs. The separable form of the pairing force is defined as 

\begin{equation}
\label{eq:Vpp}
V^{pp}(\bm{r}_{1},\bm{r}_{2},\bm{r}'_{1},\bm{r}'_{2})=-G\delta(\bm{R}-\bm{R}')
P(\bm{r})P(\bm{r}')\frac{1}{2}(1-P^{\sigma}) \, ,
\end{equation}
with the center of mass, $\bm{R}=(\bm{r_{1}}+\bm{r_{2}})/2$, and the relative coordinates, $\bm{r} = \bm{r_{1}}-\bm{r_{2}}$. The form factor $P(\bm r)$ is of Gaussian shape
\begin{equation}
\label{eq:Gaussian}
P(\bm r)=\frac{1}{(4\pi a^{2})^{3/2}}e^{-r^{2}/4a^{2}} \, .
\end{equation}
According to Tian {\it et al}.~\cite{PLB2009Tian_676_44}, the $G$ strength in Eq.~(\ref{eq:Vpp}) and the $a$ parameter in Eq.~(\ref{eq:Gaussian}) are adopted as 728~MeV$\cdot$fm$^{3}$ and 0.644~fm, respectively. These parameters were derived by fitting to the original Gogny D1S interaction, proposed by Chappert et al.~\cite{PLB2008Chappert}. The Gogny D1S interaction is a widely used high performance nuclear force model, which has been employed in various Gogny model calculations across the entire nuclide region \cite{PRC2014Agbemava,PhysRevC2023Morana,PLB2026Zietek}. In the present work, calculations based on the DD-PC1~\cite{PRC2008Niksic}, DD-MEX~\cite{PLB2020Taninah}, and DD-ME2~\cite{PRC2005Lalazissis_71_024312} effective interactions use the separable form of pairing force. Both DD-PCX and PCL3R interactions developed by Y{\"u}ksel {\it et al.}~\citep{PRC2019Yuksel} and Liu {\it et al.}~\citep{PLB2023Liu}, respectively, are coupled with further optimized $a$ and pairing strengths of $G_\mathrm{n}$ (for neutrons) and $G_\mathrm{p}$ (for protons), leading to an improved description of bulk nuclear properties.

The RHB equations are solved in the configurational space of harmonic oscillator wave functions with appropriate symmetry, whereas the densities are computed in the coordinate space. The method can be applied to axially and non-axially deformed nuclei~\cite{DIRHB}. The map of the energy curve as a function of the quadrupole deformation parameters is obtained by solving the RHB equation with constraints on the axial mass quadrupole moments of a given nucleus. The method of quadratic constraints uses an unrestricted variation of the function, 
\begin{equation}
\label{eq:quadratic_constraints}
<\hat{H}> + \sum_{\mu = 0, 2} C_{2\mu} (<\hat{Q}_{2\mu}>-q_{2\mu})^{2} \, ,
\end{equation}
where $\hat{H}$ is the total energy and $<\hat{Q}_{2\mu}>$ denotes the expectation value of the mass quadrupole operators, 
\begin{equation}
\label{eq:quadratic_operators}
\hat{Q}_{20} = 2z^{2}-x^{2}-y^{2}~~~{\rm and}~~~\hat{Q}_{22} = x^{2}-y^{2} \, .
\end{equation}
$q_{2\mu}$ is the constrained value of the multipole moment and $C_{2\mu}$ is the corresponding stiffness constant~\cite{ManybodyProb1980}.

\section{Results and Discussions}
\label{Sec:Results}

We use the self-consistent axially-symmetric and spherically-symmetric RHB approaches with either meson-exchange or point-coupling effective interactions for the present study, i.e., PC-L3R~\cite{PLB2023Liu}, DD-PCX~\cite{PRC2019Yuksel}, DD-PC1~\cite{PRC2008Niksic}, DD-MEX~\cite{PLB2020Taninah}, and DD-ME2~\cite{PRC2005Lalazissis_71_024312}, to study the shape evolution along the Kr isotopic chain. These interactions are capable of describing the ground state and excited state properties over the entire nuclear landscape and revealing the single-particle structures that dominate the contribution to shape evolution, binding energies, and one- and two-neutron separation energies.

\begin{figure*}%
\centering
\begin{minipage}{0.225\linewidth}
\centering
\includegraphics[width=\linewidth, angle=0]{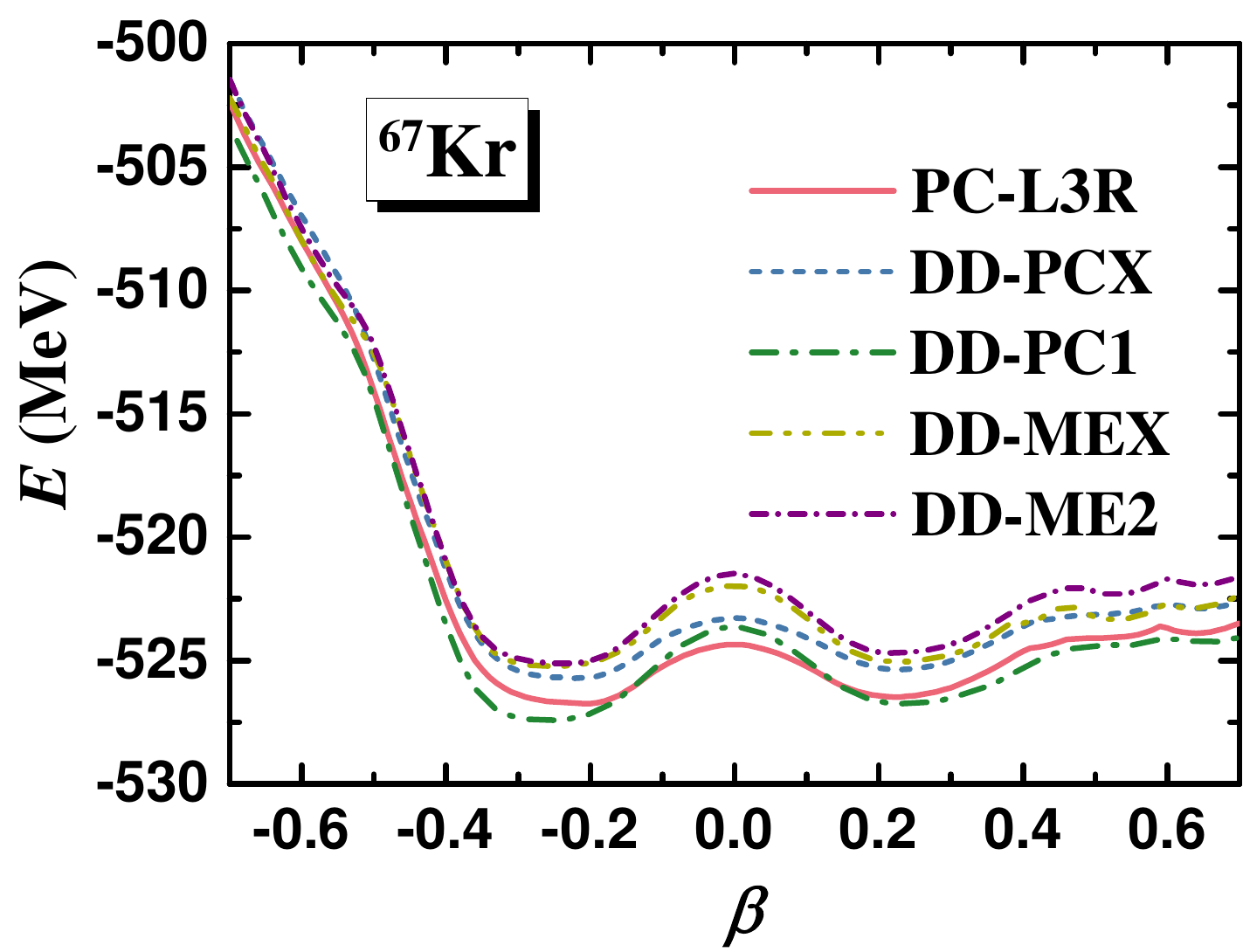}
\end{minipage}%
\begin{minipage}{0.23\linewidth}
\includegraphics[width=\linewidth, angle=0]{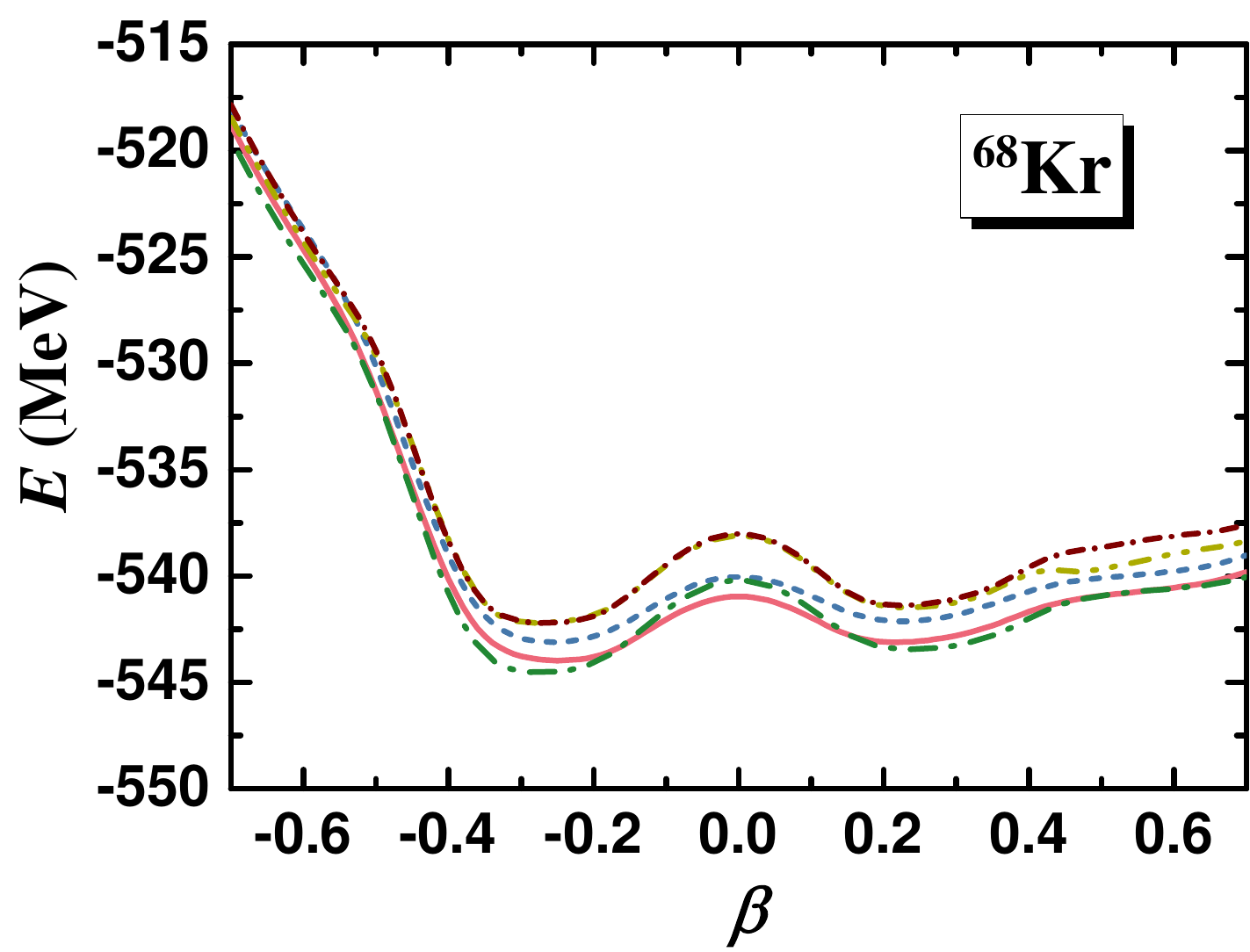}
\end{minipage}
\begin{minipage}{0.23\linewidth}
\includegraphics[width=\linewidth, angle=0]{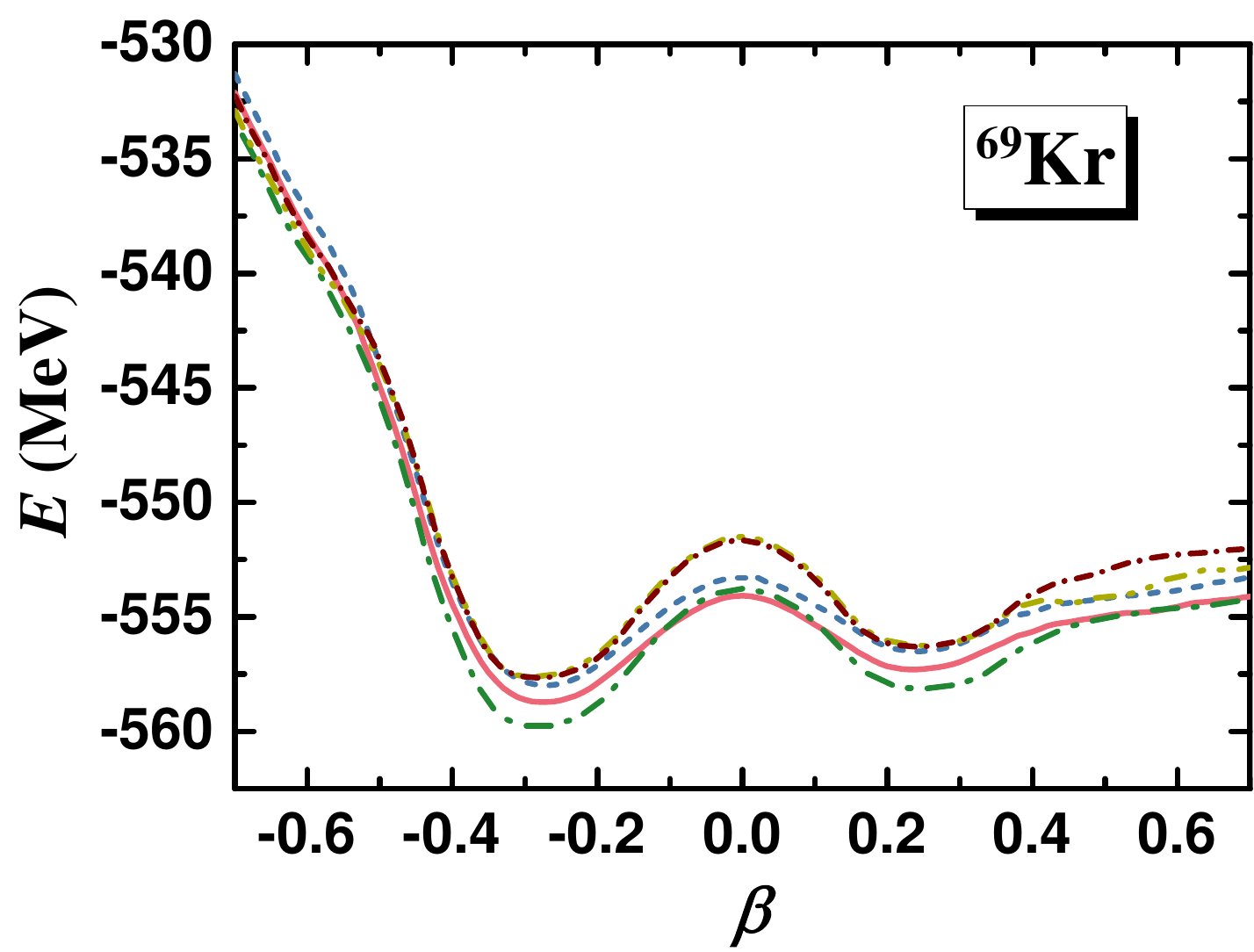}
\end{minipage}
\begin{minipage}{0.23\linewidth}
\includegraphics[width=\linewidth, angle=0]{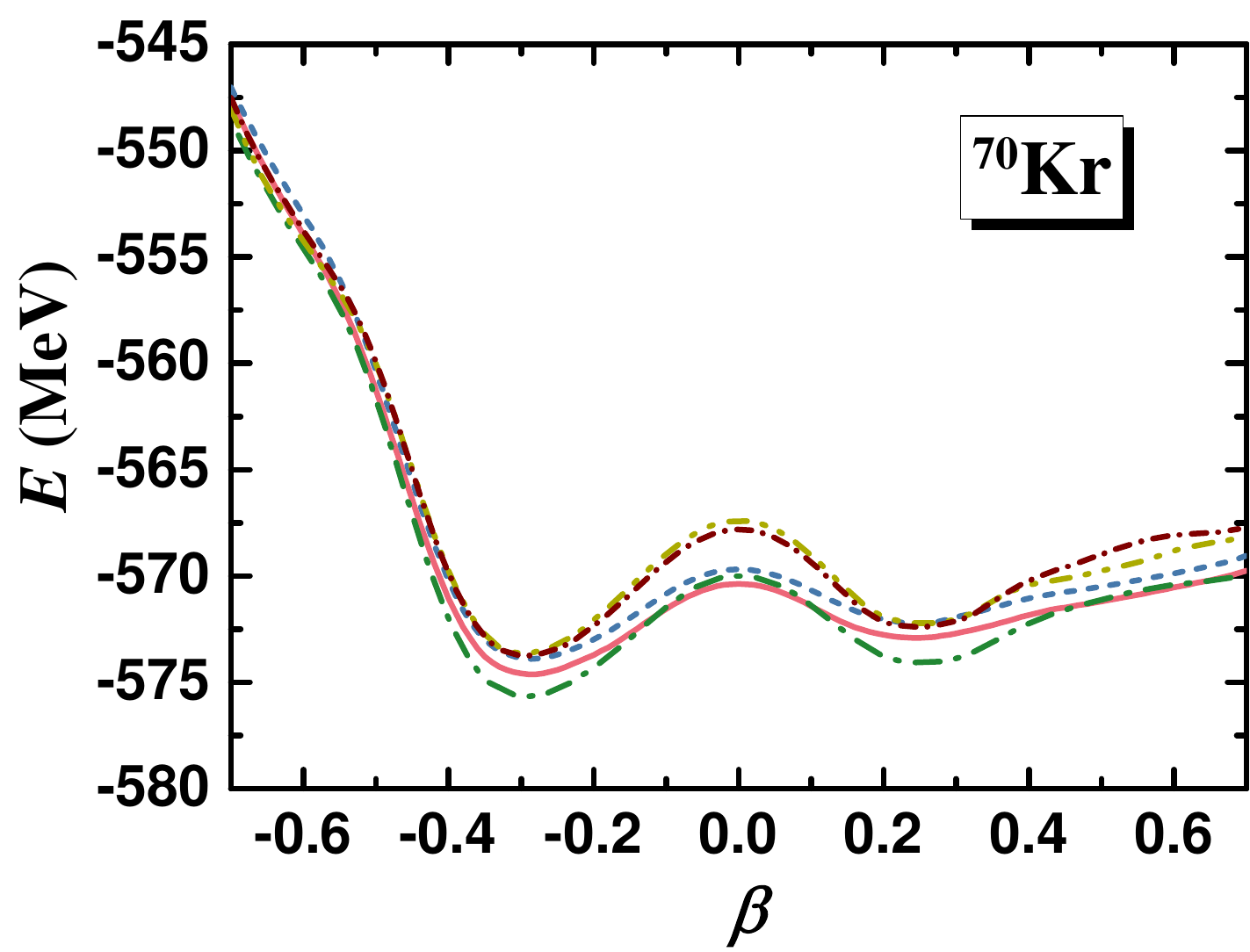}
\end{minipage}%
\vspace{-1mm}
\begin{minipage}{0.23\linewidth}
\centering
\includegraphics[width=\linewidth, angle=0]{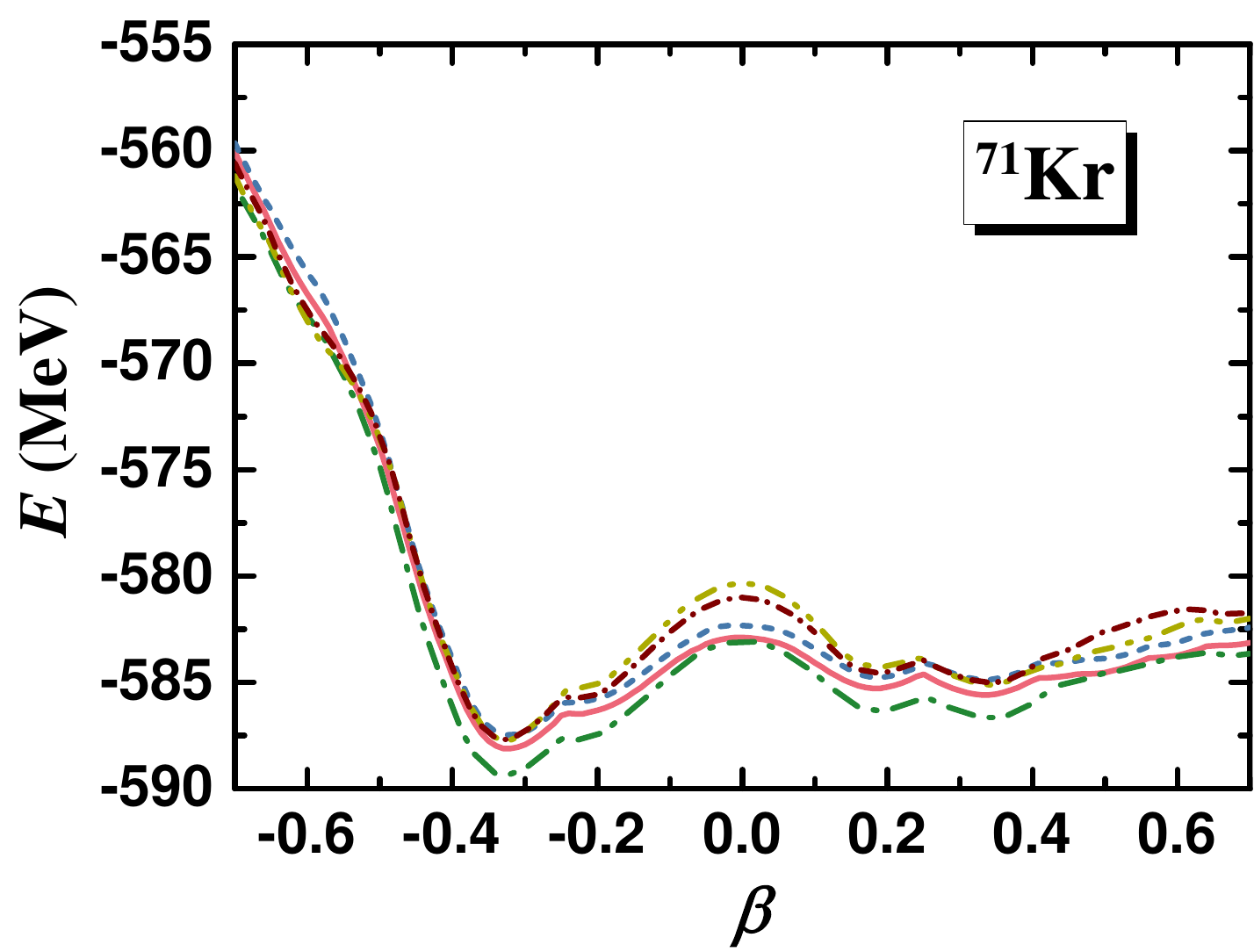}
\end{minipage}%
\begin{minipage}{0.23\linewidth}
\includegraphics[width=\linewidth, angle=0]{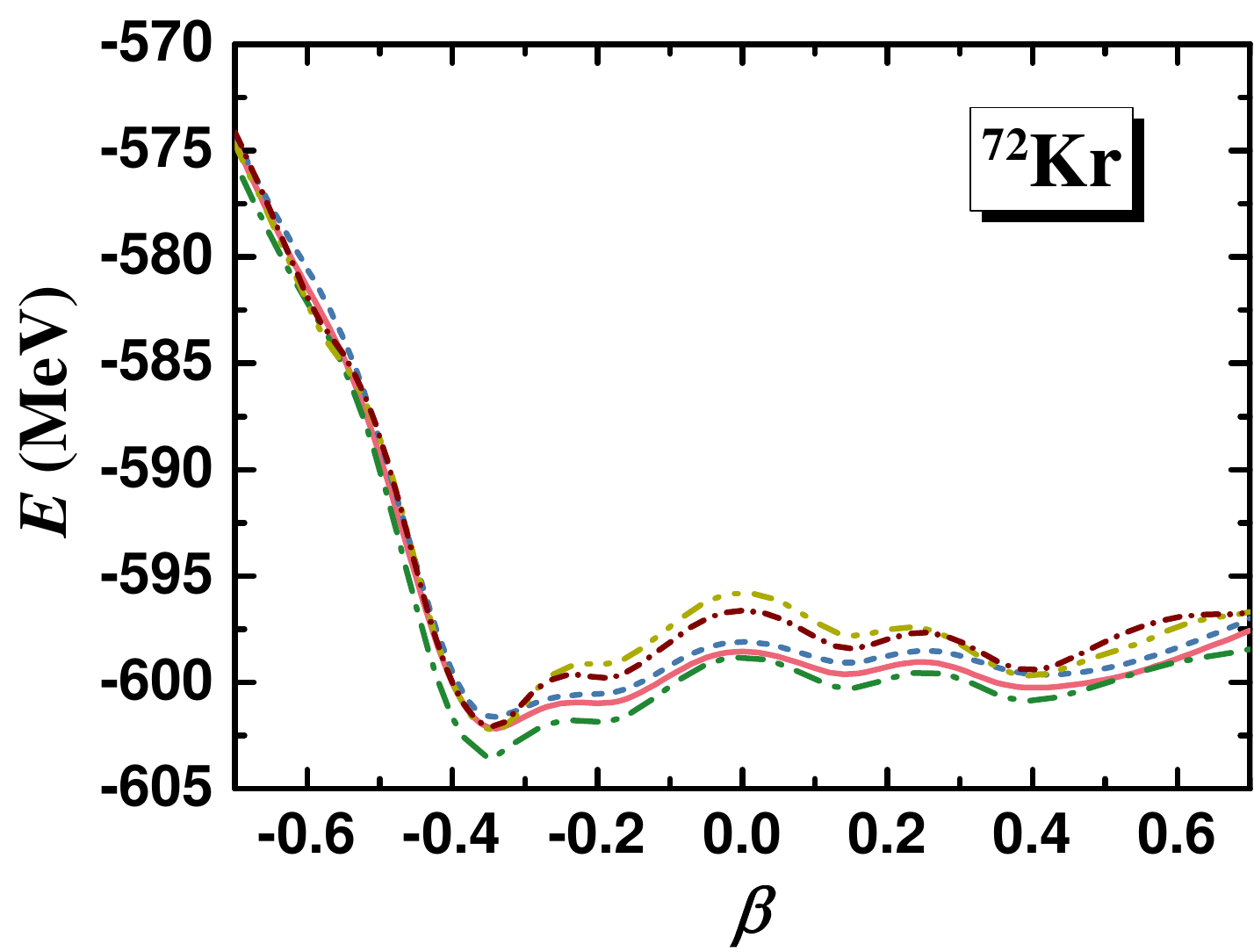}
\end{minipage}
\begin{minipage}{0.23\linewidth}
\includegraphics[width=\linewidth, angle=0]{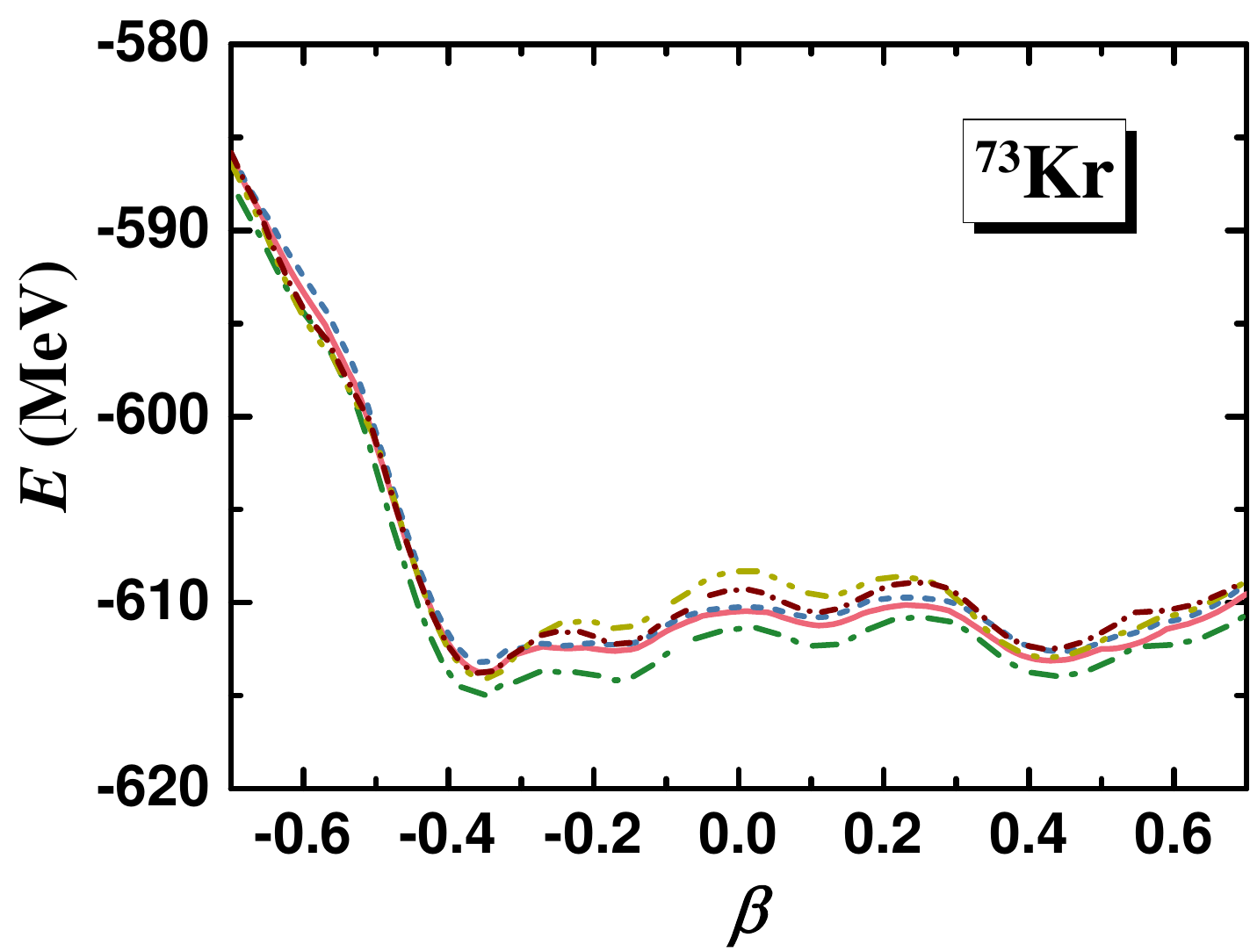}
\end{minipage}
\begin{minipage}{0.23\linewidth}
\includegraphics[width=\linewidth, angle=0]{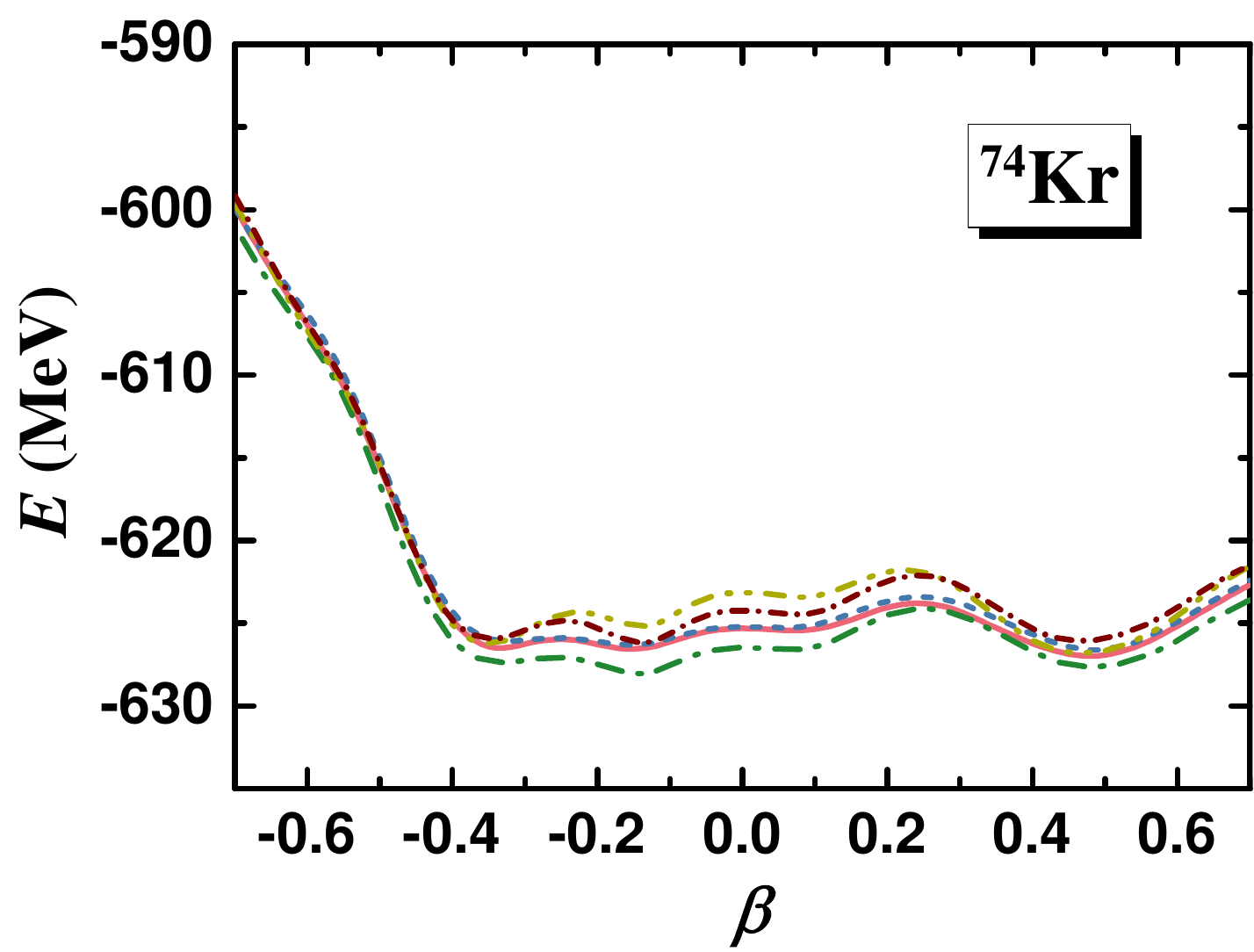}
\end{minipage}%
\vspace{-1mm}
\begin{minipage}{0.23\linewidth}
\centering
\includegraphics[width=\linewidth, angle=0]{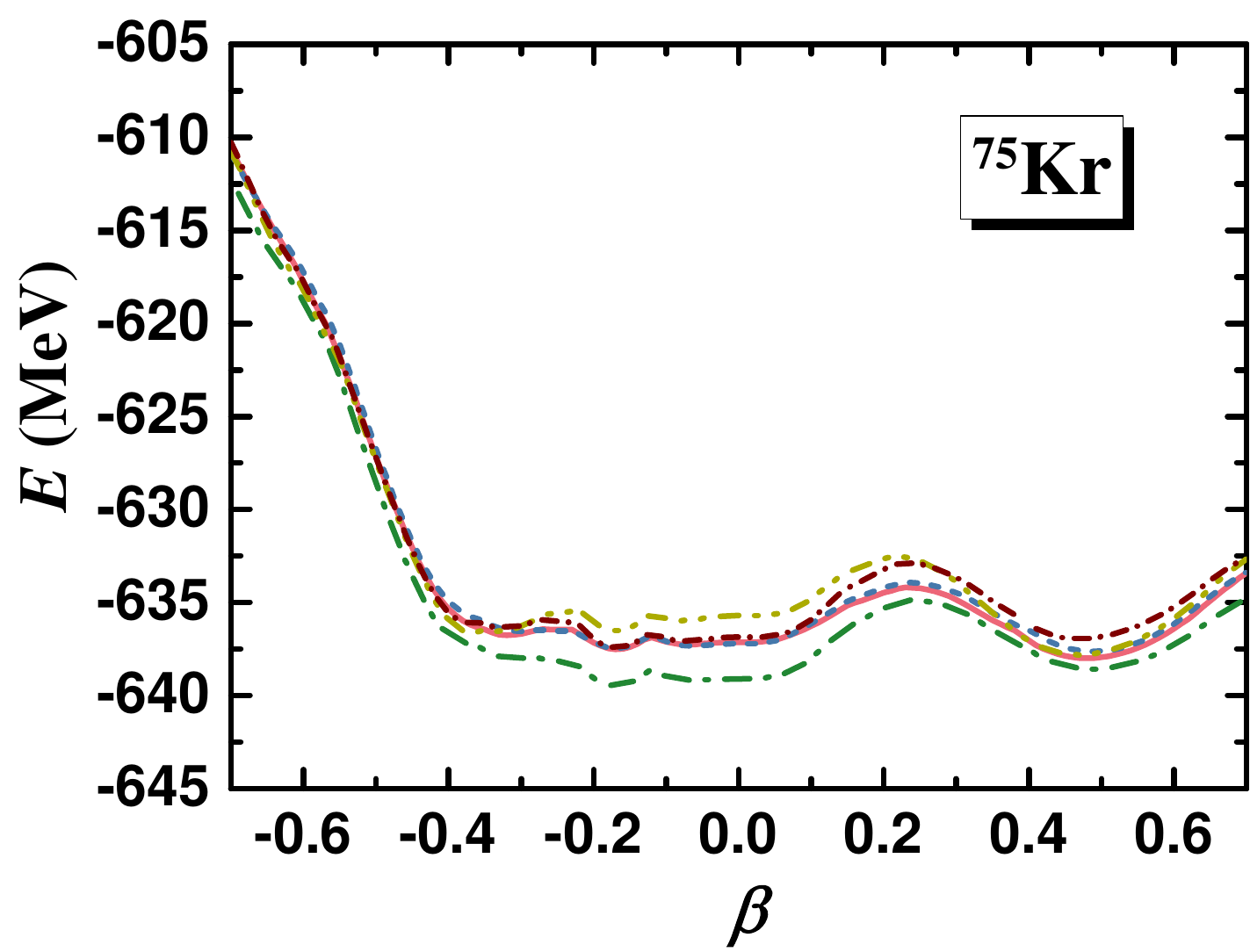}
\end{minipage}%
\begin{minipage}{0.23\linewidth}
\includegraphics[width=\linewidth, angle=0]{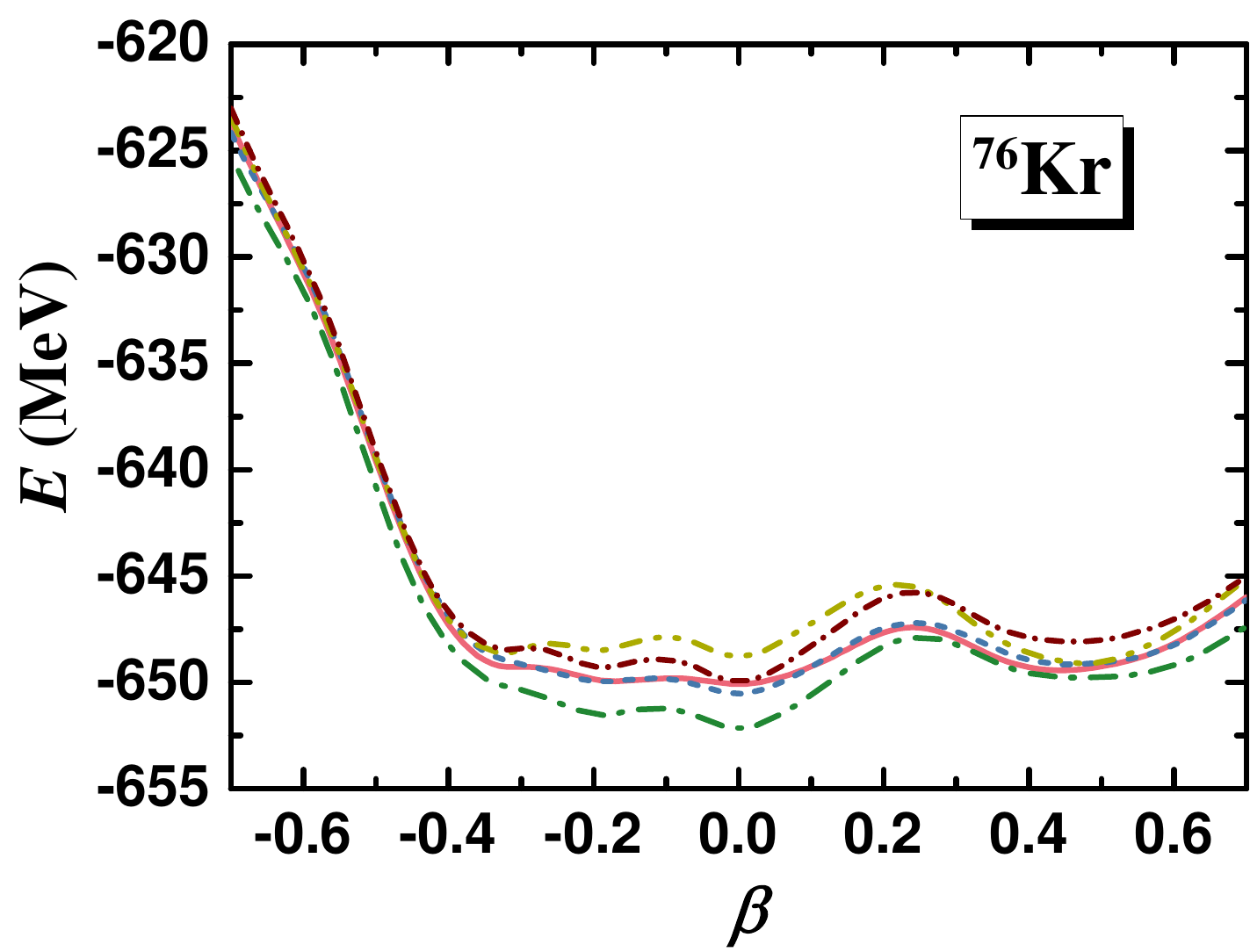}
\end{minipage}
\begin{minipage}{0.23\linewidth}
\includegraphics[width=\linewidth, angle=0]{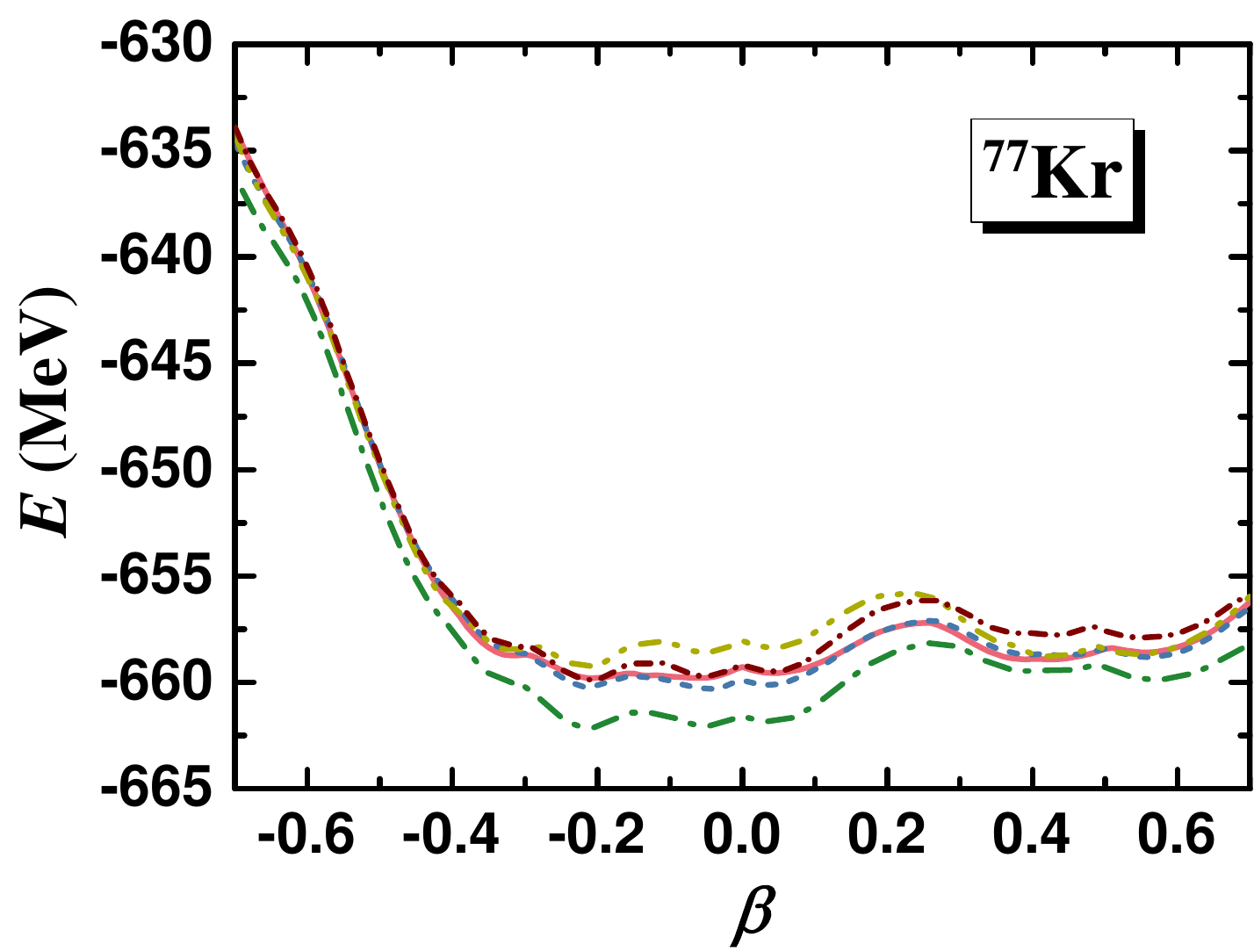}
\end{minipage}
\begin{minipage}{0.23\linewidth}
\includegraphics[width=\linewidth, angle=0]{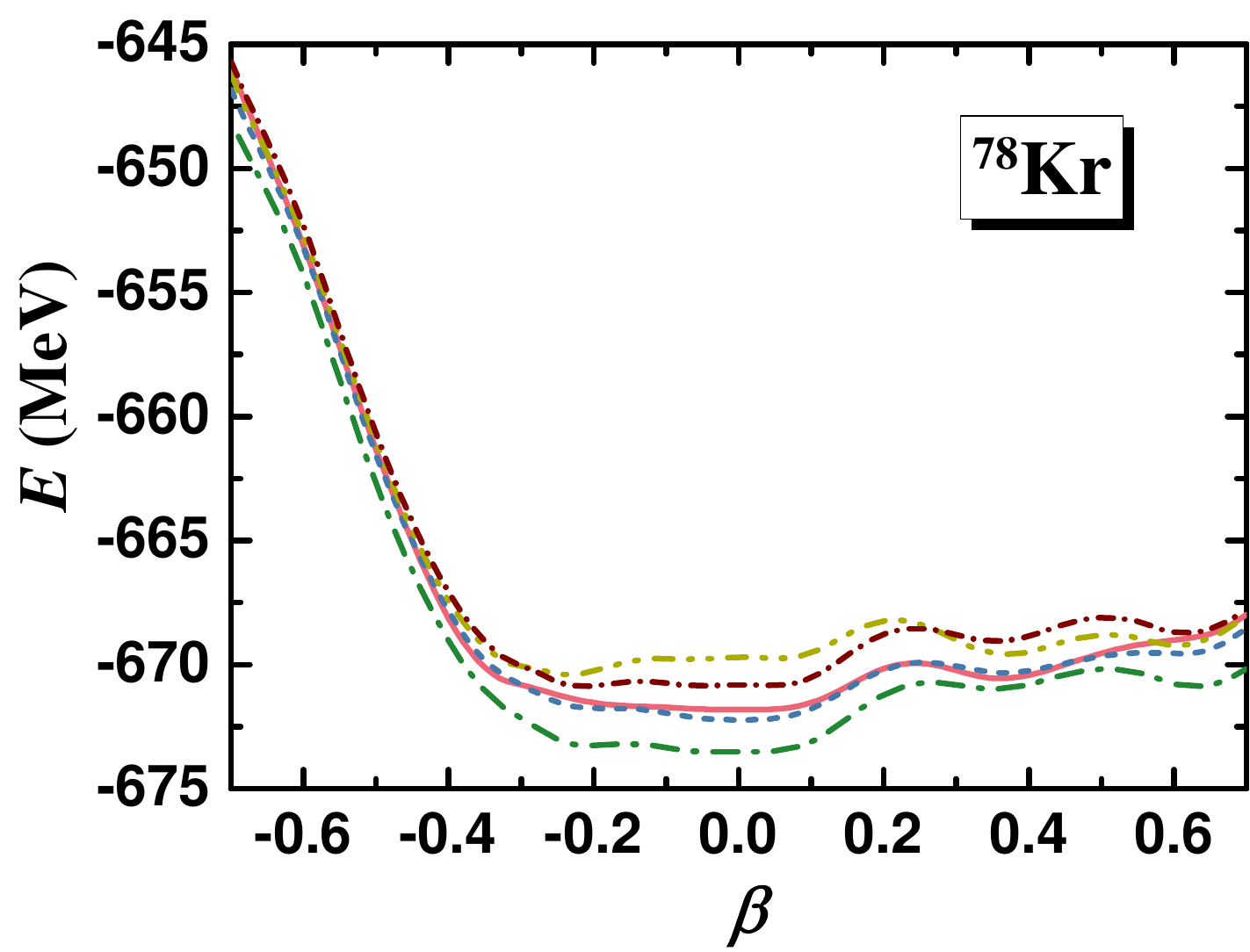}
\end{minipage}%
\vspace{-1mm}
\begin{minipage}{0.23\linewidth}
\centering
\includegraphics[width=\linewidth, angle=0]{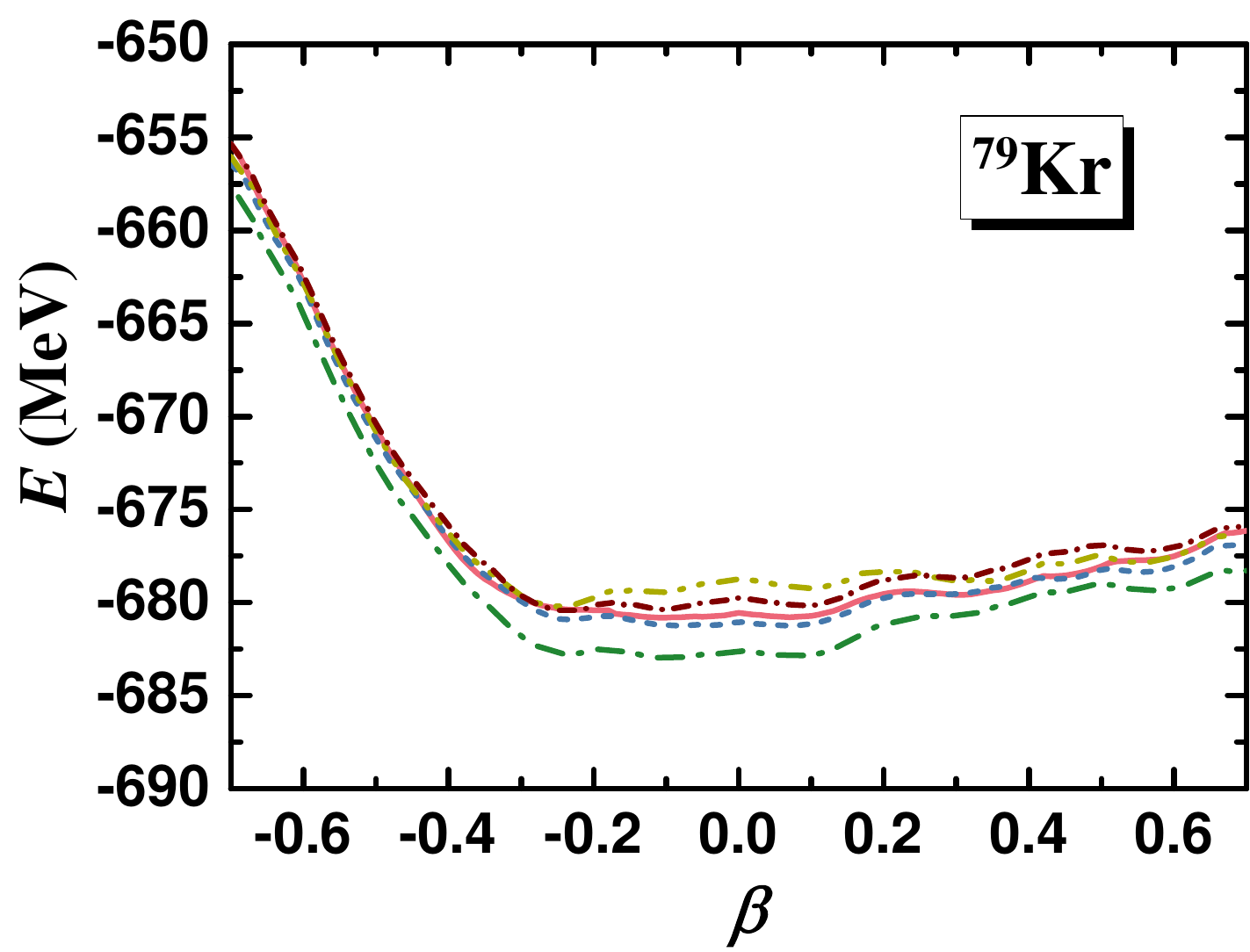}
\end{minipage}%
\begin{minipage}{0.23\linewidth}
\includegraphics[width=\linewidth, angle=0]{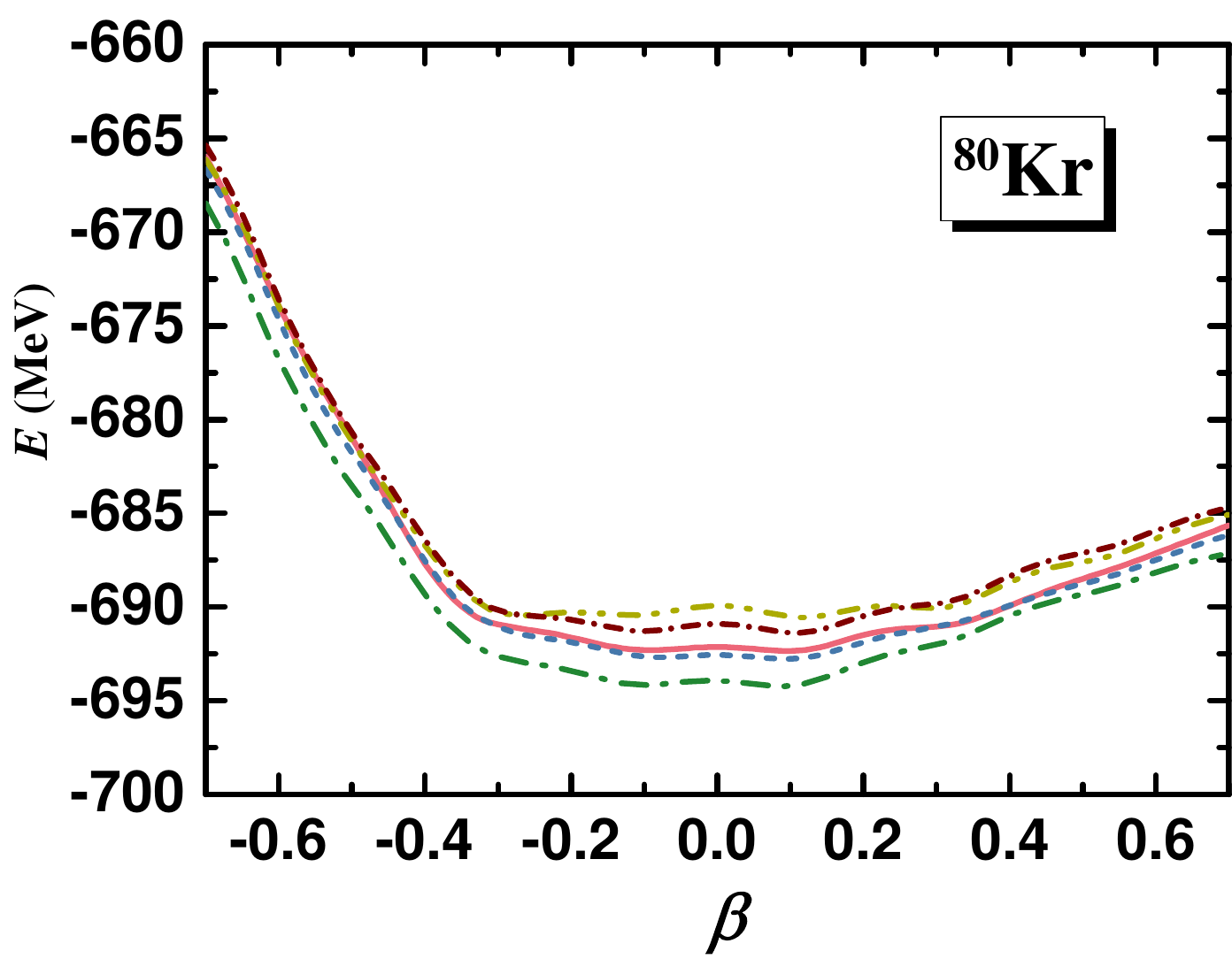}
\end{minipage}
\begin{minipage}{0.23\linewidth}
\includegraphics[width=\linewidth, angle=0]{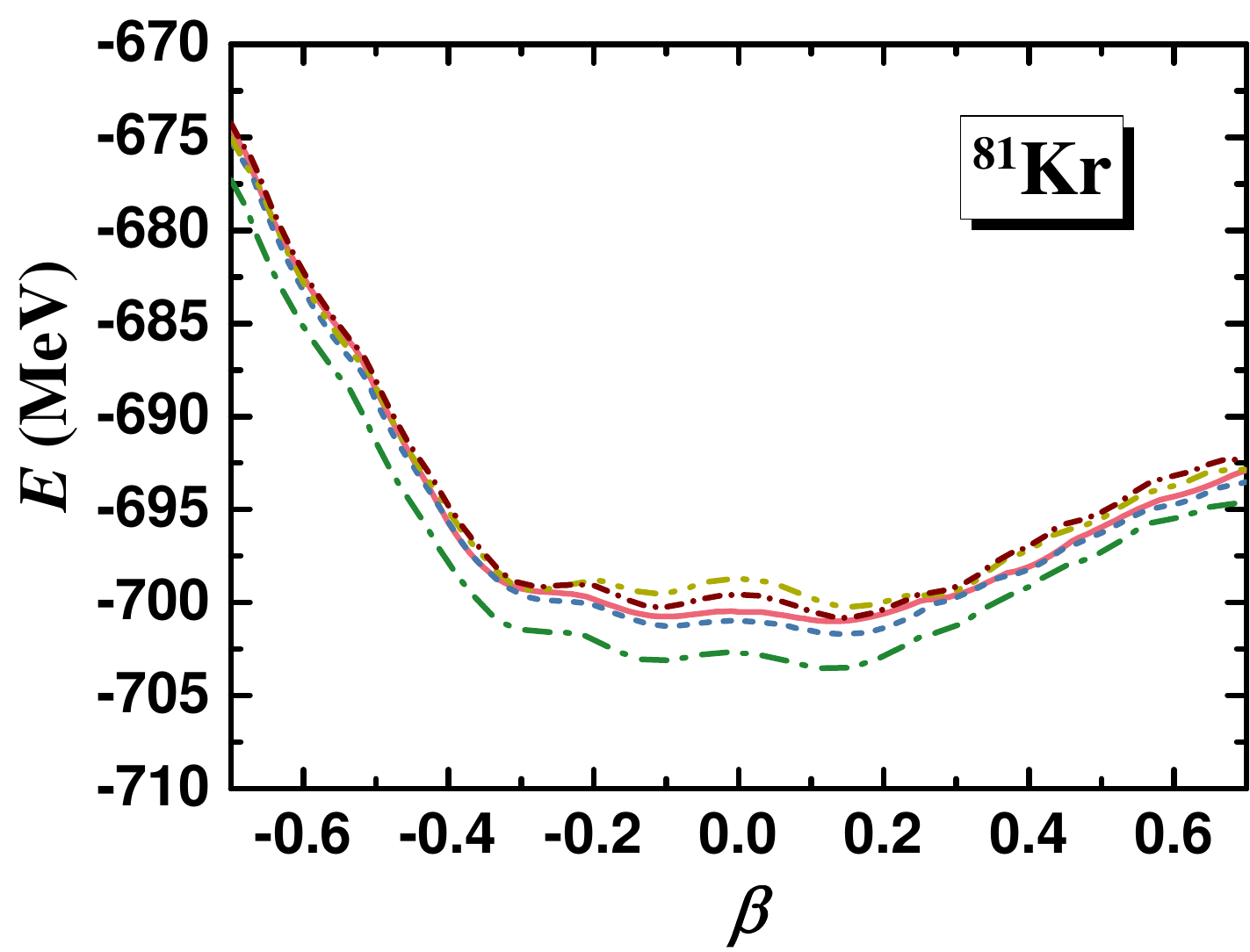}
\end{minipage}
\begin{minipage}{0.23\linewidth}
\includegraphics[width=\linewidth, angle=0]{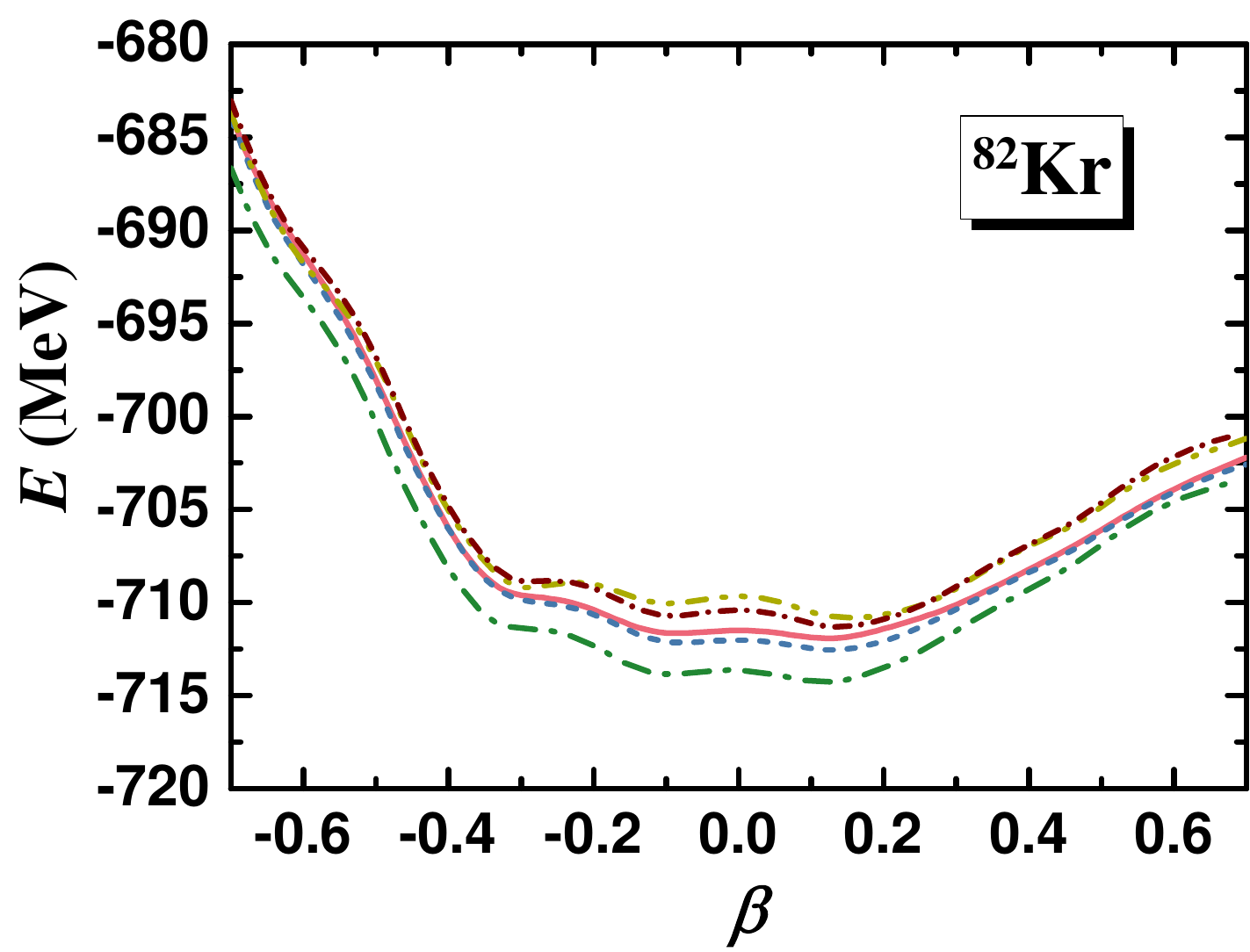}
\end{minipage}%
\vspace{-1mm}
\begin{minipage}{0.23\linewidth}
\centering
\includegraphics[width=\linewidth, angle=0]{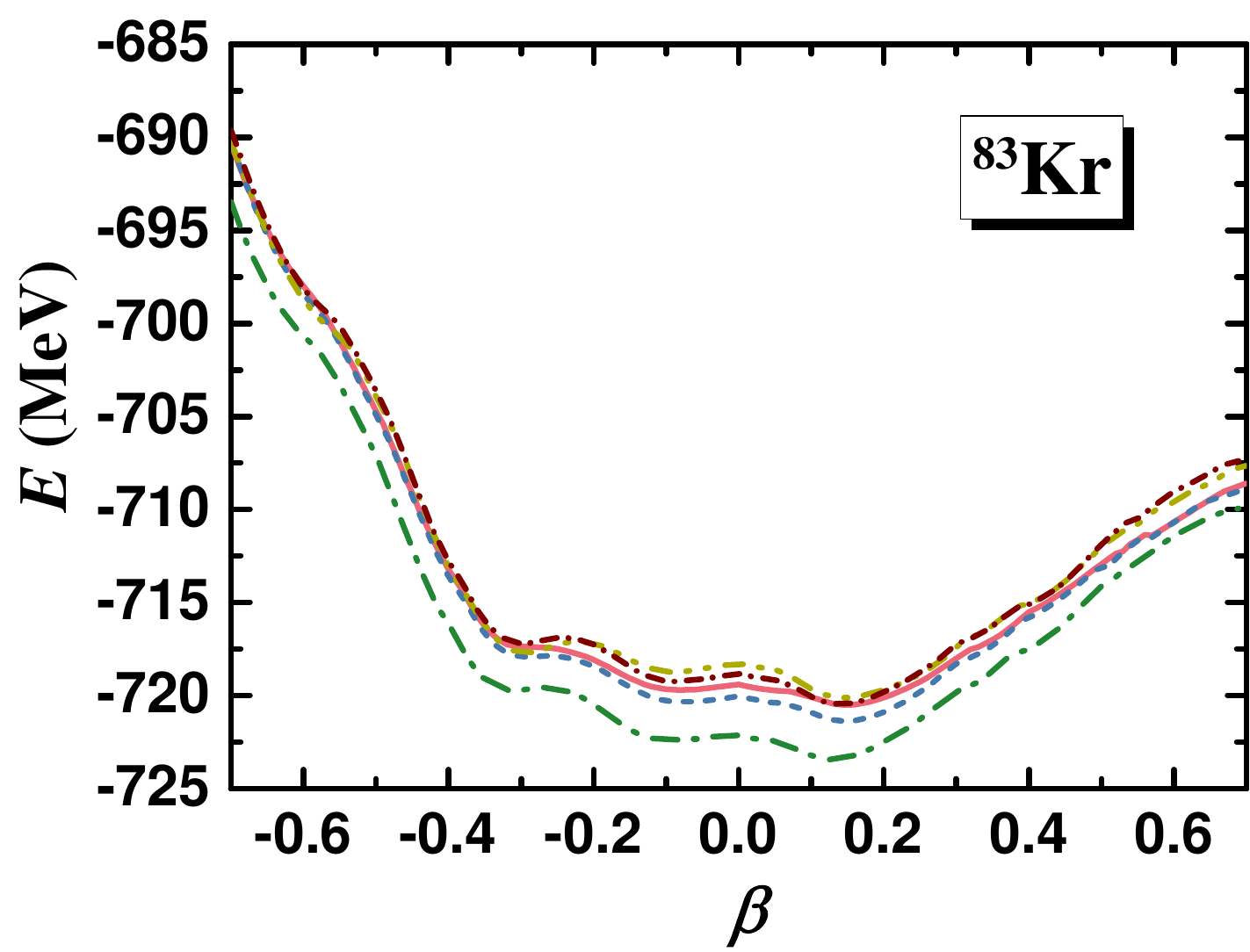}
\end{minipage}%
\begin{minipage}{0.23\linewidth}
\includegraphics[width=\linewidth, angle=0]{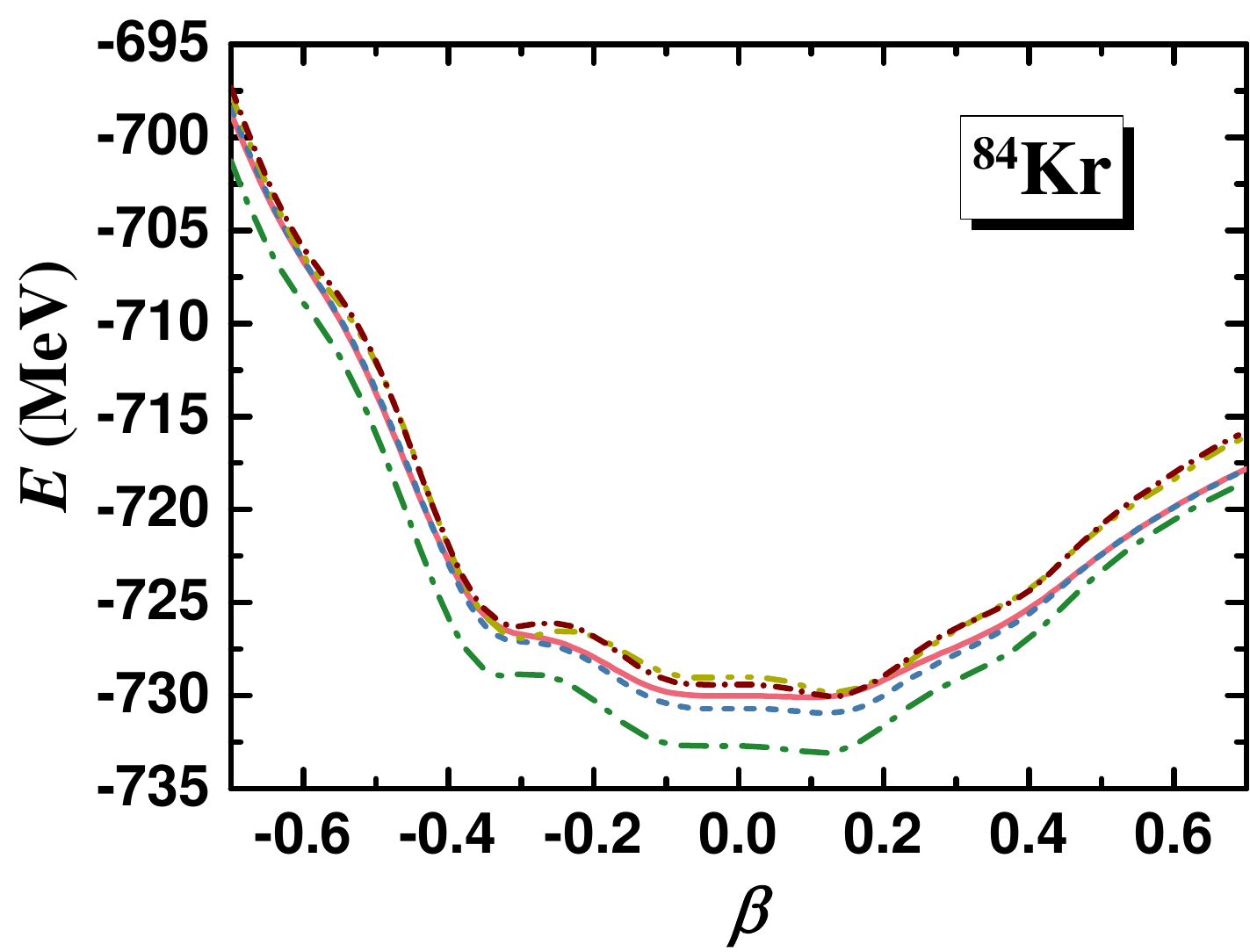}
\end{minipage}
\begin{minipage}{0.23\linewidth}
\includegraphics[width=\linewidth, angle=0]{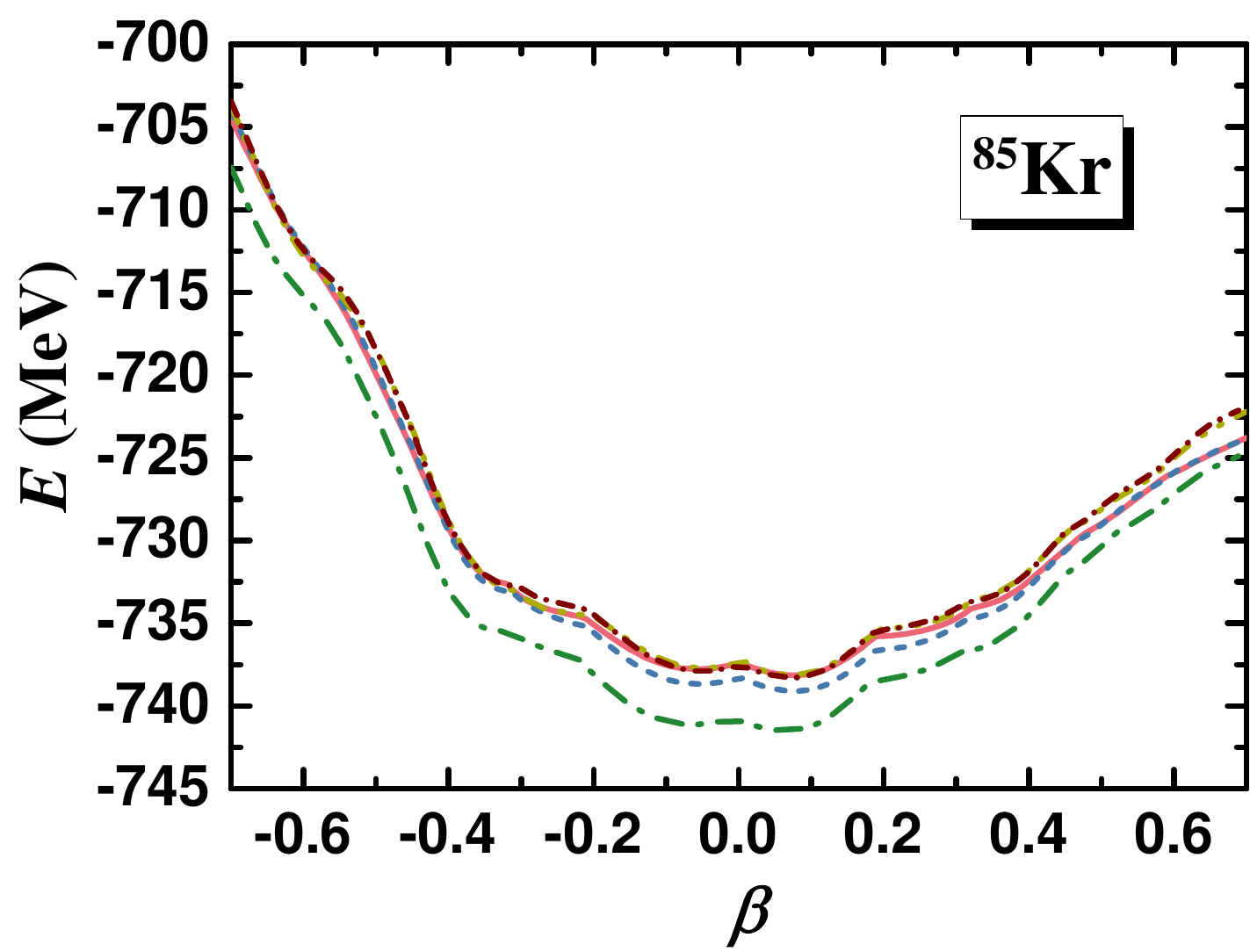}
\end{minipage}
\begin{minipage}{0.23\linewidth}
\includegraphics[width=\linewidth, angle=0]{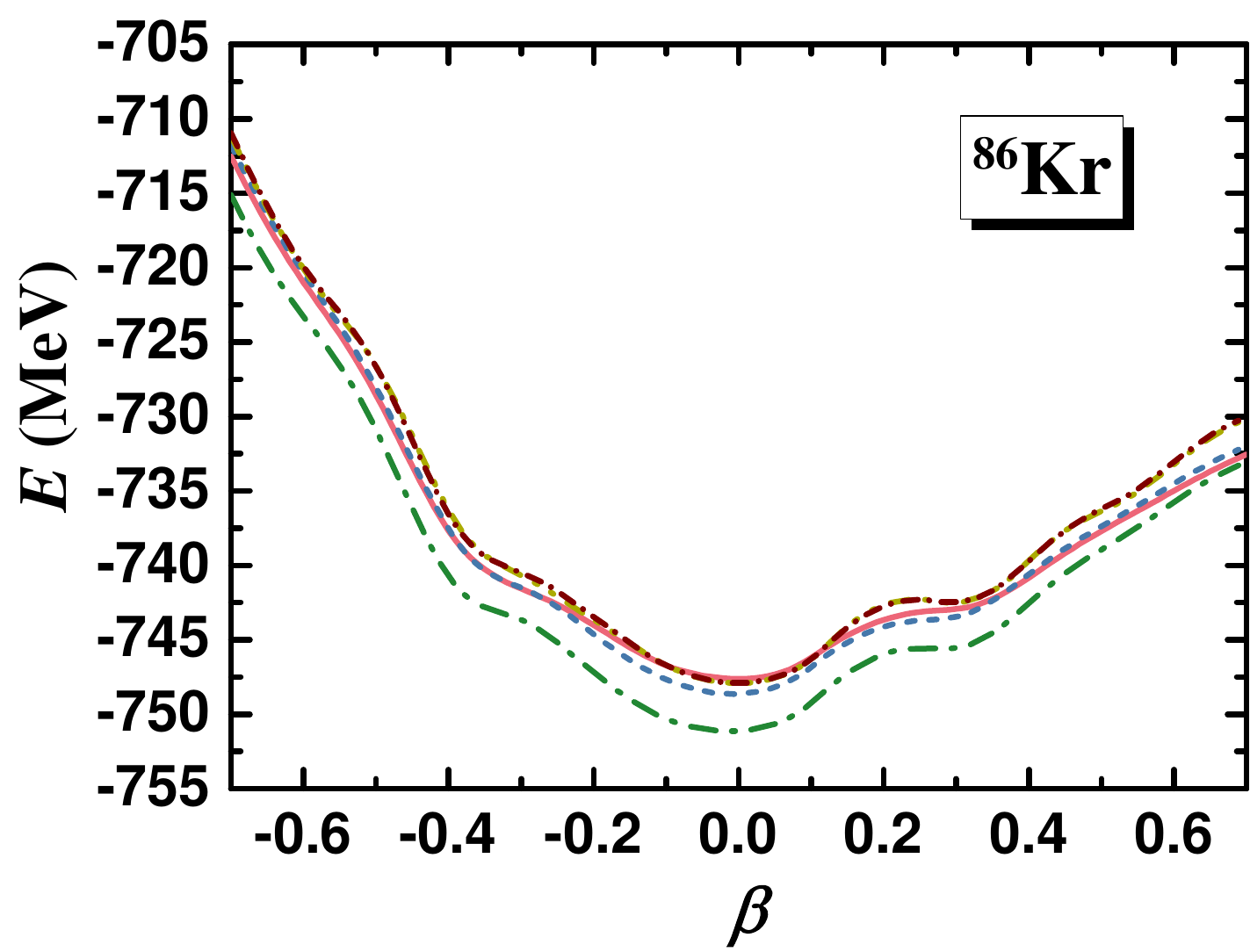}
\end{minipage}%
\vspace{-1mm}
\begin{minipage}{0.23\linewidth}
\centering
\includegraphics[width=\linewidth, angle=0]{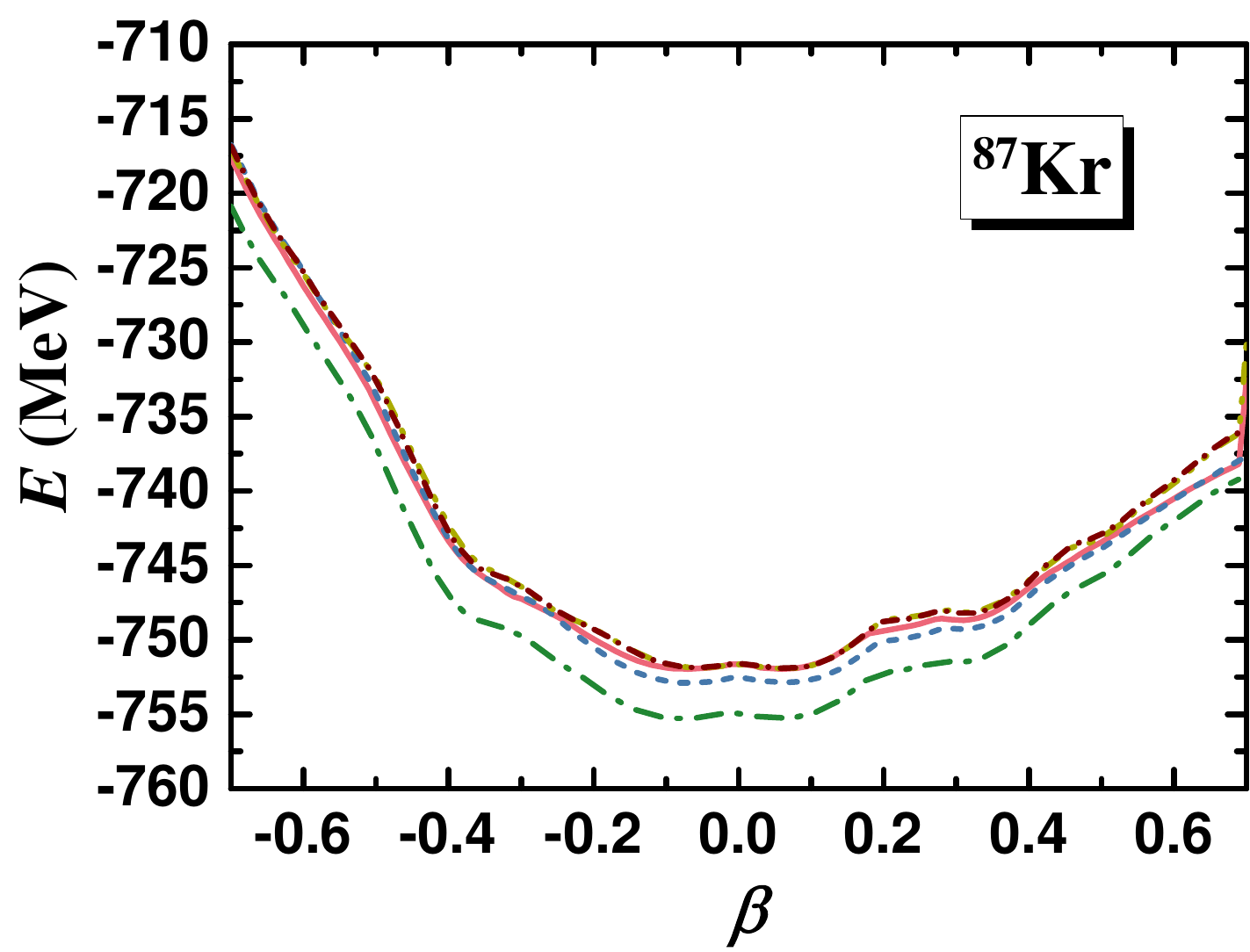}
\end{minipage}%
\begin{minipage}{0.23\linewidth}
\includegraphics[width=\linewidth, angle=0]{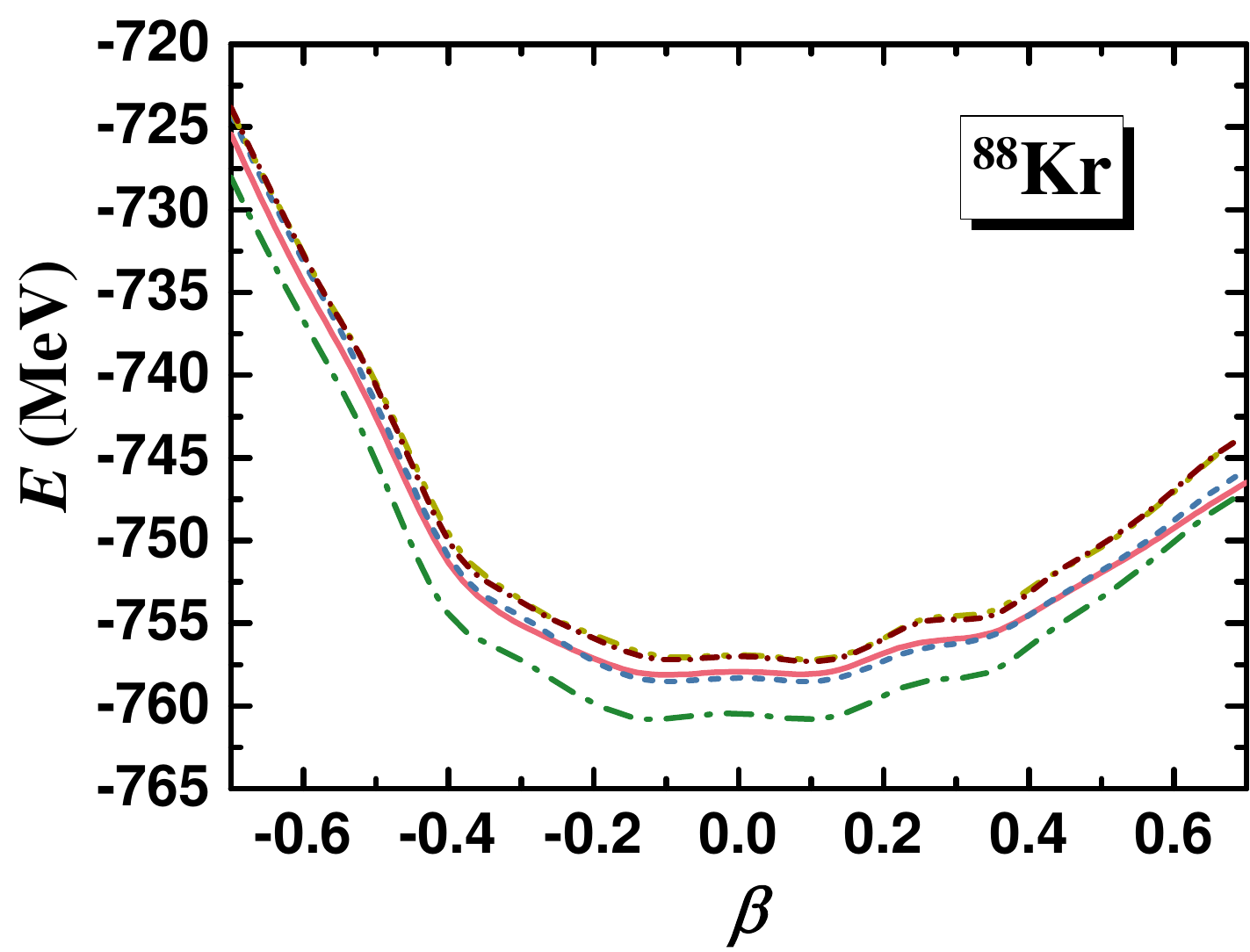}
\end{minipage}
\begin{minipage}{0.23\linewidth}
\includegraphics[width=\linewidth, angle=0]{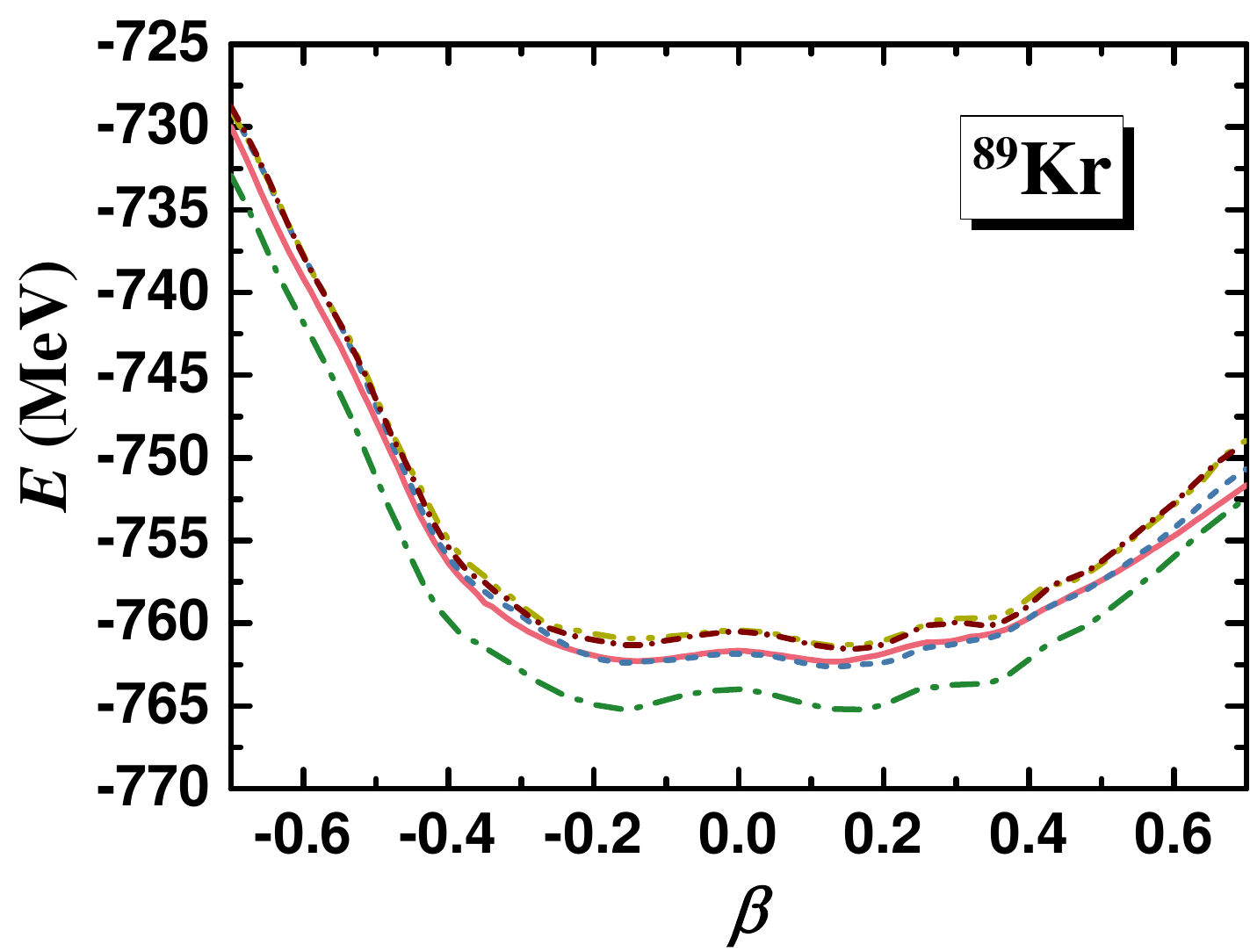}
\end{minipage}
\begin{minipage}{0.23\linewidth}
\includegraphics[width=\linewidth, angle=0]{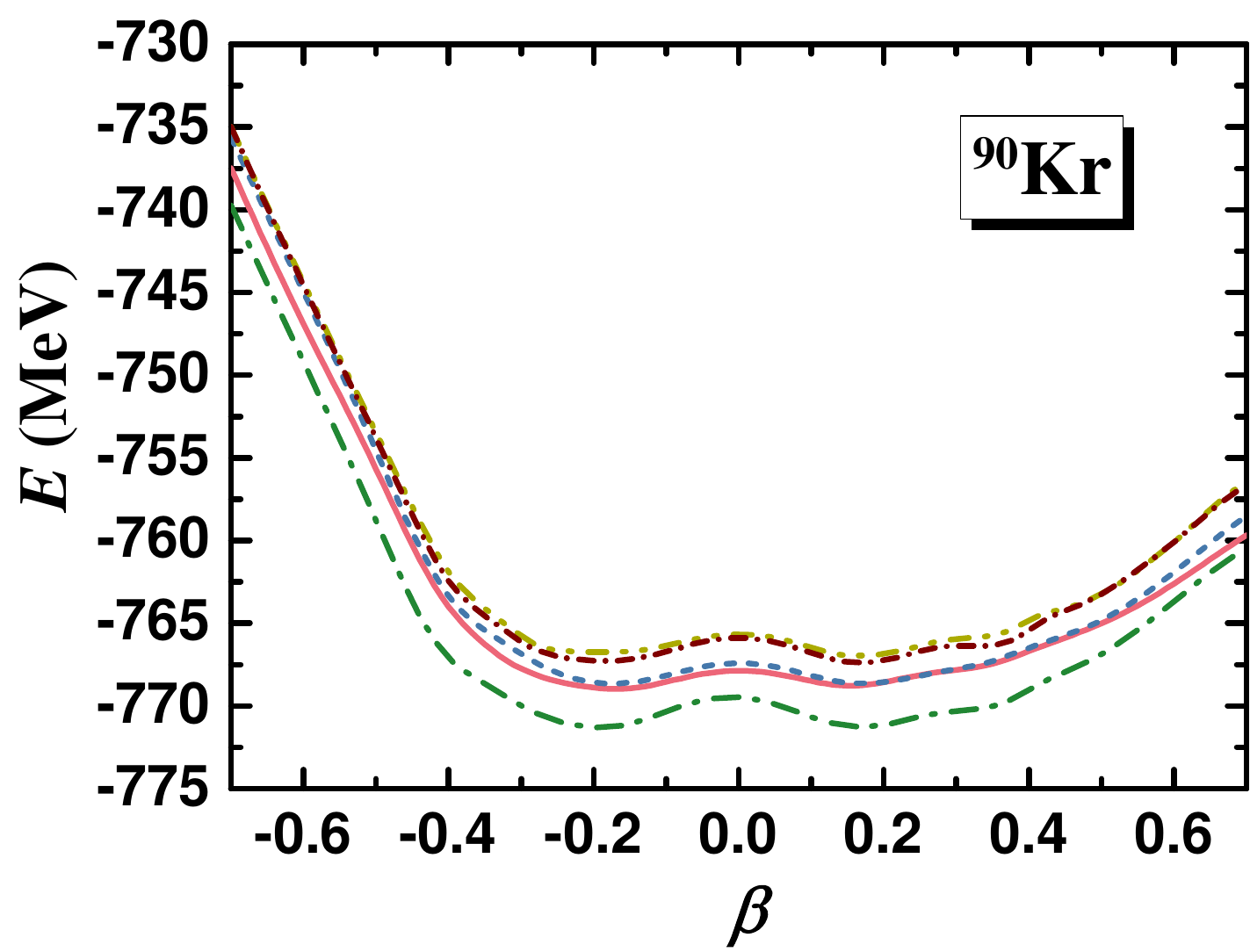}
\end{minipage}%
\vspace{-1mm}
\begin{minipage}{0.23\linewidth}
\centering
\includegraphics[width=\linewidth, angle=0]{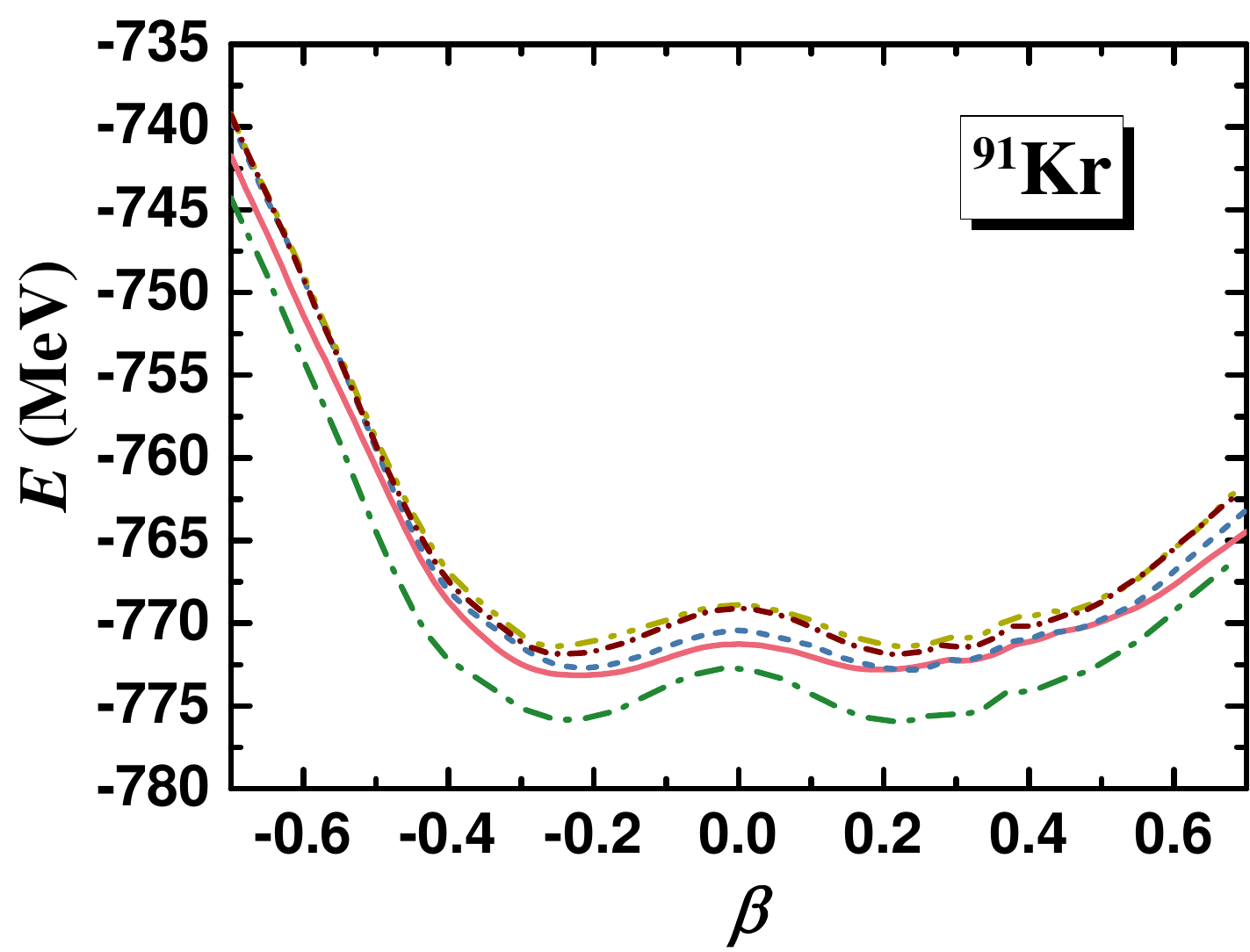}
\end{minipage}%
\begin{minipage}{0.23\linewidth}
\includegraphics[width=\linewidth, angle=0]{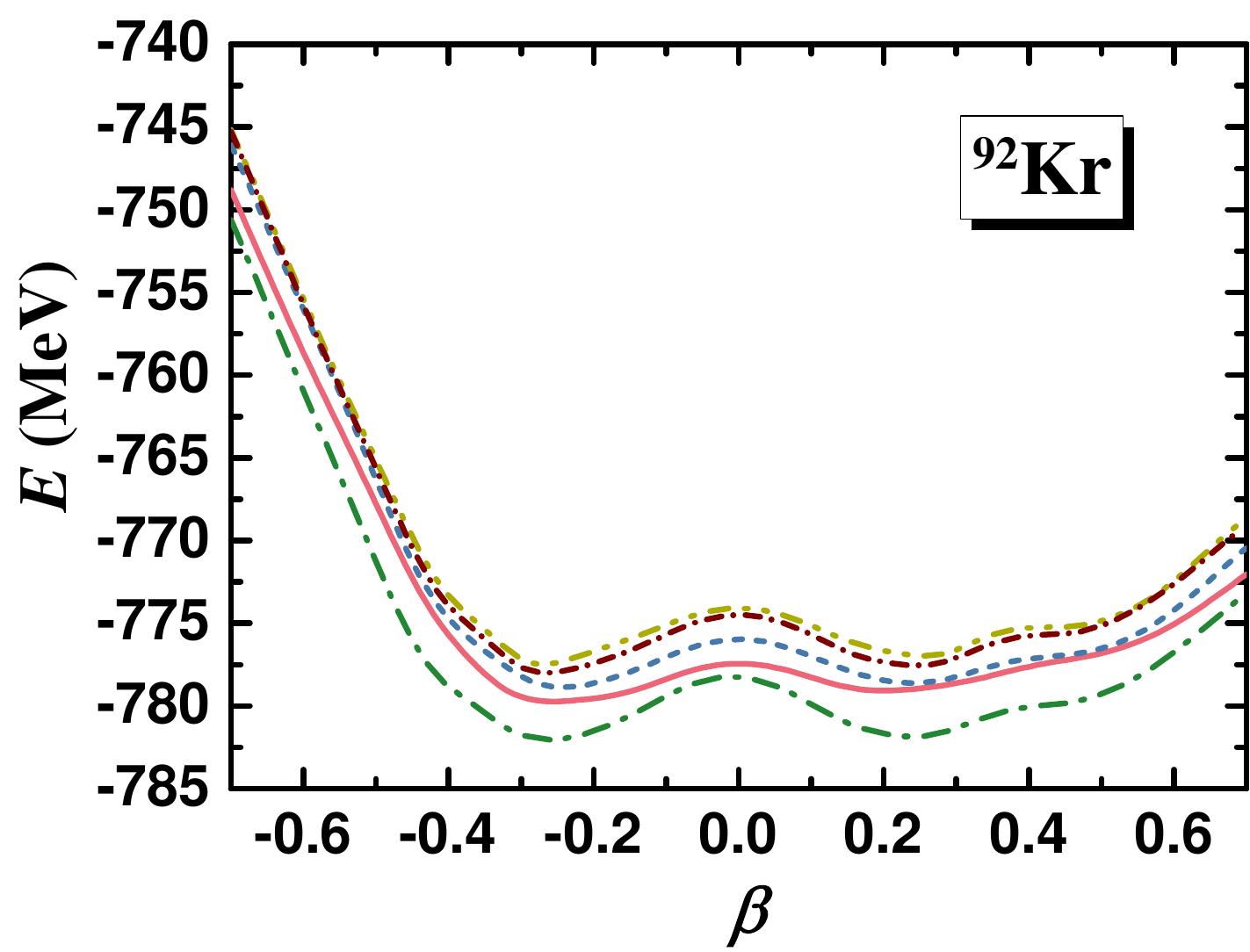}
\end{minipage}
\begin{minipage}{0.23\linewidth}
\includegraphics[width=\linewidth, angle=0]{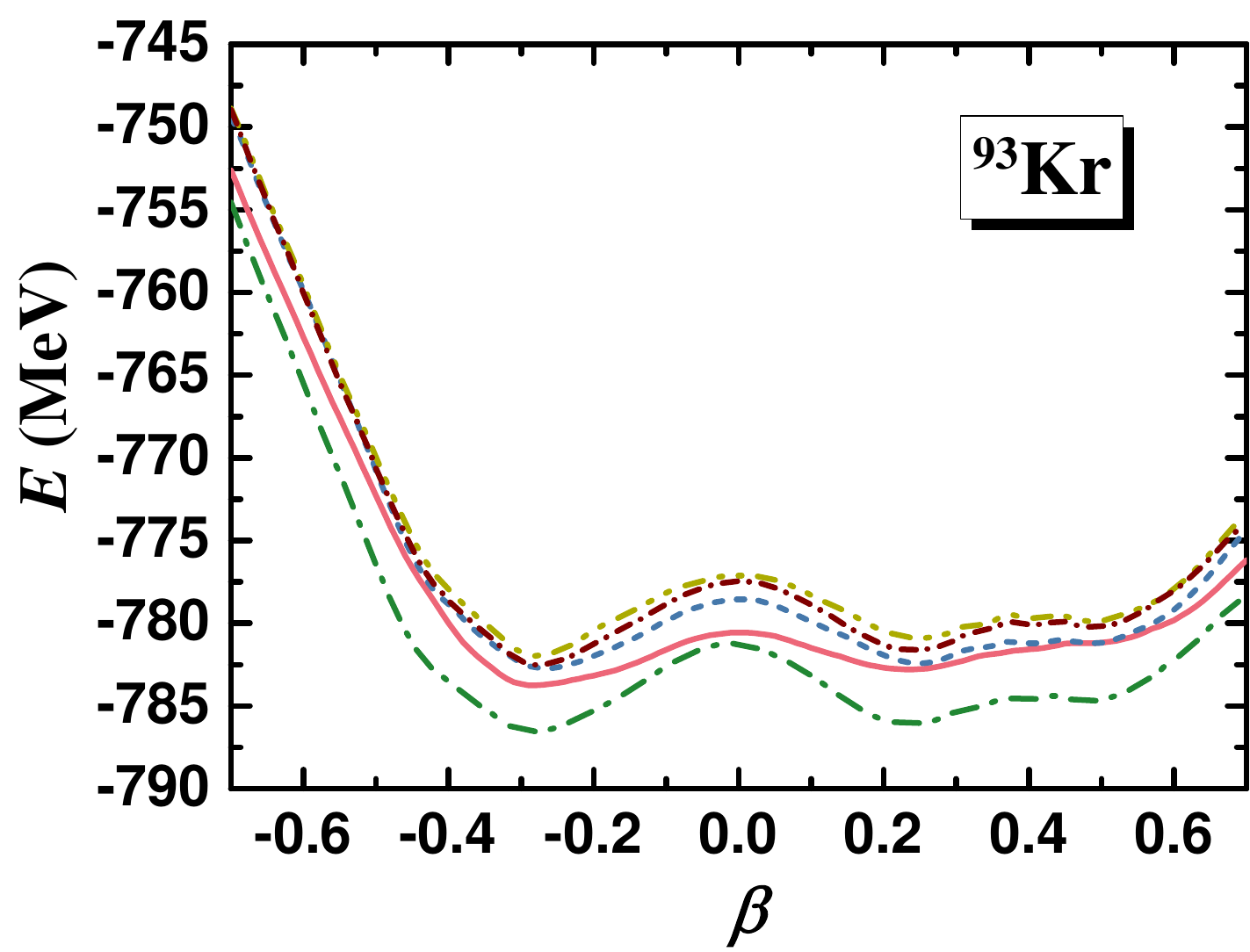}
\end{minipage}
\begin{minipage}{0.23\linewidth}
\includegraphics[width=\linewidth, angle=0]{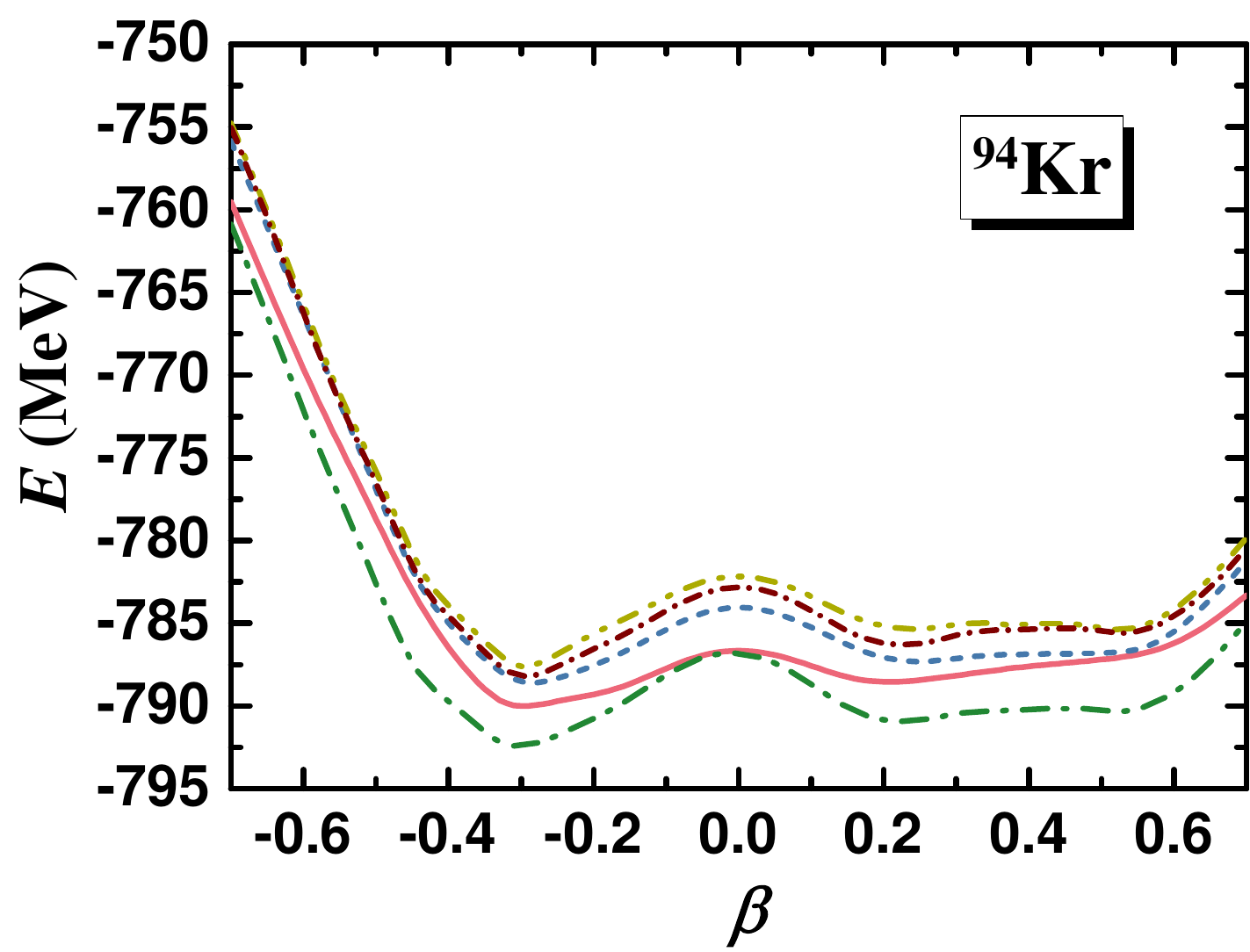}
\end{minipage}%
\vspace{-2mm}
\caption{Potential energy curves as functions of the quadrupole $\beta_{2}$ deformation for the $^{67\mathrm{-}94}$Kr isotopes computed with the relativistic Hartree-Bogoliubov approach with axially-symmetrical basis implementing the PC-L3R, DD-PCX, DD-PC1, DD-MEX, and DD-ME2 effective interactions and the respective separable pairing force of finite range.}
\label{fig:Kr-PES1}
\end{figure*}

\begin{figure*}%
\centering
\begin{minipage}{0.225\linewidth}
\centering
\includegraphics[width=\linewidth, angle=0]{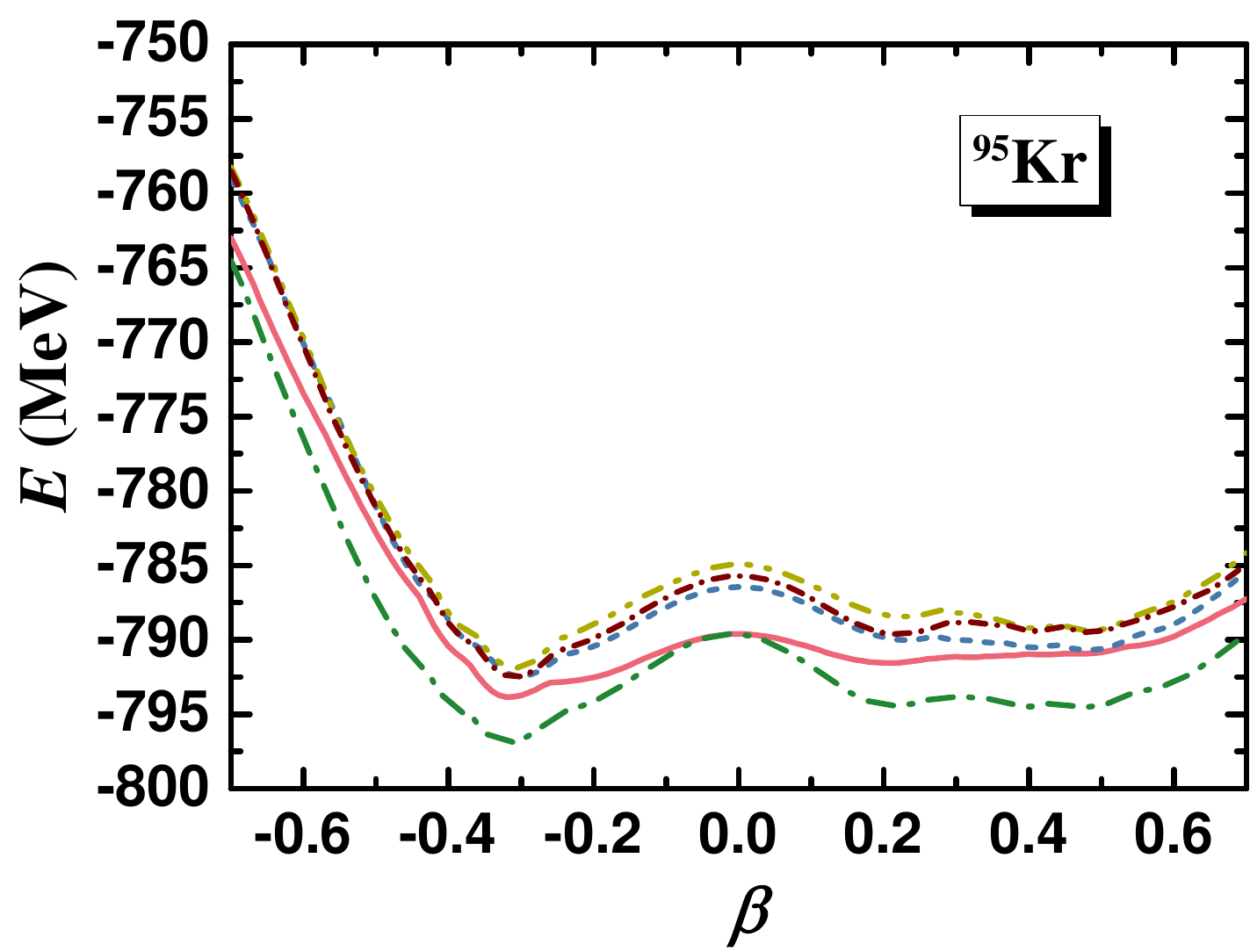}
\end{minipage}%
\begin{minipage}{0.23\linewidth}
\includegraphics[width=\linewidth, angle=0]{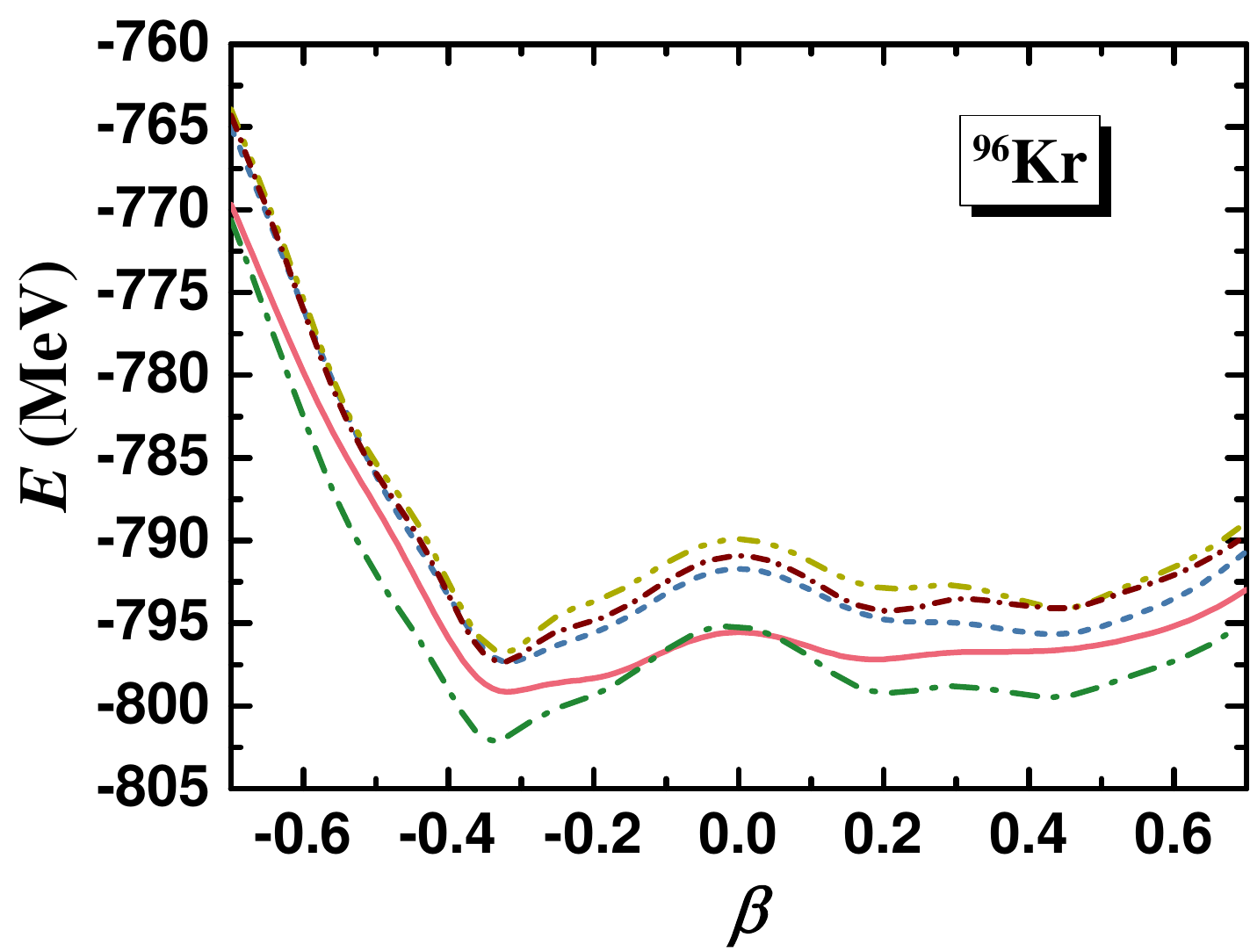}
\end{minipage}
\begin{minipage}{0.23\linewidth}
\includegraphics[width=\linewidth, angle=0]{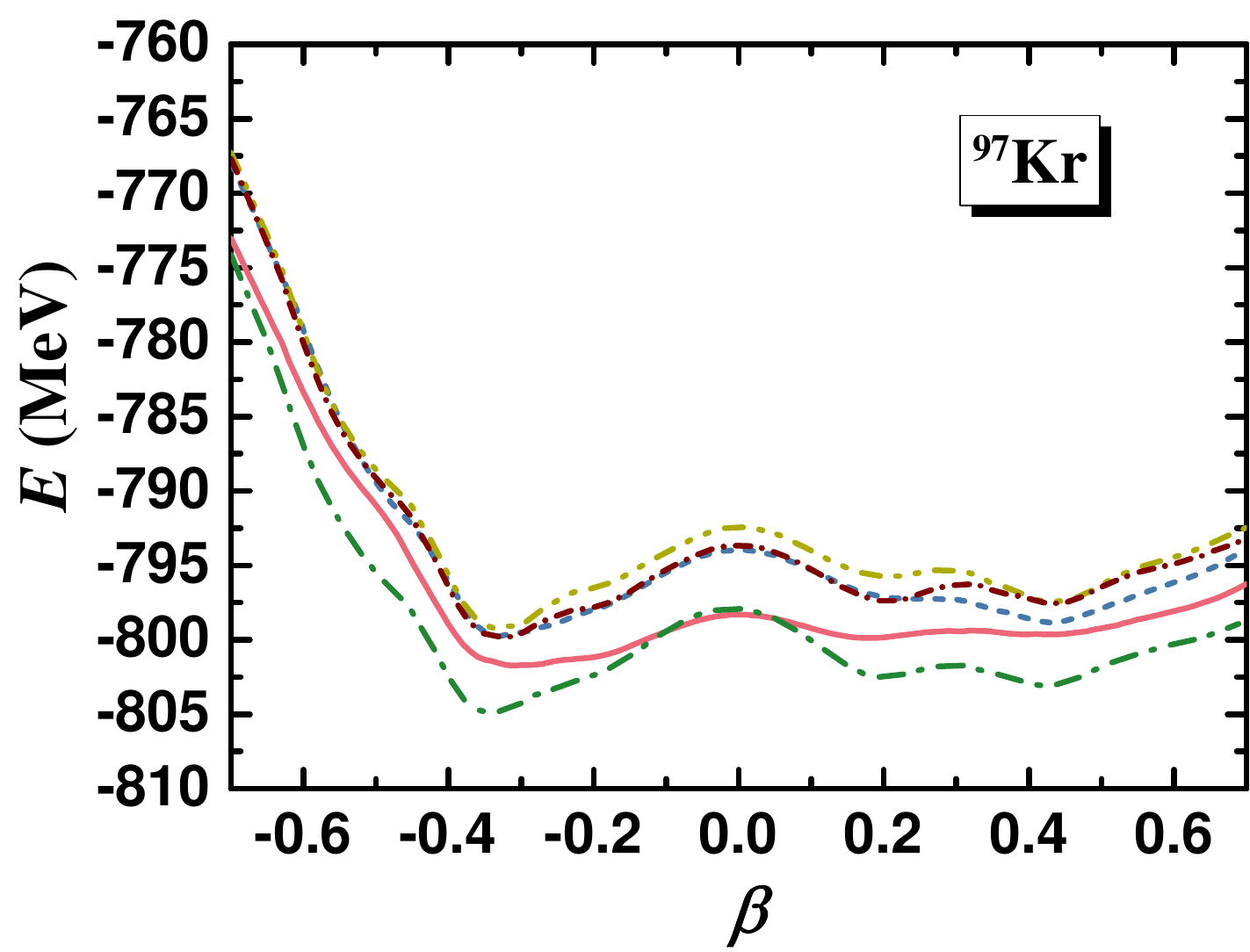}
\end{minipage}
\begin{minipage}{0.23\linewidth}
\includegraphics[width=\linewidth, angle=0]{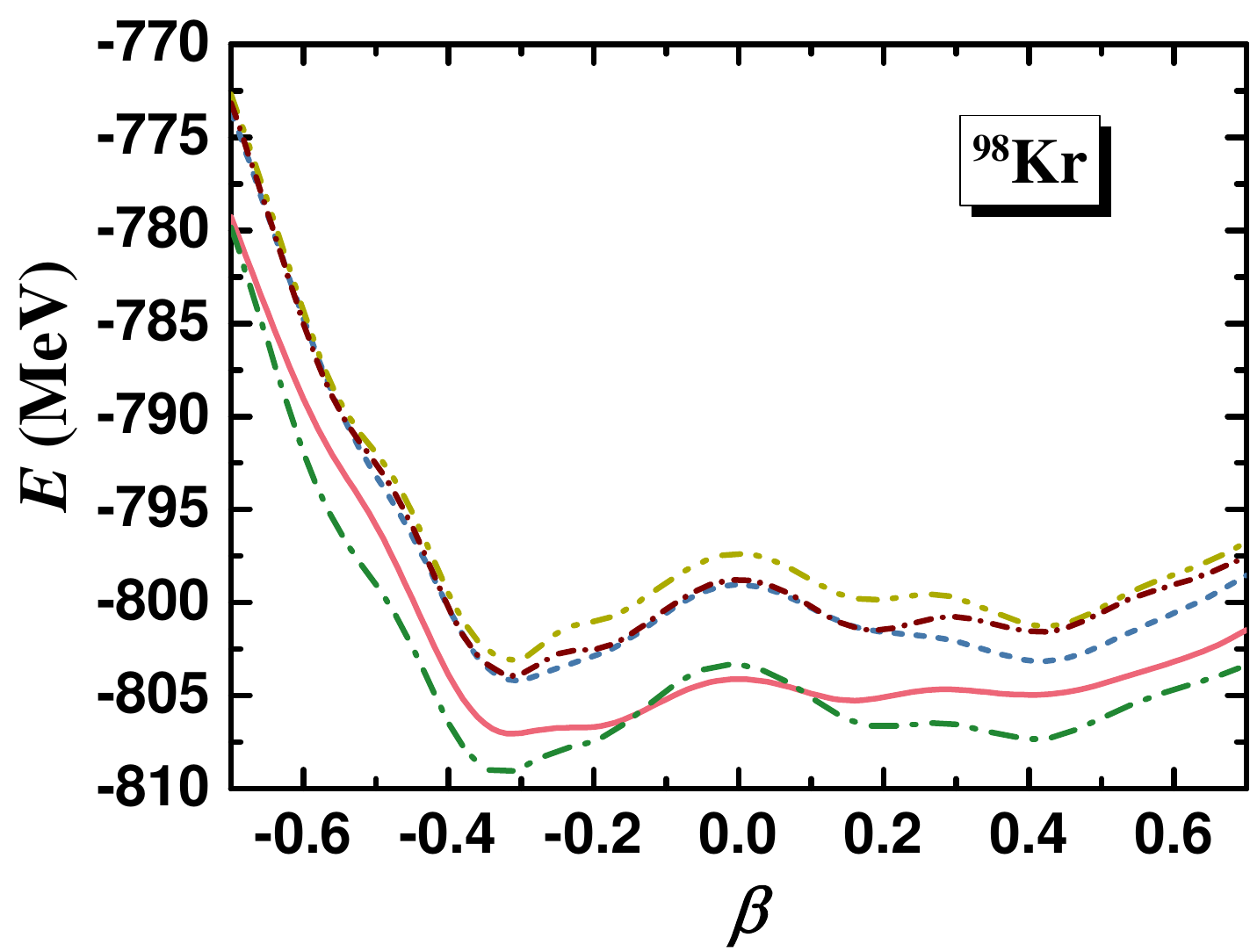}
\end{minipage}%
\vspace{-1mm}
\begin{minipage}{0.23\linewidth}
\centering
\includegraphics[width=\linewidth, angle=0]{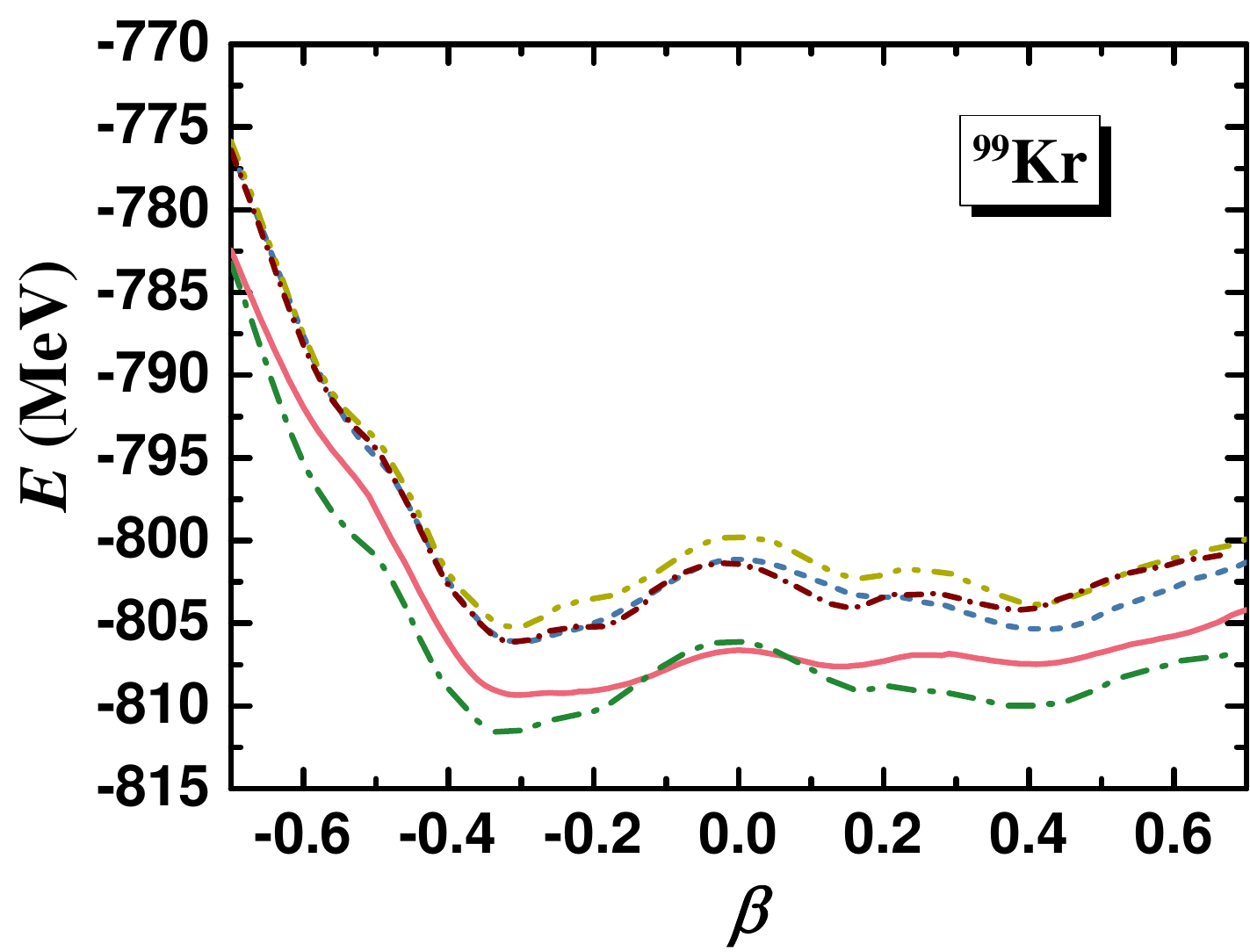}
\end{minipage}%
\begin{minipage}{0.23\linewidth}
\includegraphics[width=\linewidth, angle=0]{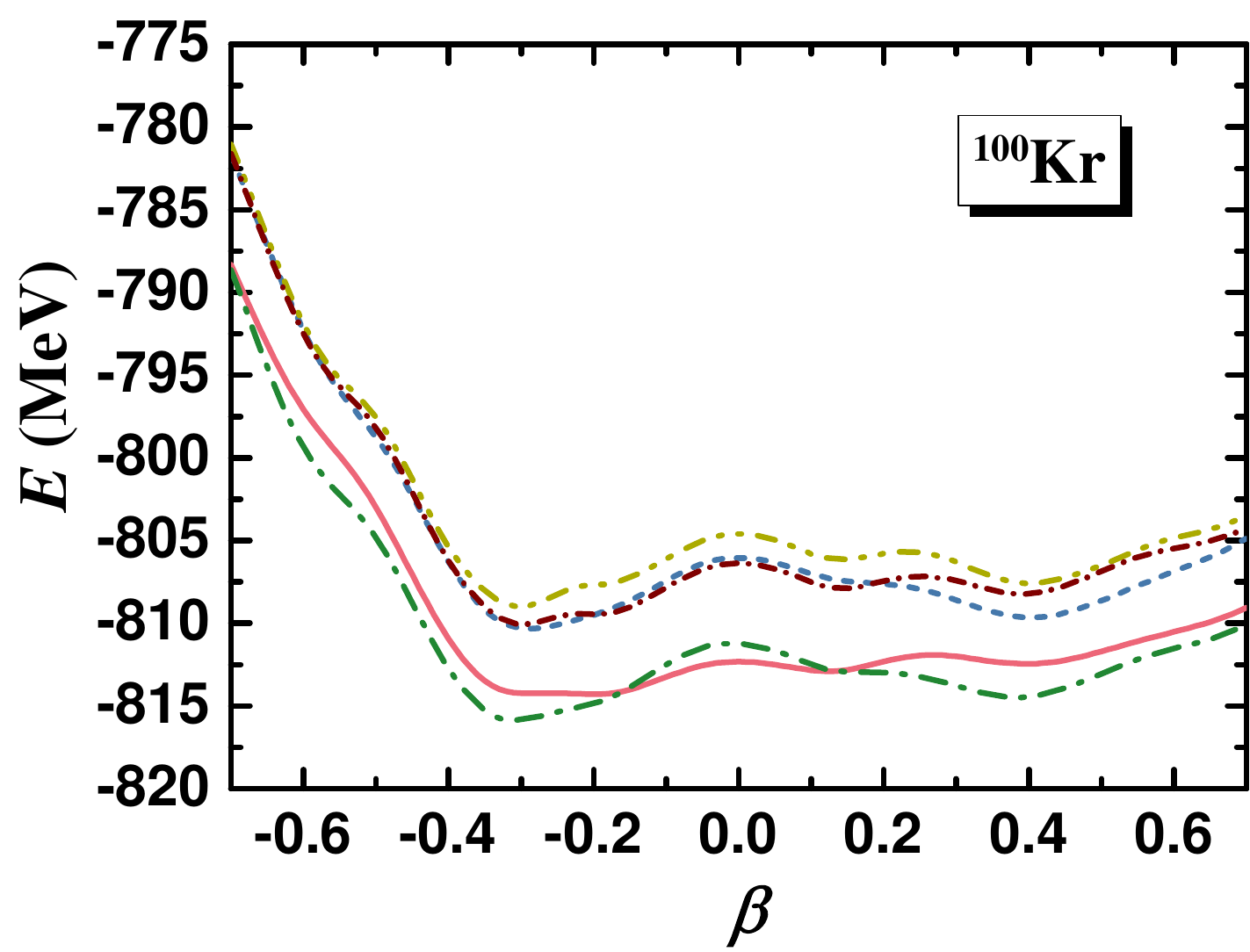}
\end{minipage}
\begin{minipage}{0.23\linewidth}
\includegraphics[width=\linewidth, angle=0]{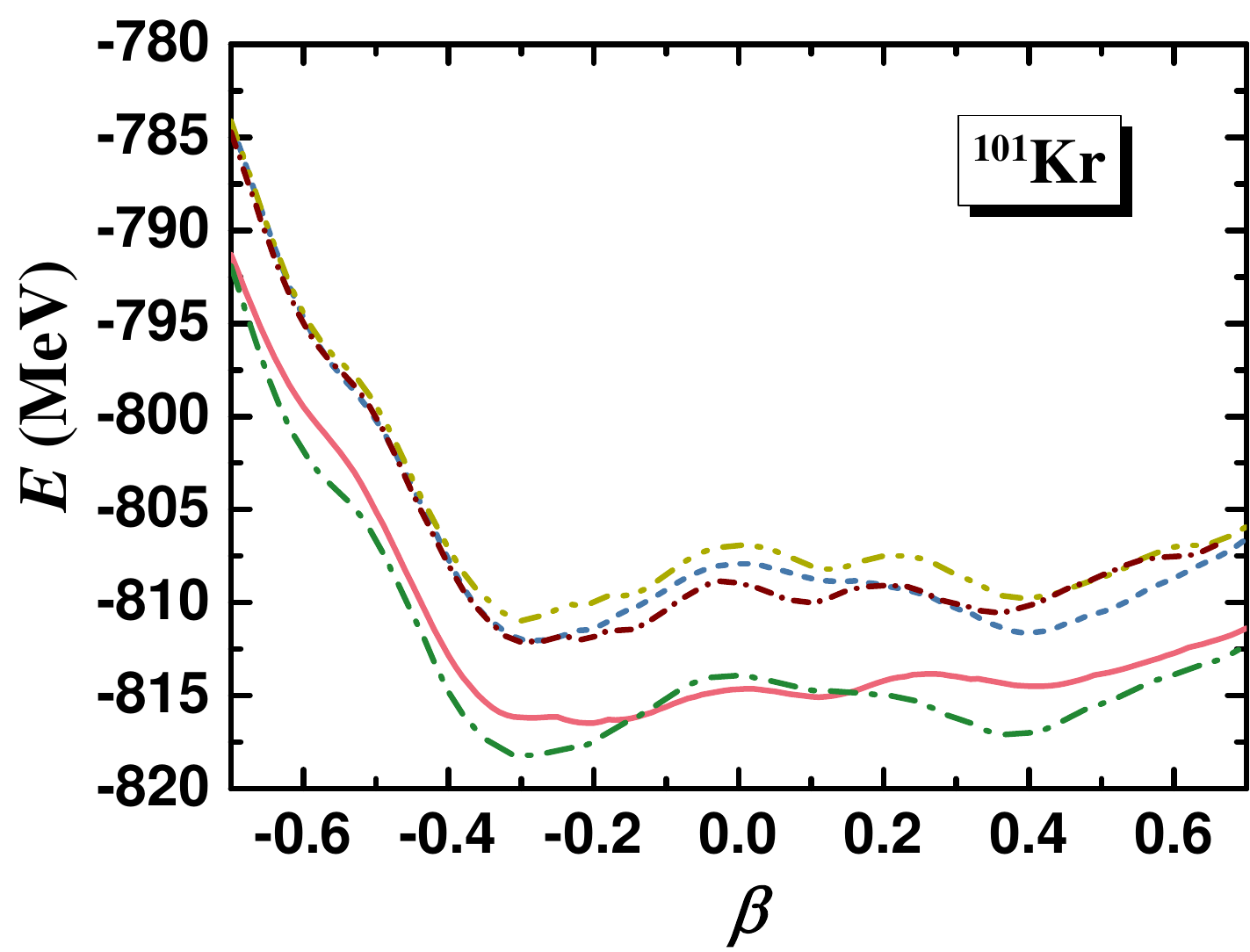}
\end{minipage}
\begin{minipage}{0.23\linewidth}
\includegraphics[width=\linewidth, angle=0]{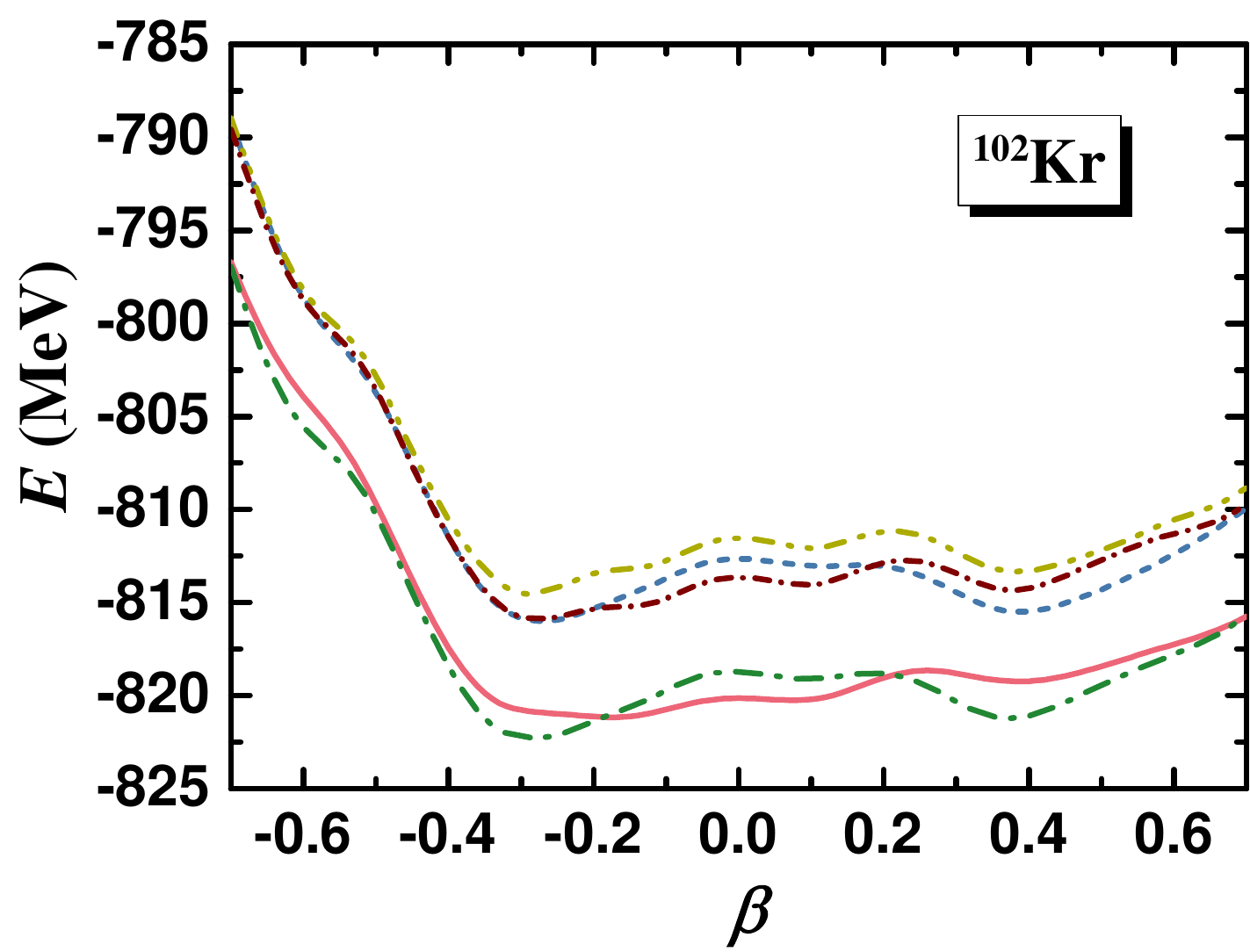}
\end{minipage}%
\vspace{-1mm}
\begin{minipage}{0.23\linewidth}
\centering
\includegraphics[width=\linewidth, angle=0]{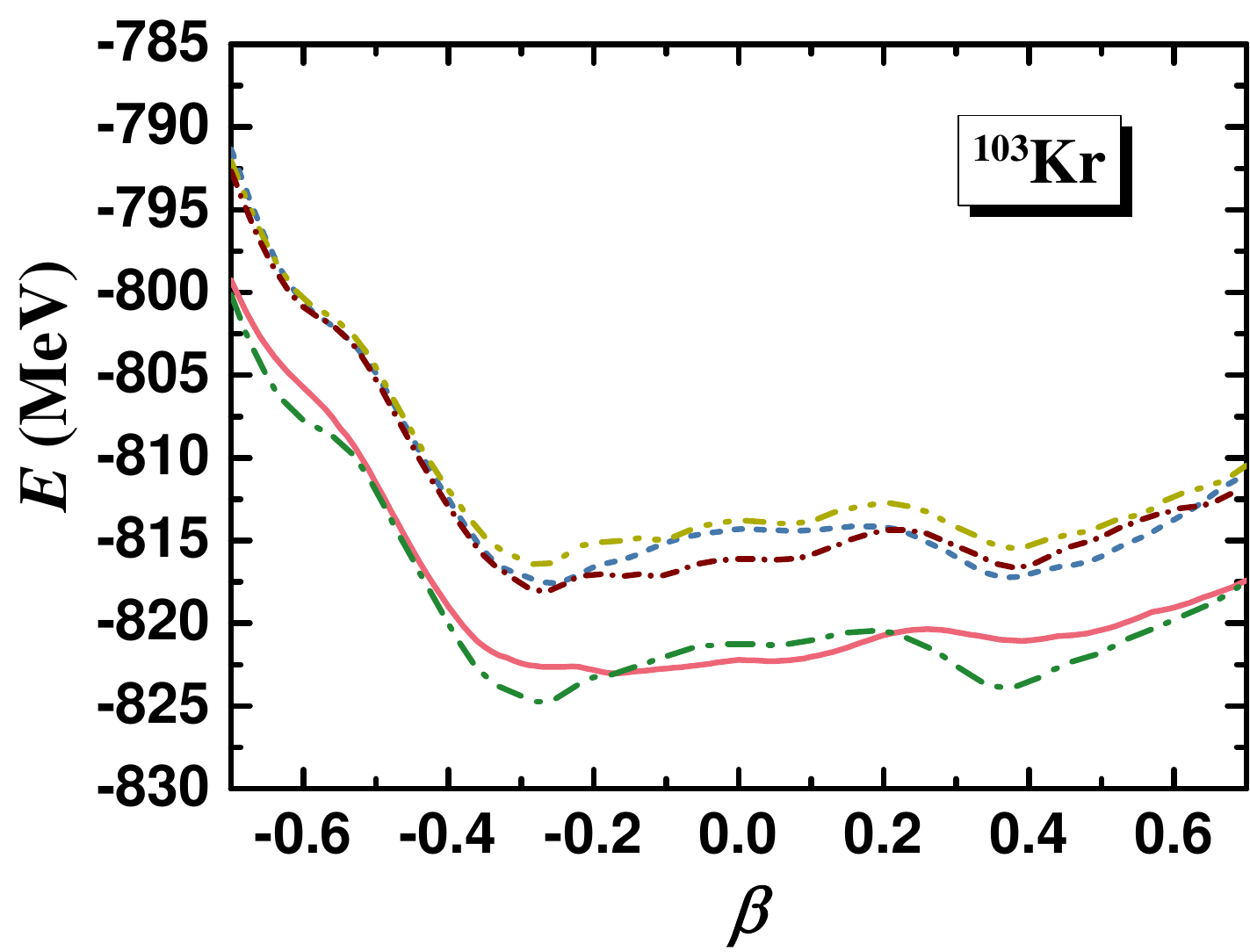}
\end{minipage}%
\begin{minipage}{0.23\linewidth}
\includegraphics[width=\linewidth, angle=0]{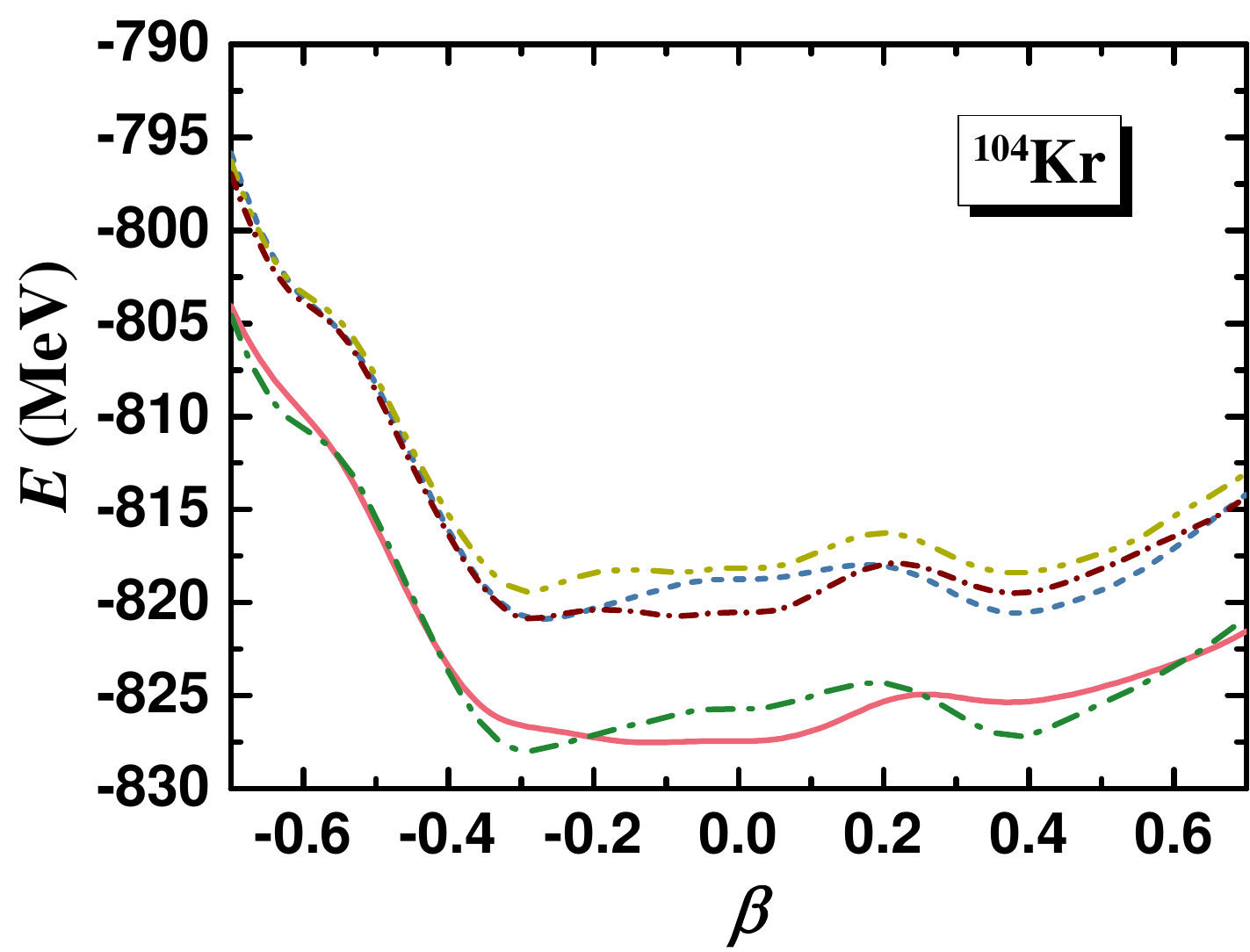}
\end{minipage}
\begin{minipage}{0.23\linewidth}
\includegraphics[width=\linewidth, angle=0]{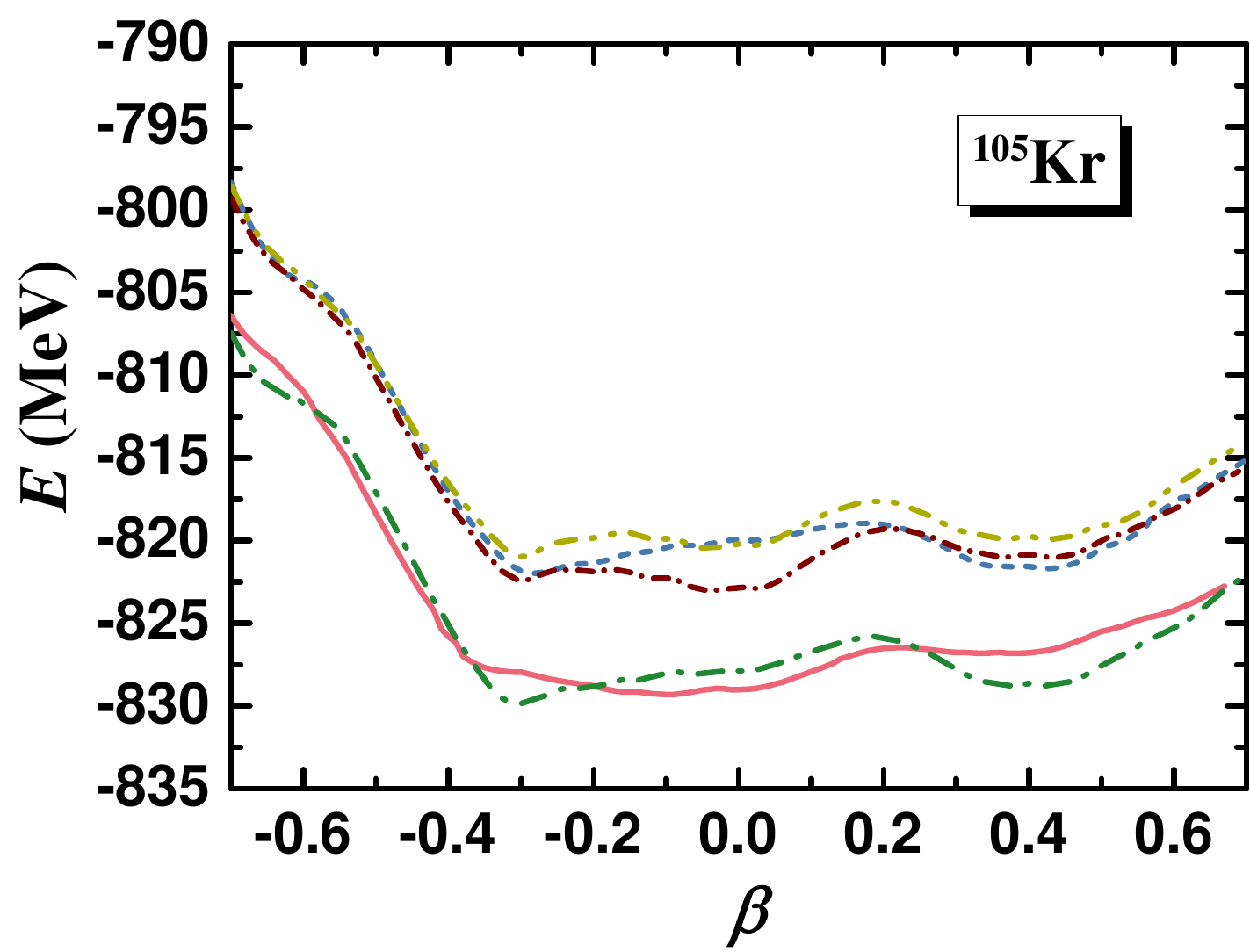}
\end{minipage}
\begin{minipage}{0.23\linewidth}
\includegraphics[width=\linewidth, angle=0]{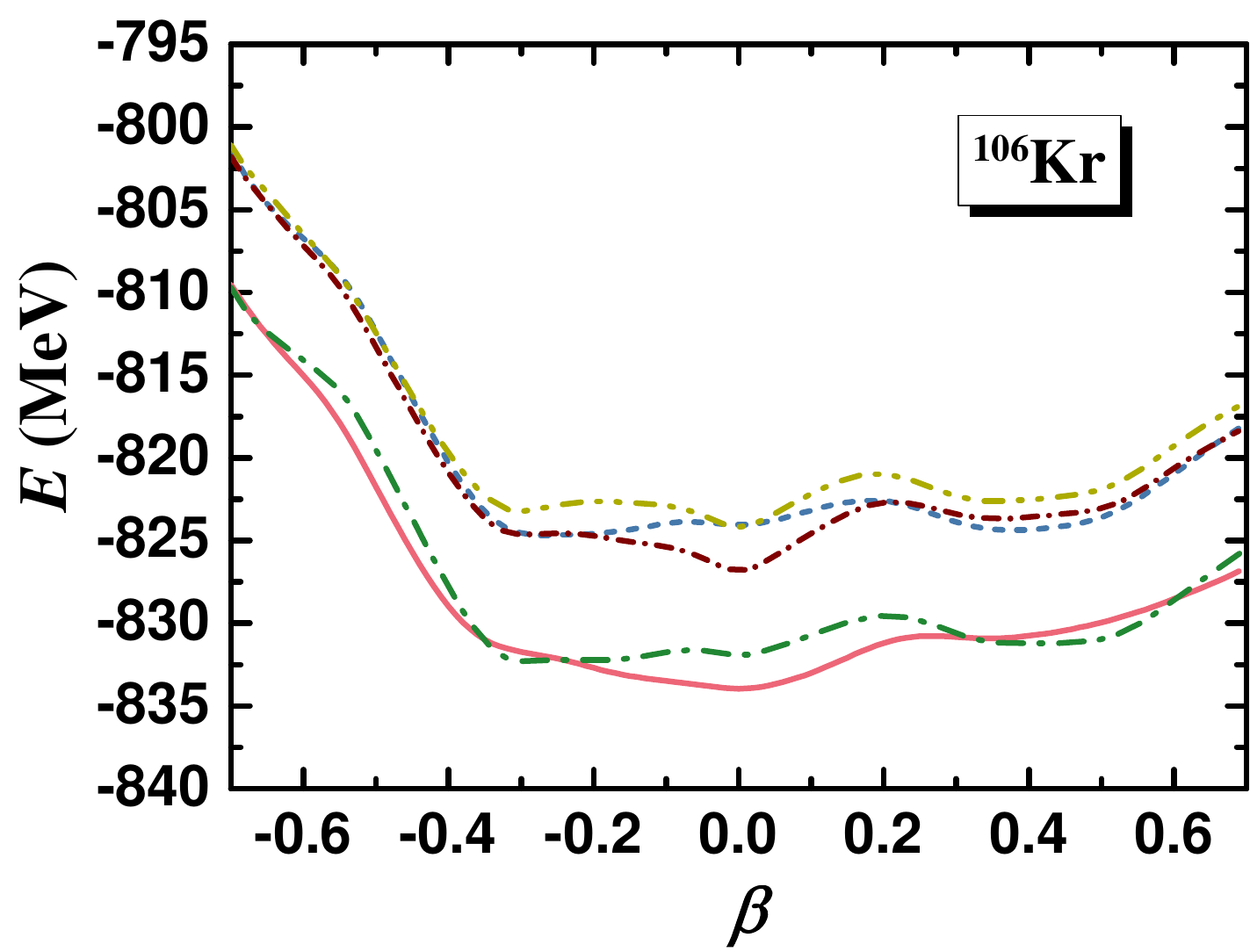}
\end{minipage}%
\vspace{-1mm}
\begin{minipage}{0.23\linewidth}
\centering
\includegraphics[width=\linewidth, angle=0]{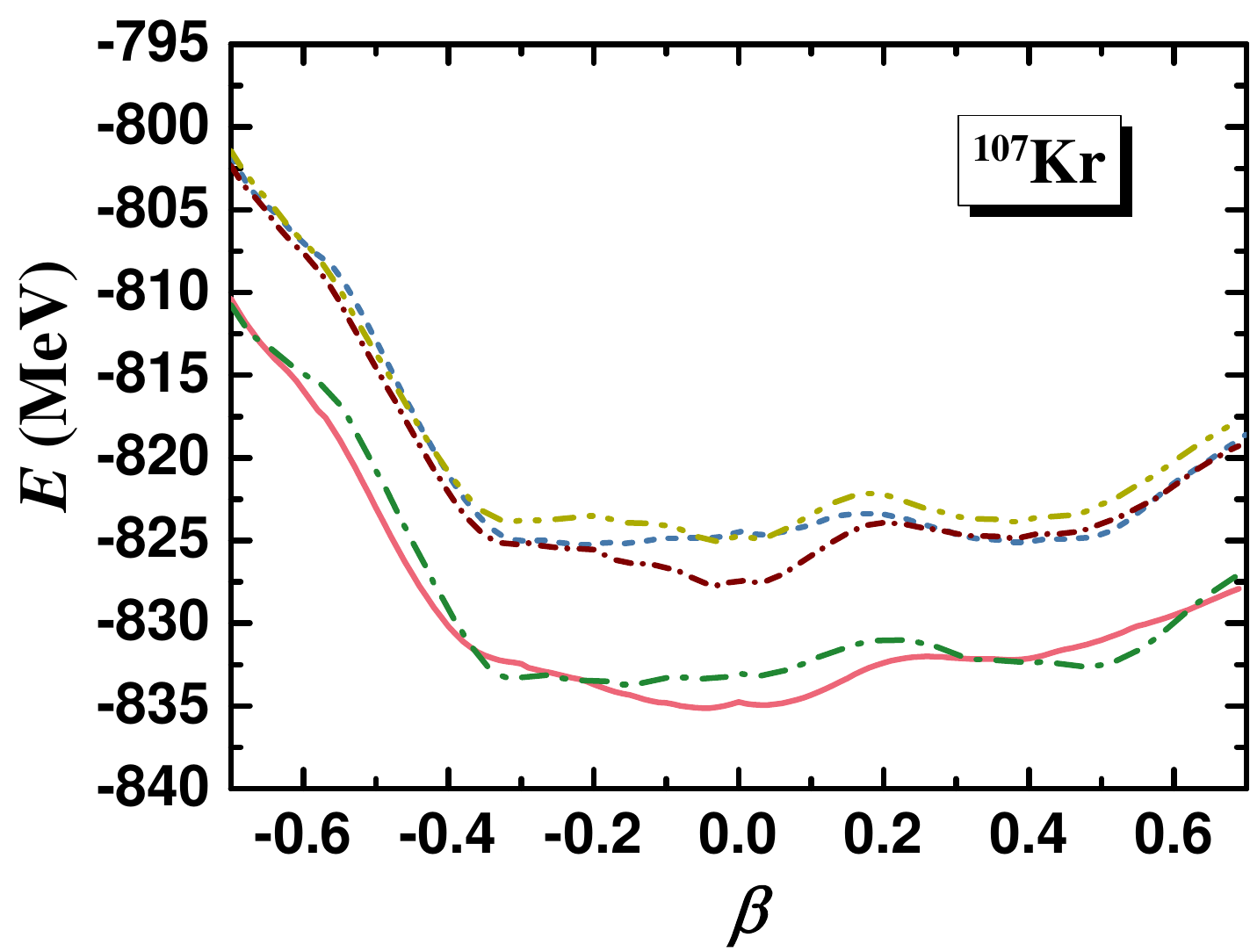}
\end{minipage}%
\begin{minipage}{0.23\linewidth}
\includegraphics[width=\linewidth, angle=0]{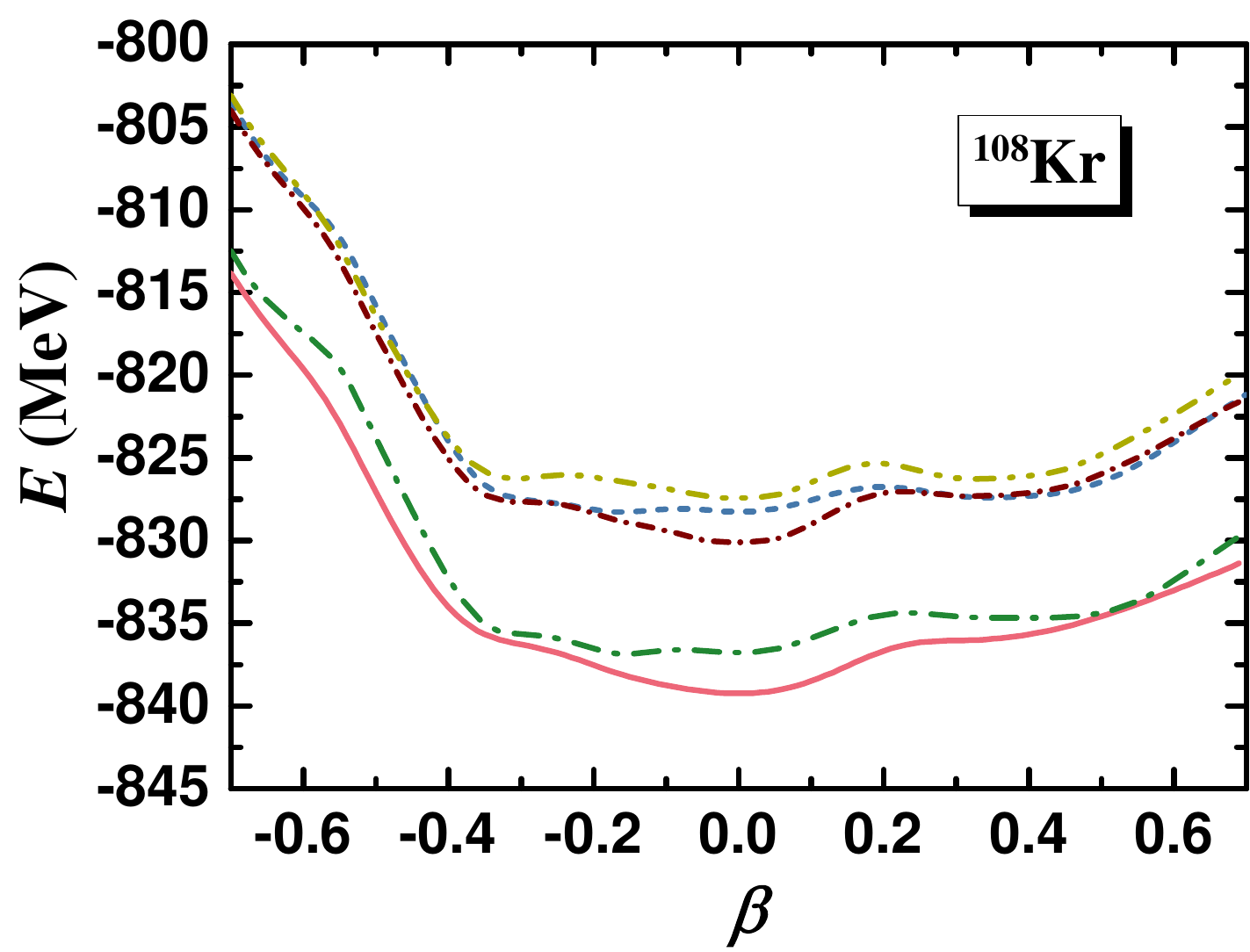}
\end{minipage}
\begin{minipage}{0.23\linewidth}
\includegraphics[width=\linewidth, angle=0]{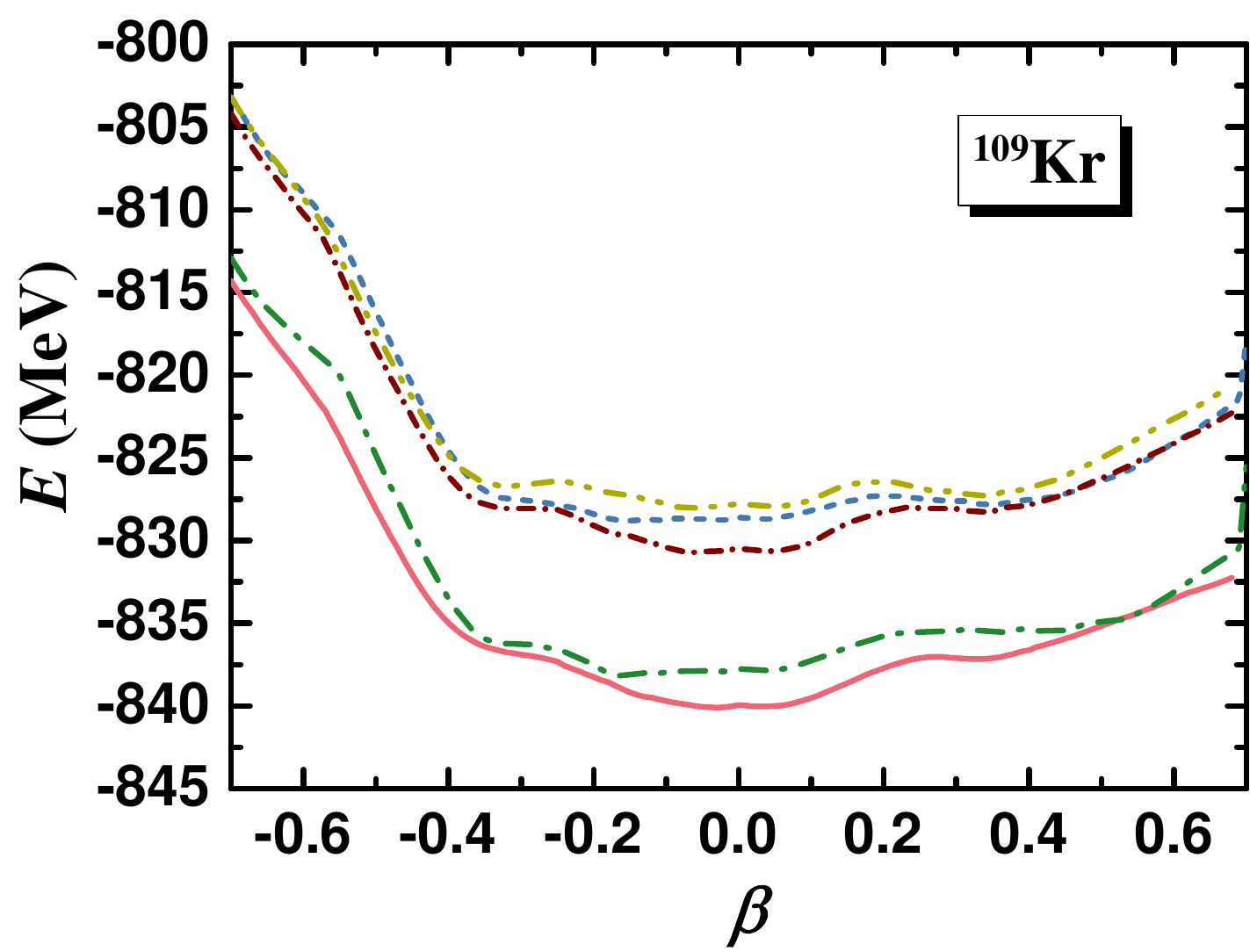}
\end{minipage}
\begin{minipage}{0.23\linewidth}
\includegraphics[width=\linewidth, angle=0]{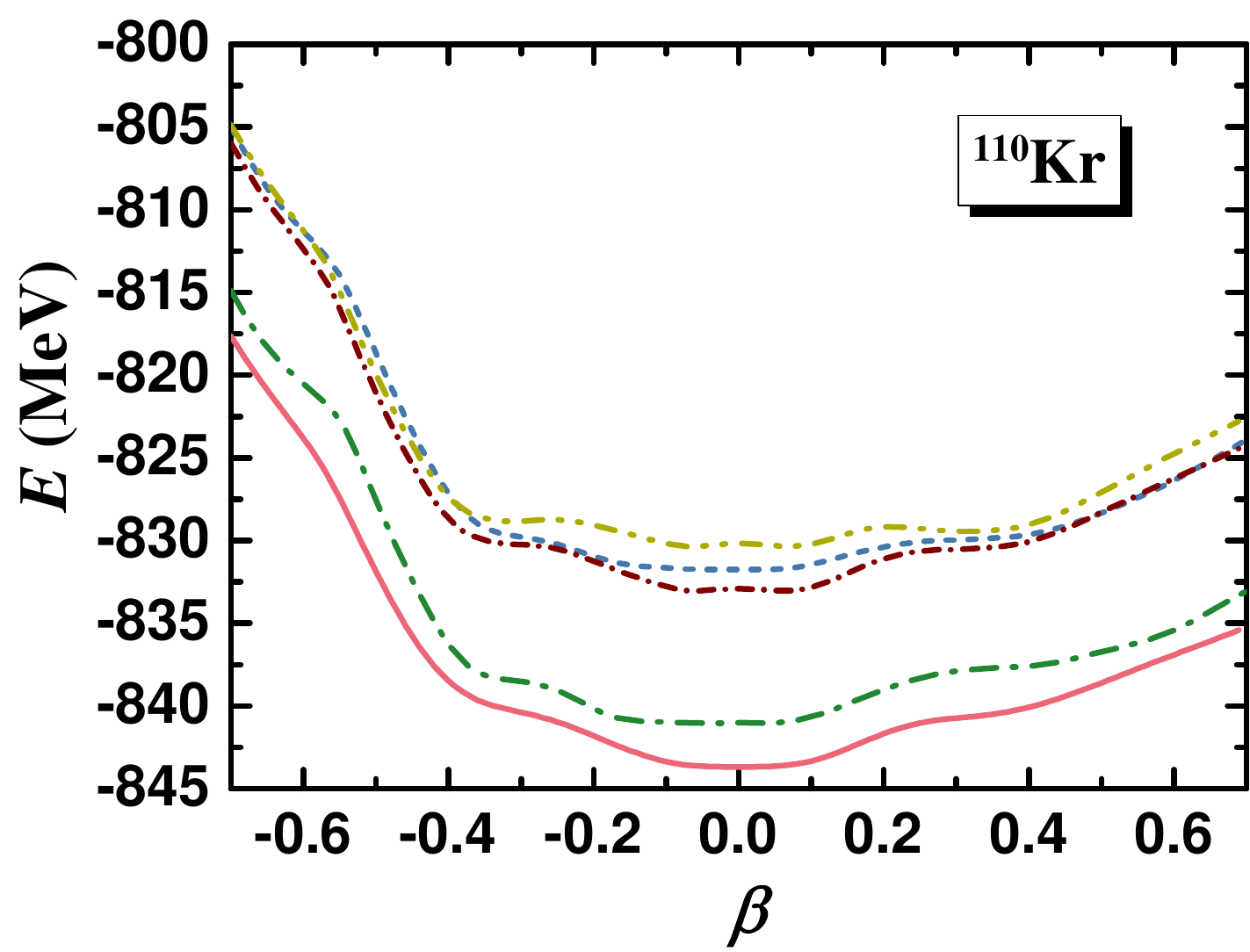}
\end{minipage}%
\vspace{-1mm}
\begin{minipage}{0.23\linewidth}
\centering
\includegraphics[width=\linewidth, angle=0]{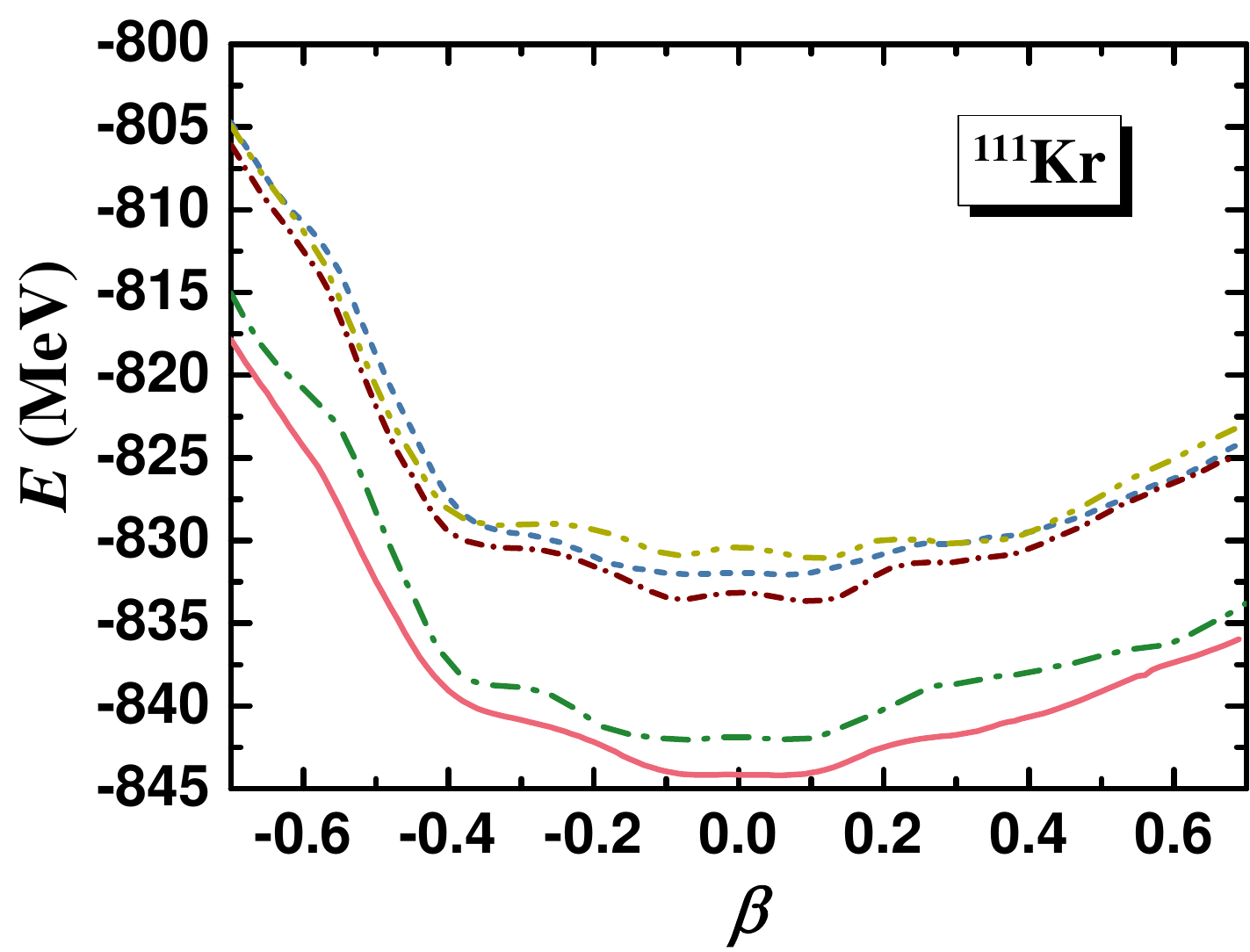}
\end{minipage}%
\begin{minipage}{0.23\linewidth}
\includegraphics[width=\linewidth, angle=0]{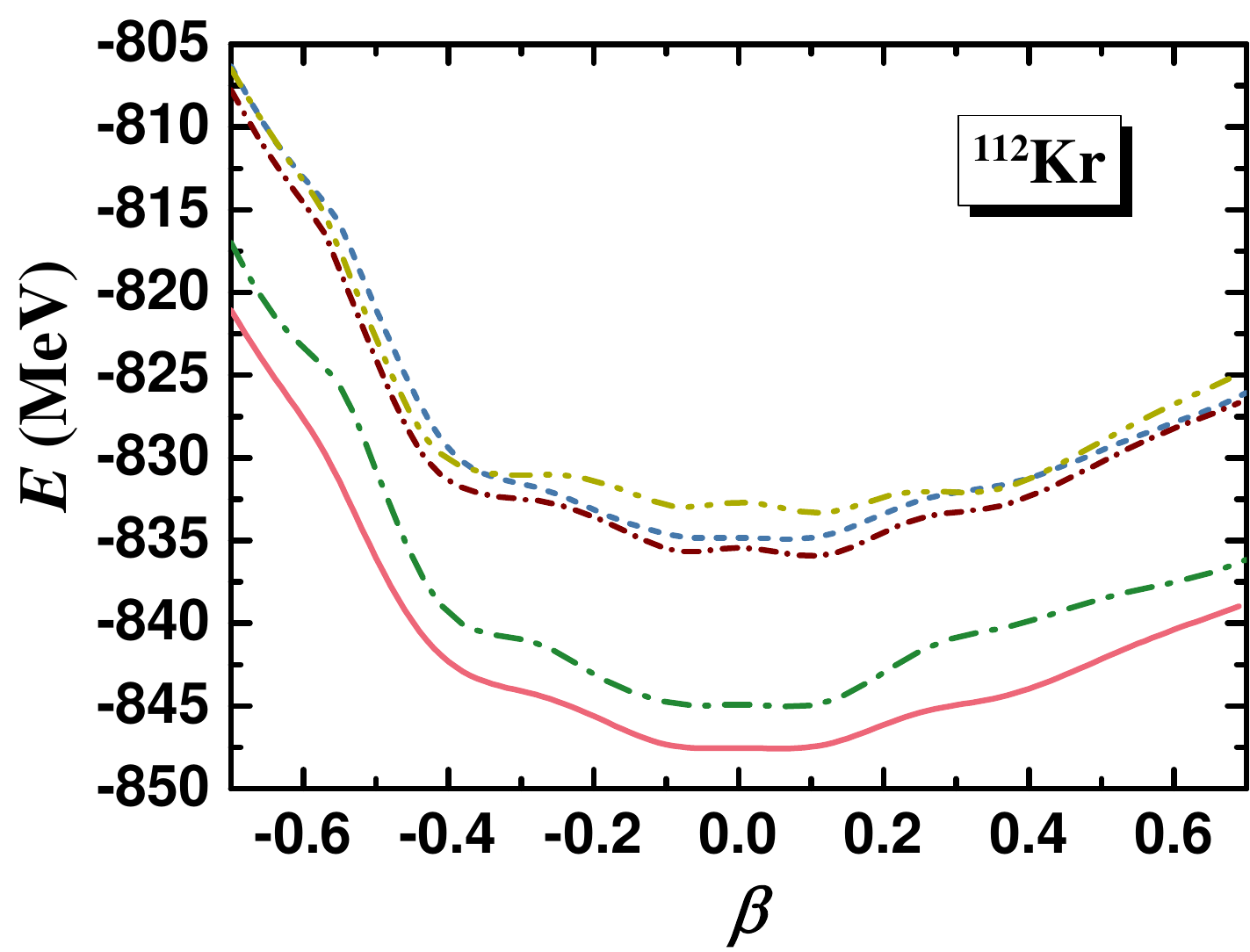}
\end{minipage}
\begin{minipage}{0.23\linewidth}
\includegraphics[width=\linewidth, angle=0]{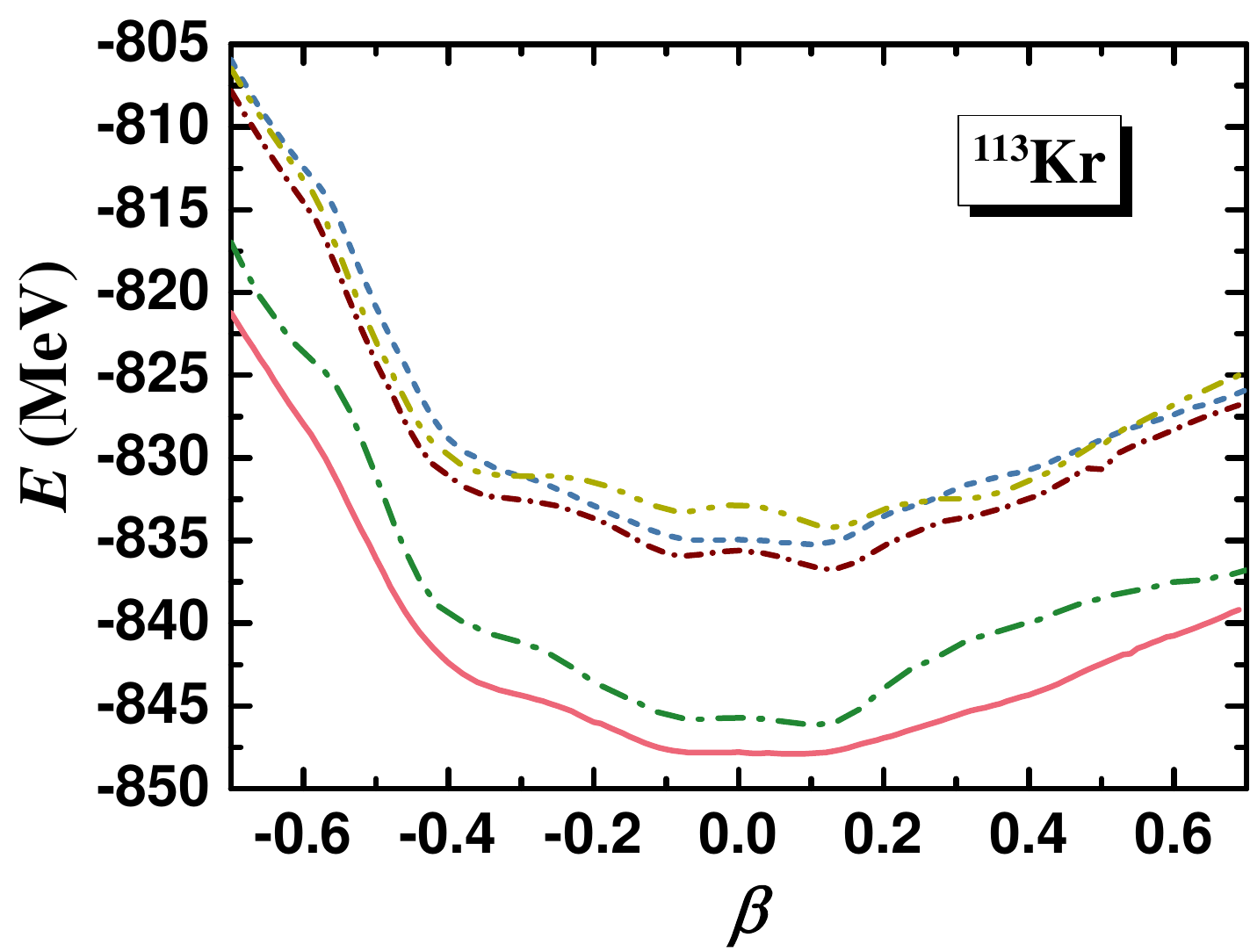}
\end{minipage}
\begin{minipage}{0.23\linewidth}
\includegraphics[width=\linewidth, angle=0]{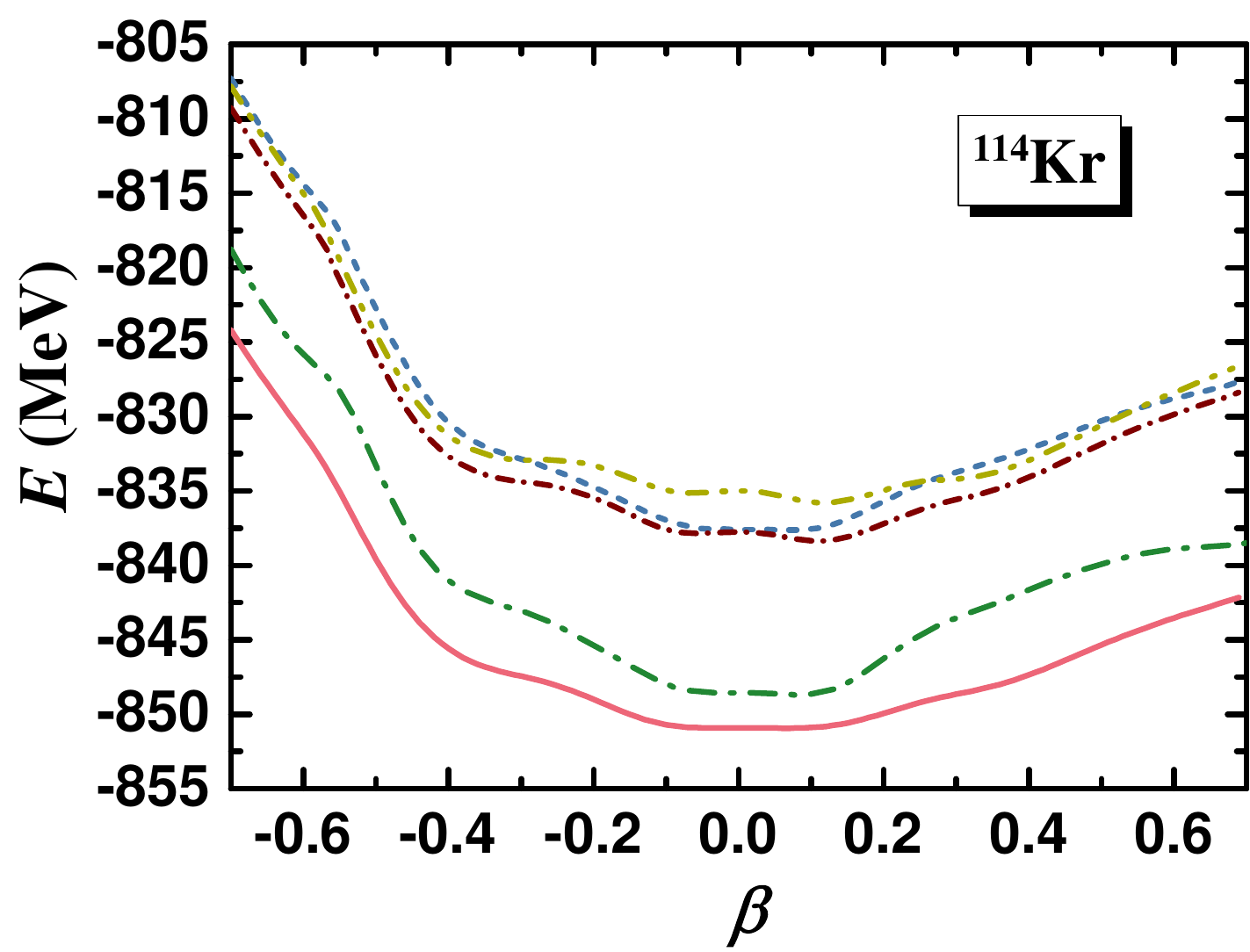}
\end{minipage}%
\vspace{-1mm}
\begin{minipage}{0.23\linewidth}
\centering
\includegraphics[width=\linewidth, angle=0]{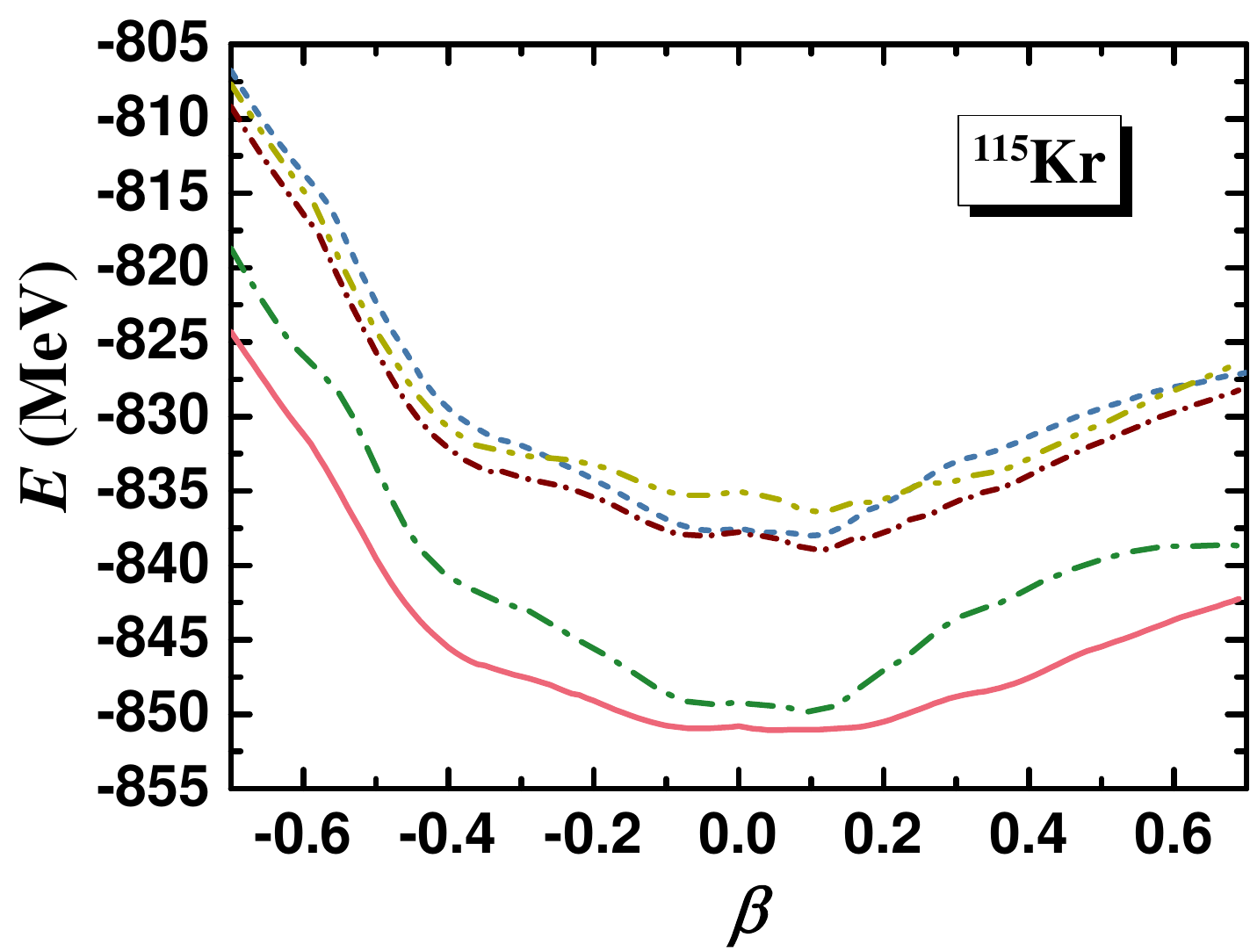}
\end{minipage}%
\begin{minipage}{0.23\linewidth}
\includegraphics[width=\linewidth, angle=0]{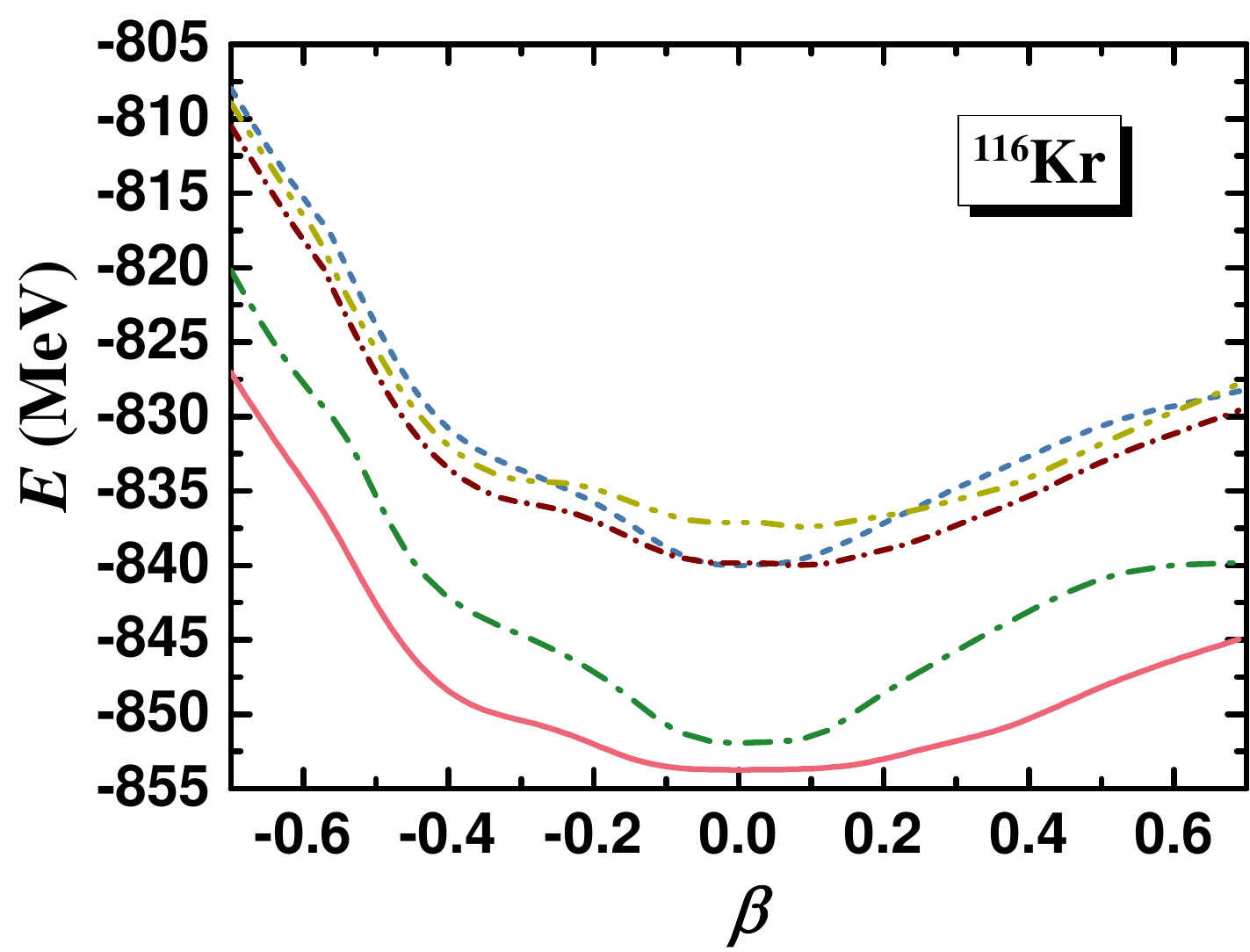}
\end{minipage}
\begin{minipage}{0.23\linewidth}
\includegraphics[width=\linewidth, angle=0]{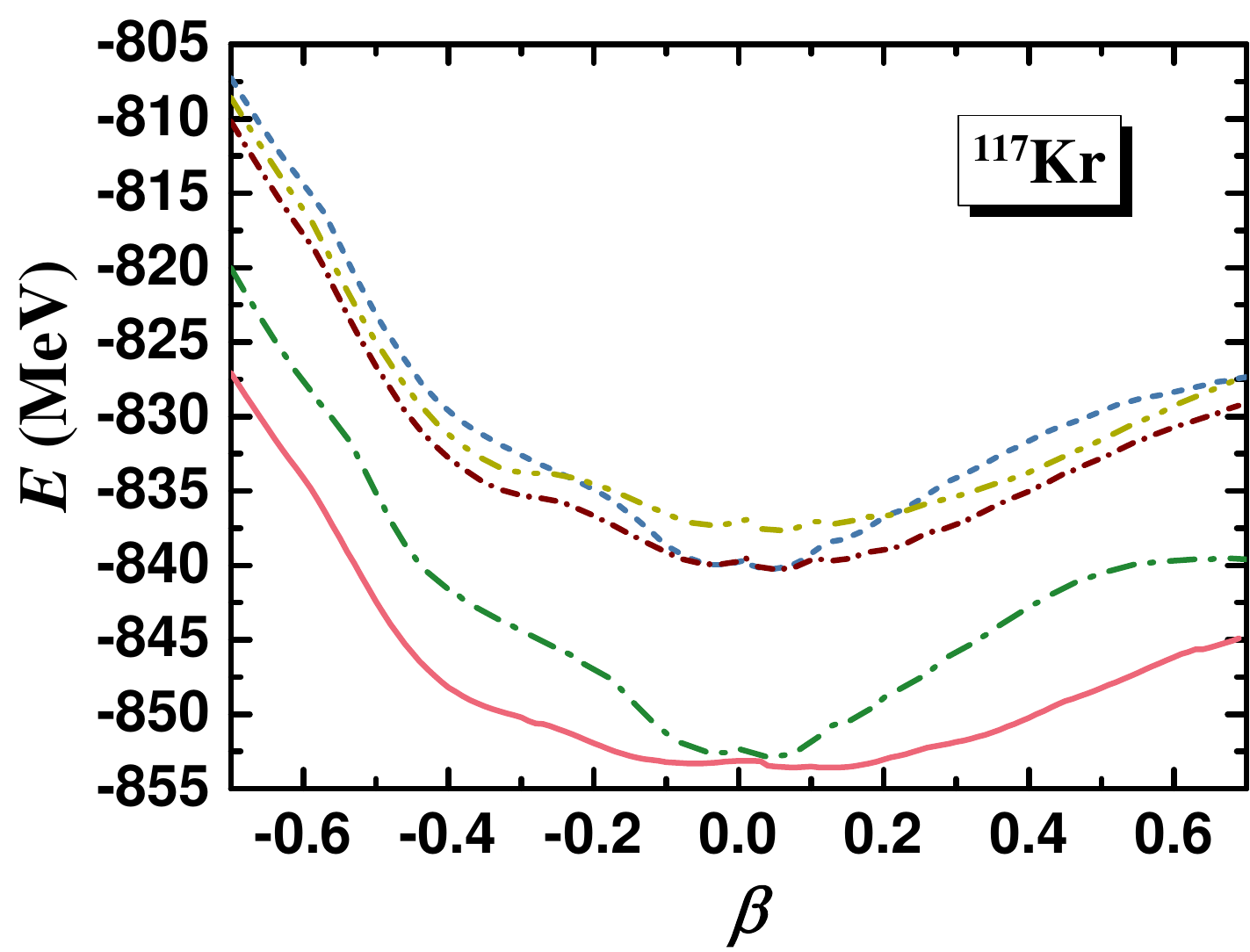}
\end{minipage}
\begin{minipage}{0.23\linewidth}
\includegraphics[width=\linewidth, angle=0]{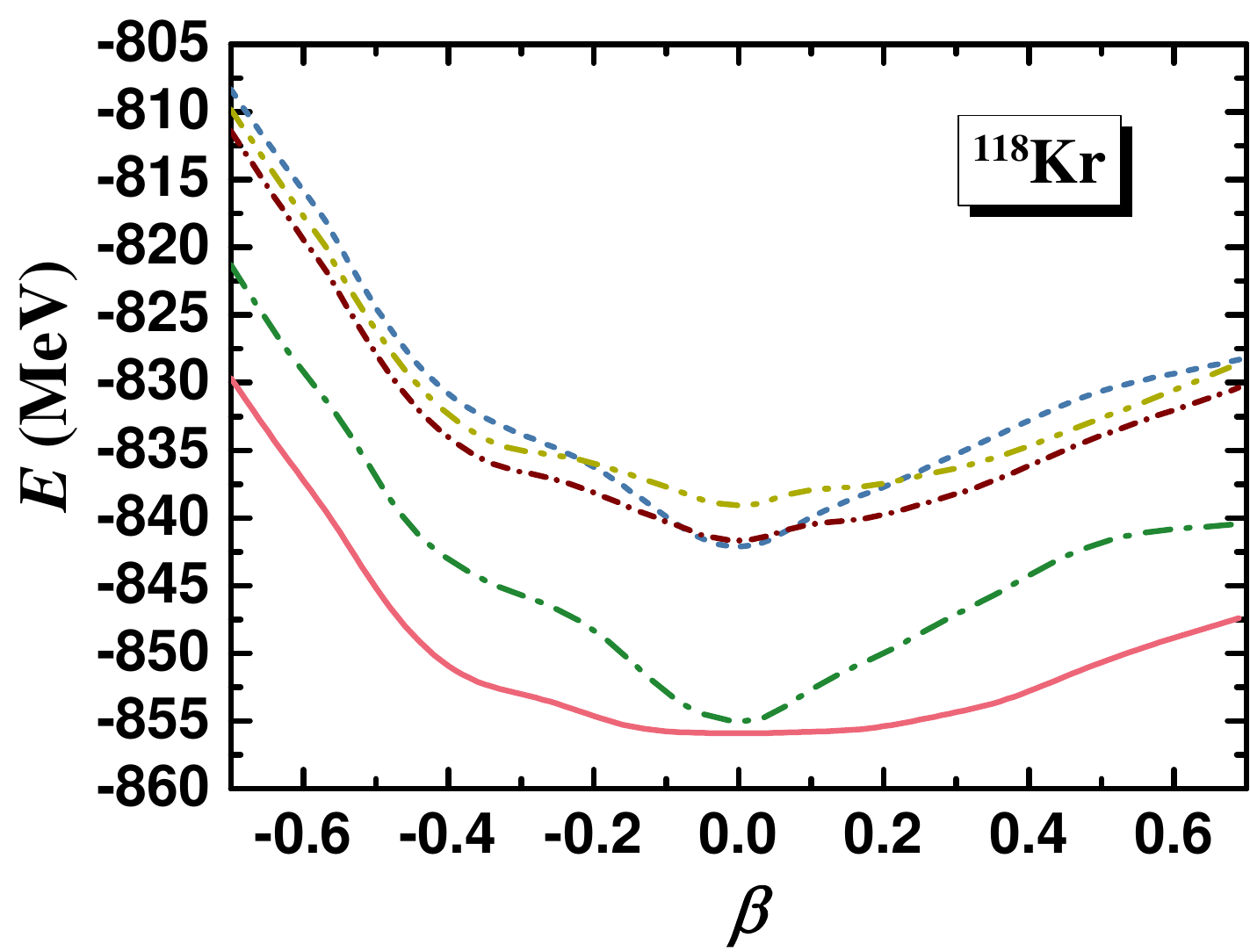}
\end{minipage}%
\vspace{-1mm}
\begin{minipage}{0.23\linewidth}
\centering
\includegraphics[width=\linewidth, angle=0]{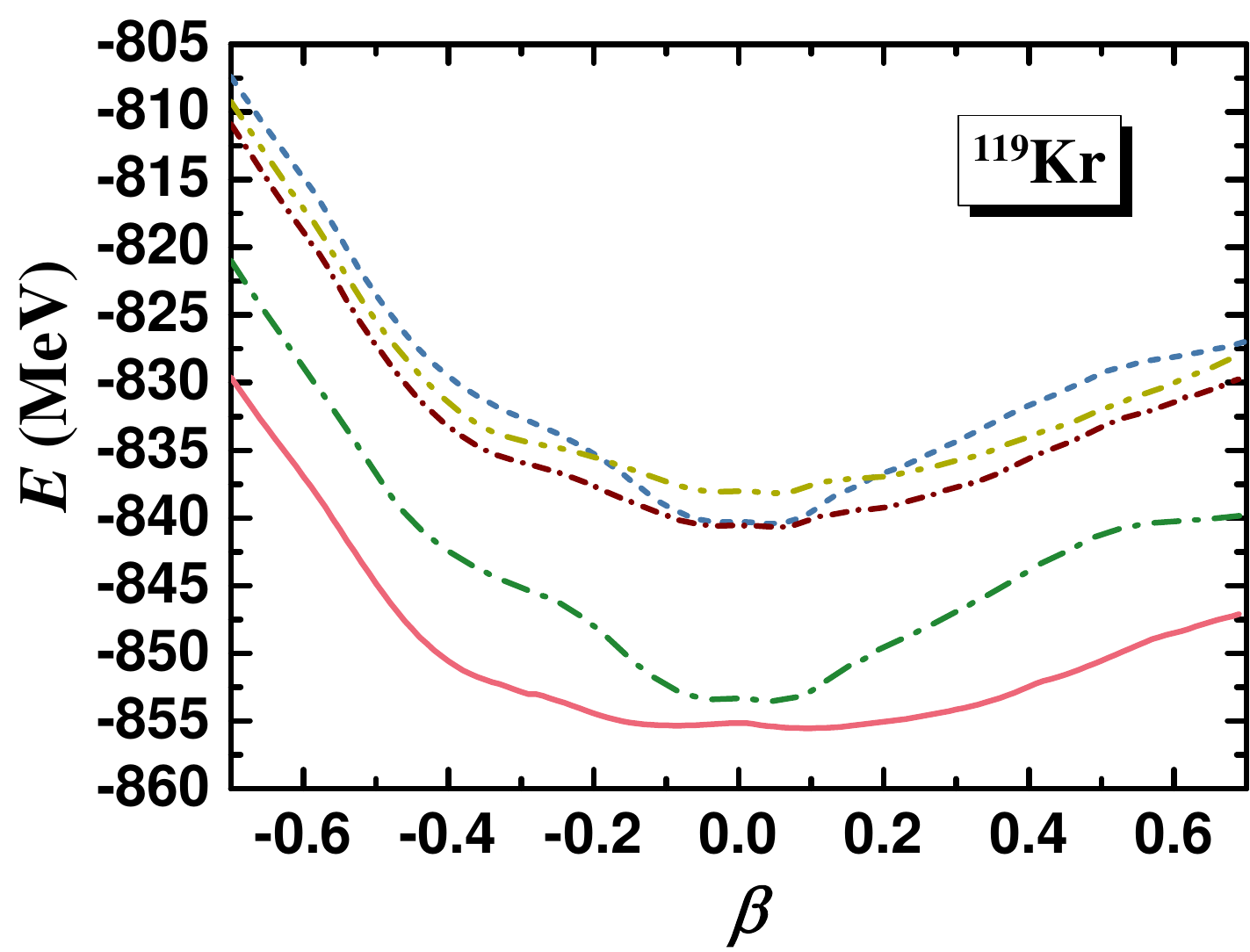}
\end{minipage}%
\begin{minipage}{0.23\linewidth}
\includegraphics[width=\linewidth, angle=0]{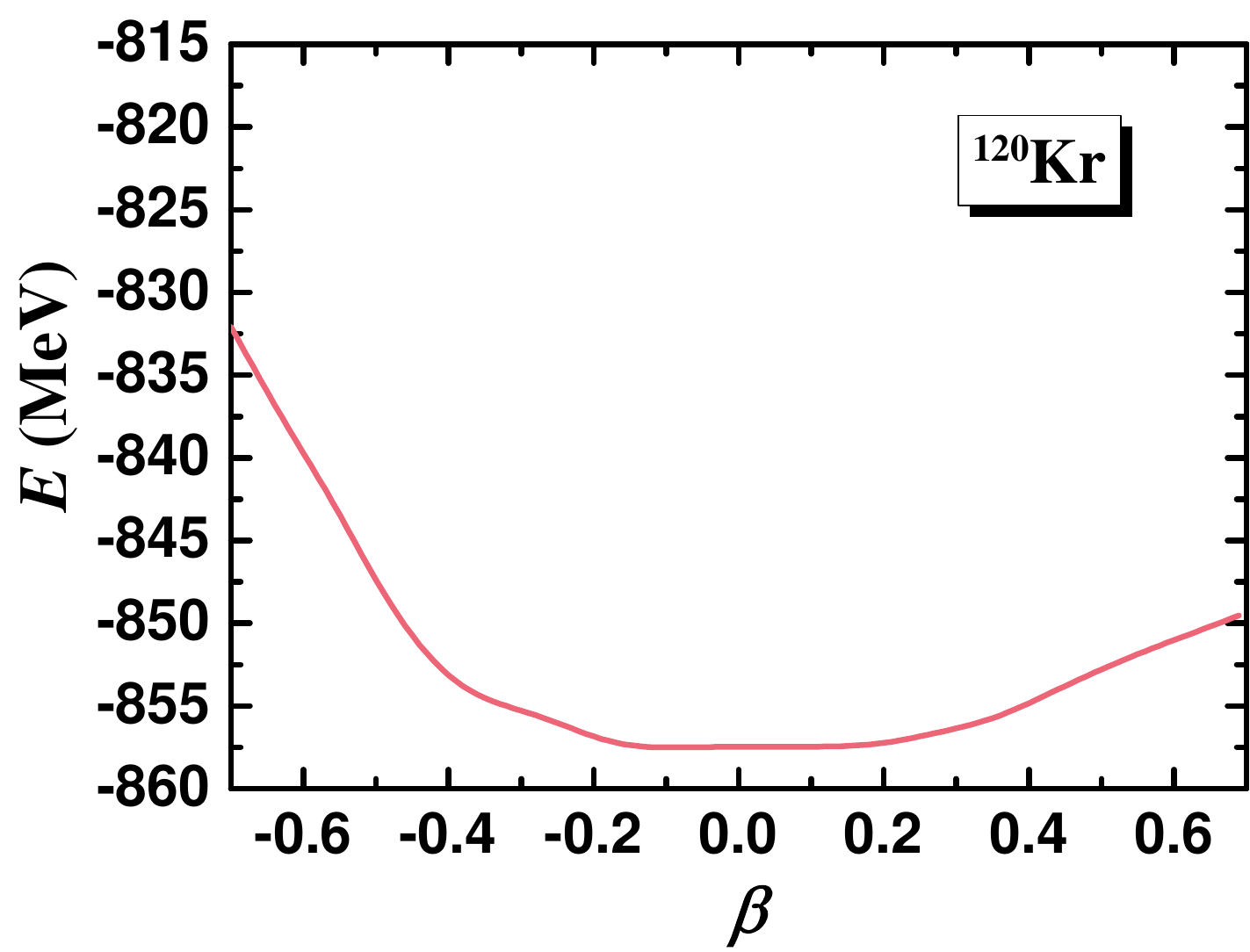}
\end{minipage}
\begin{minipage}{0.23\linewidth}
\includegraphics[width=\linewidth, angle=0]{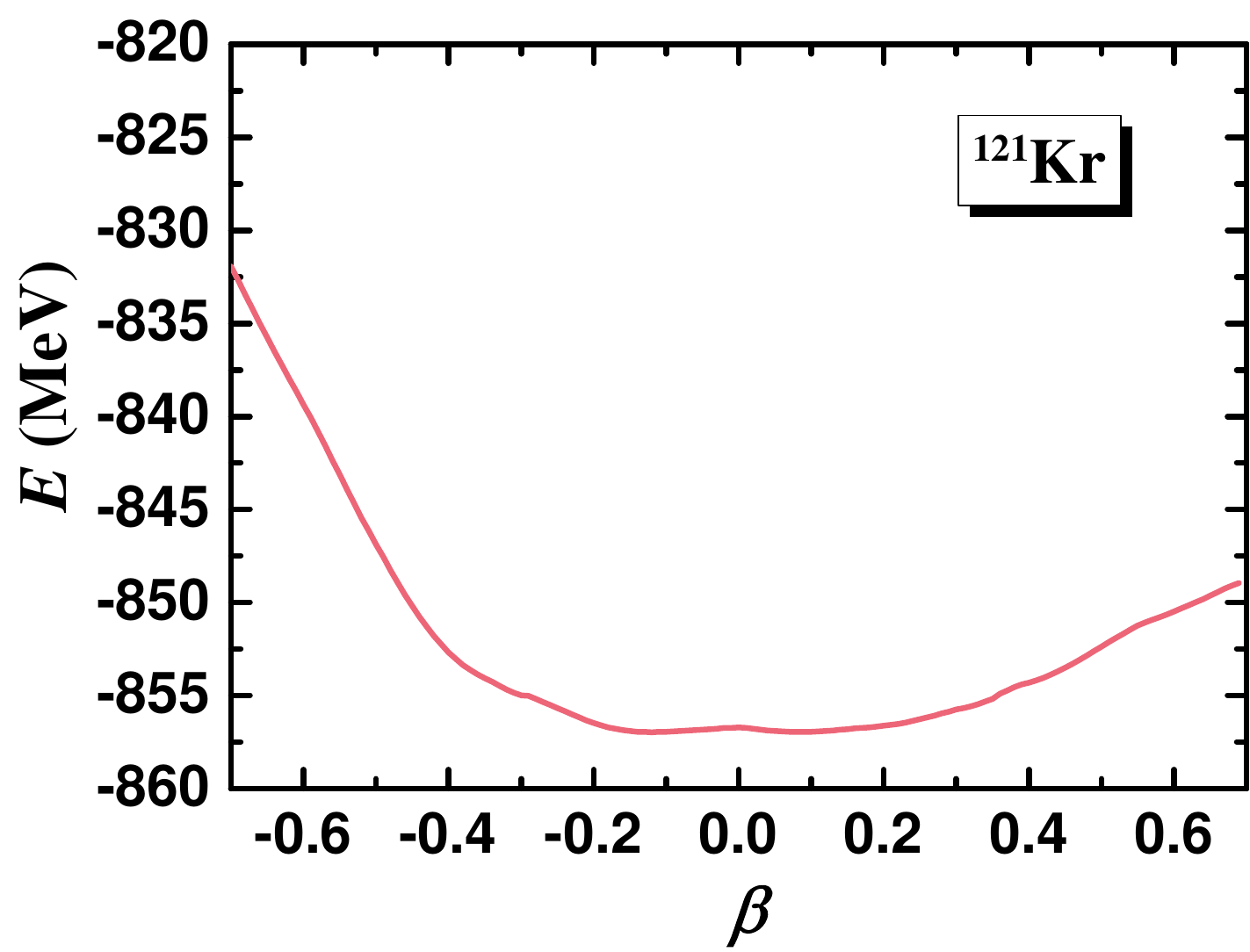}
\end{minipage}
\begin{minipage}{0.23\linewidth}
\includegraphics[width=\linewidth, angle=0]{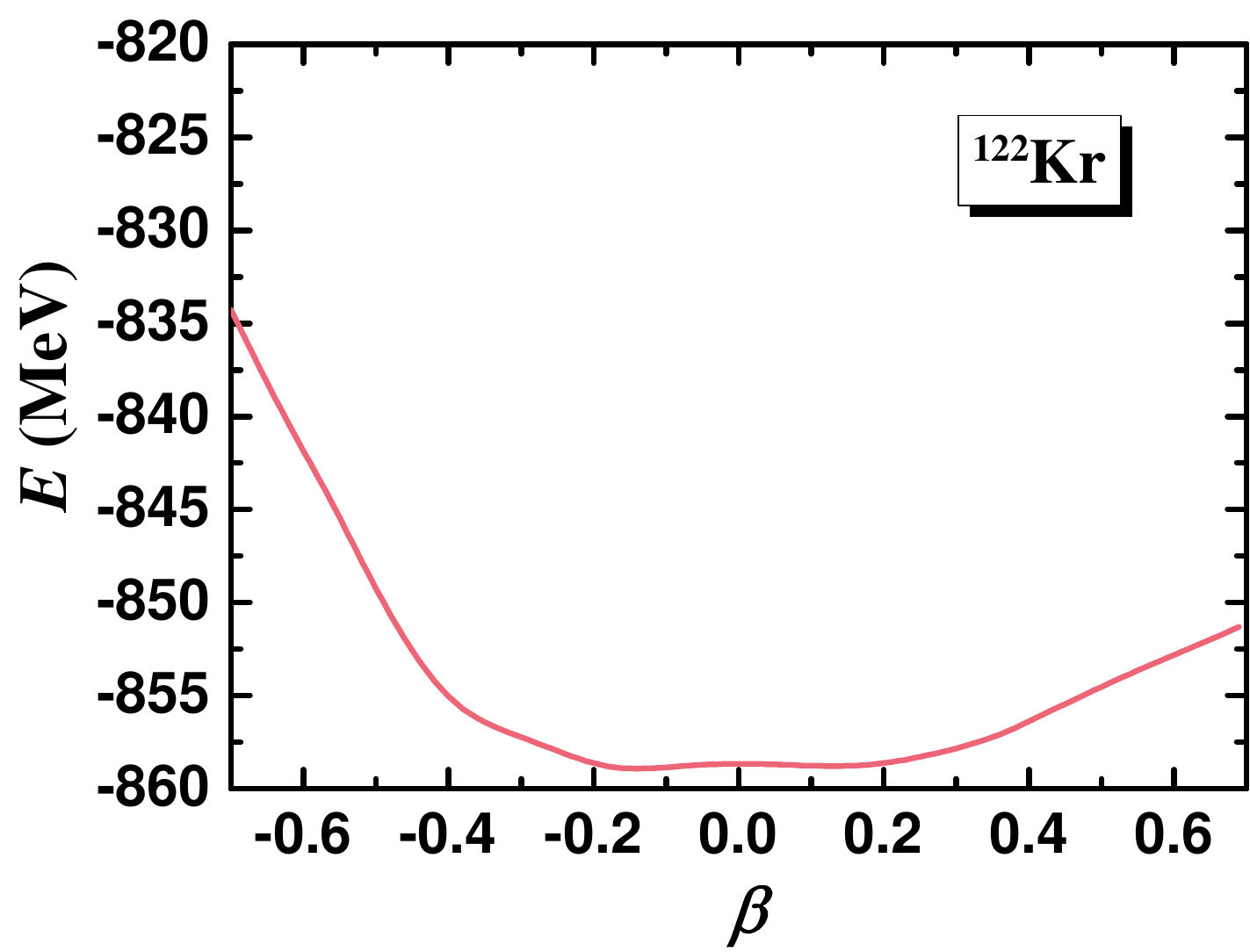}
\end{minipage}%
\vspace{-2mm}
\caption{Potential energy curves as functions of the quadrupole $\beta_{2}$ deformation for the $^{95-122}$Kr isotopes computed with the relativistic Hartree-Bogoliubov approach with axially-symmetrical basis employing the PC-L3R, DD-PCX, DD-PC1, DD-MEX, and DD-ME2 effective interactions and the respective separable pairing force of finite range.}
\label{fig:Kr-PES2}
\end{figure*}

\begin{figure*}%
\centering
\begin{minipage}{0.23\linewidth}
\centering
\includegraphics[width=\linewidth, angle=0]{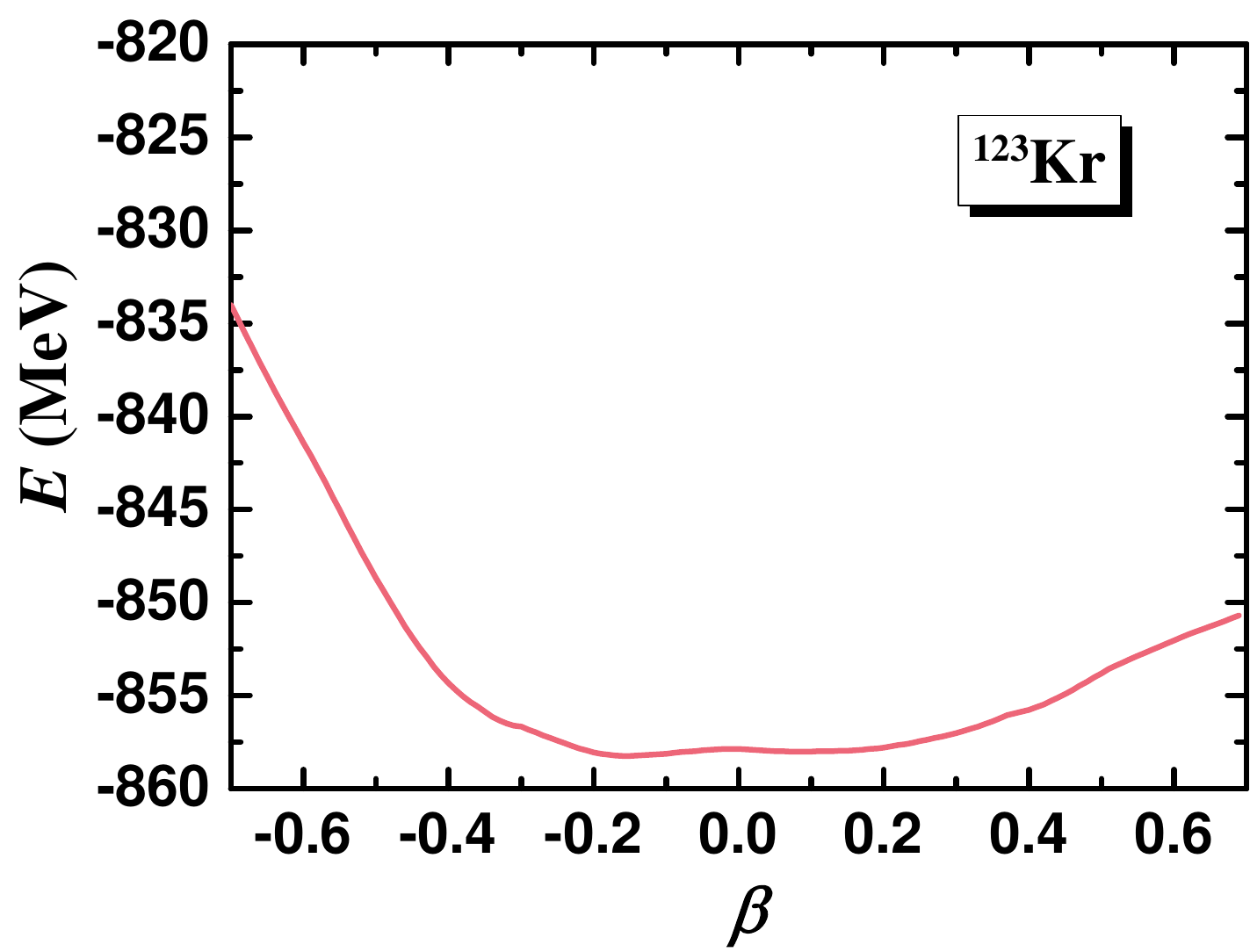}
\end{minipage}%
\begin{minipage}{0.23\linewidth}
\includegraphics[width=\linewidth, angle=0]{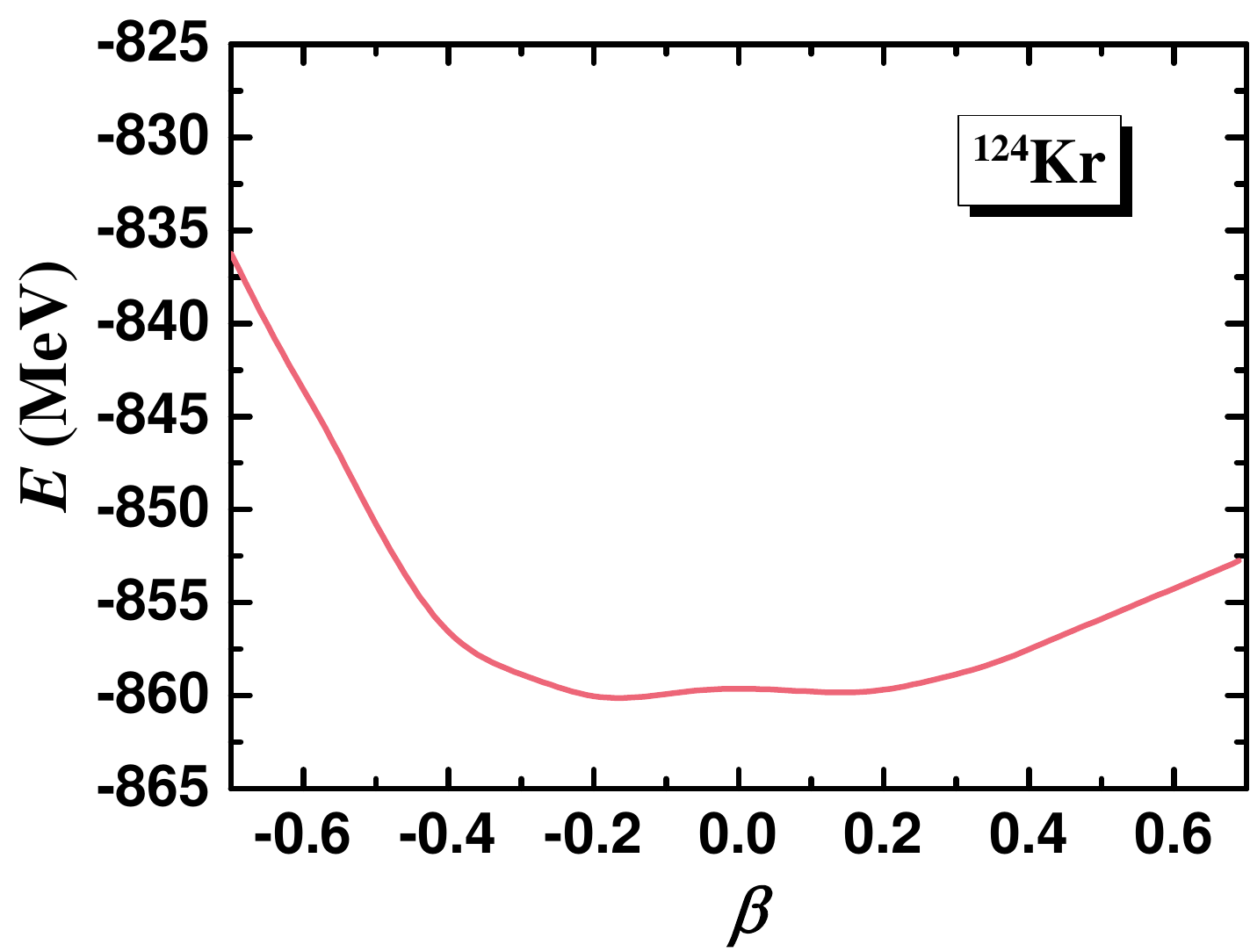}
\end{minipage}
\begin{minipage}{0.23\linewidth}
\includegraphics[width=\linewidth, angle=0]{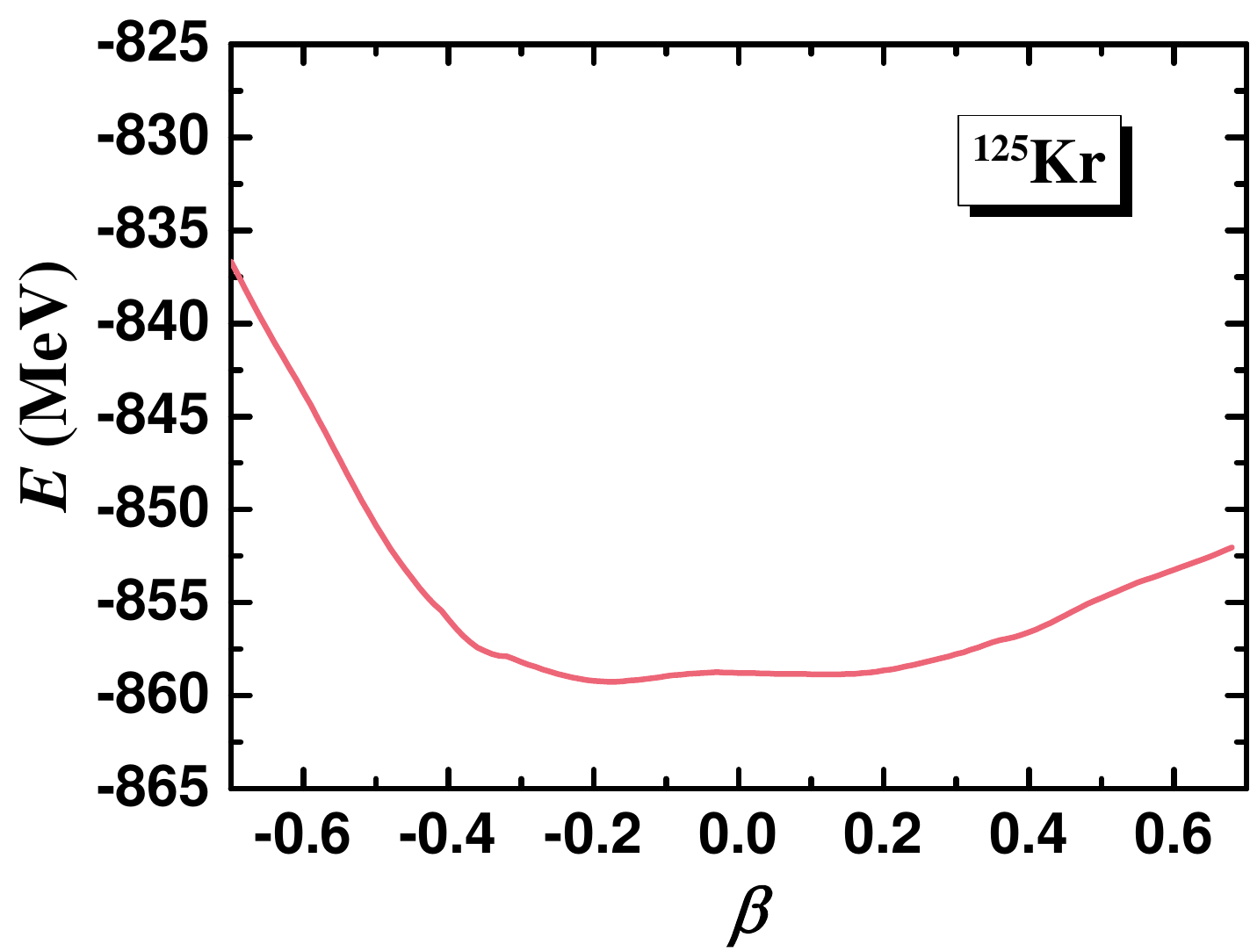}
\end{minipage}
\begin{minipage}{0.23\linewidth}
\includegraphics[width=\linewidth, angle=0]{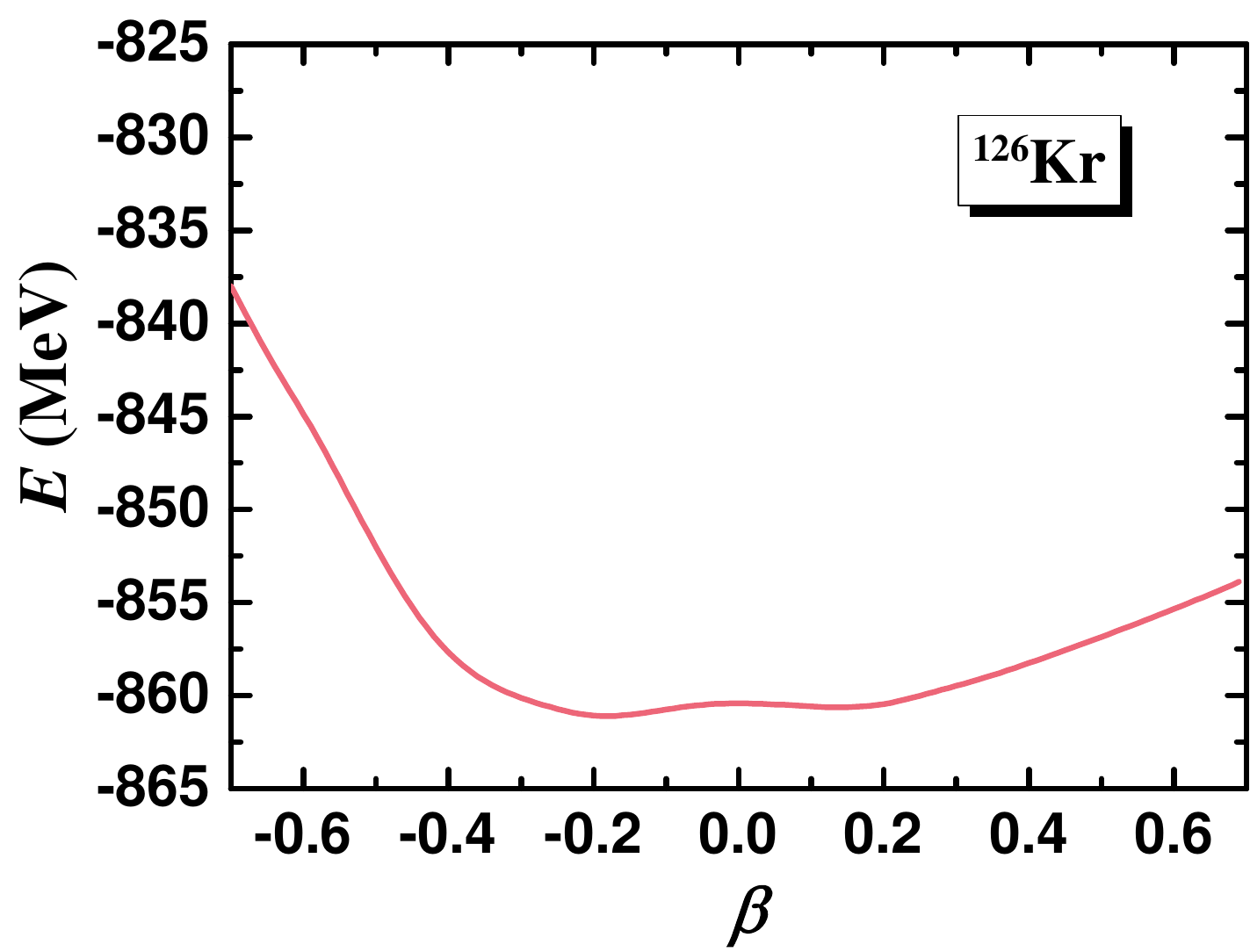}
\end{minipage}
\vspace{-1mm}
\begin{minipage}{0.23\linewidth}
\centering
\includegraphics[width=\linewidth, angle=0]{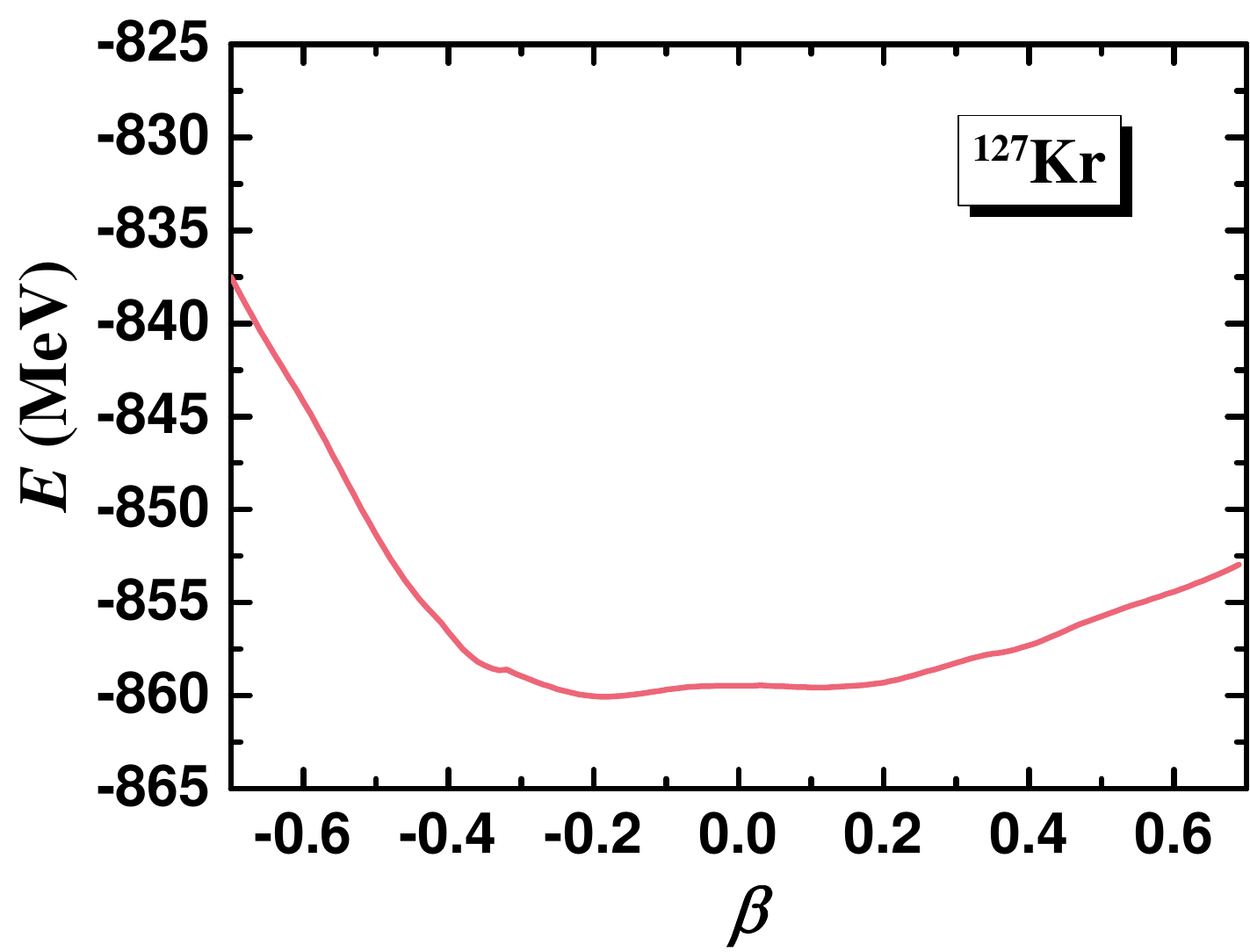}
\end{minipage}%
\begin{minipage}{0.23\linewidth}
\includegraphics[width=\linewidth, angle=0]{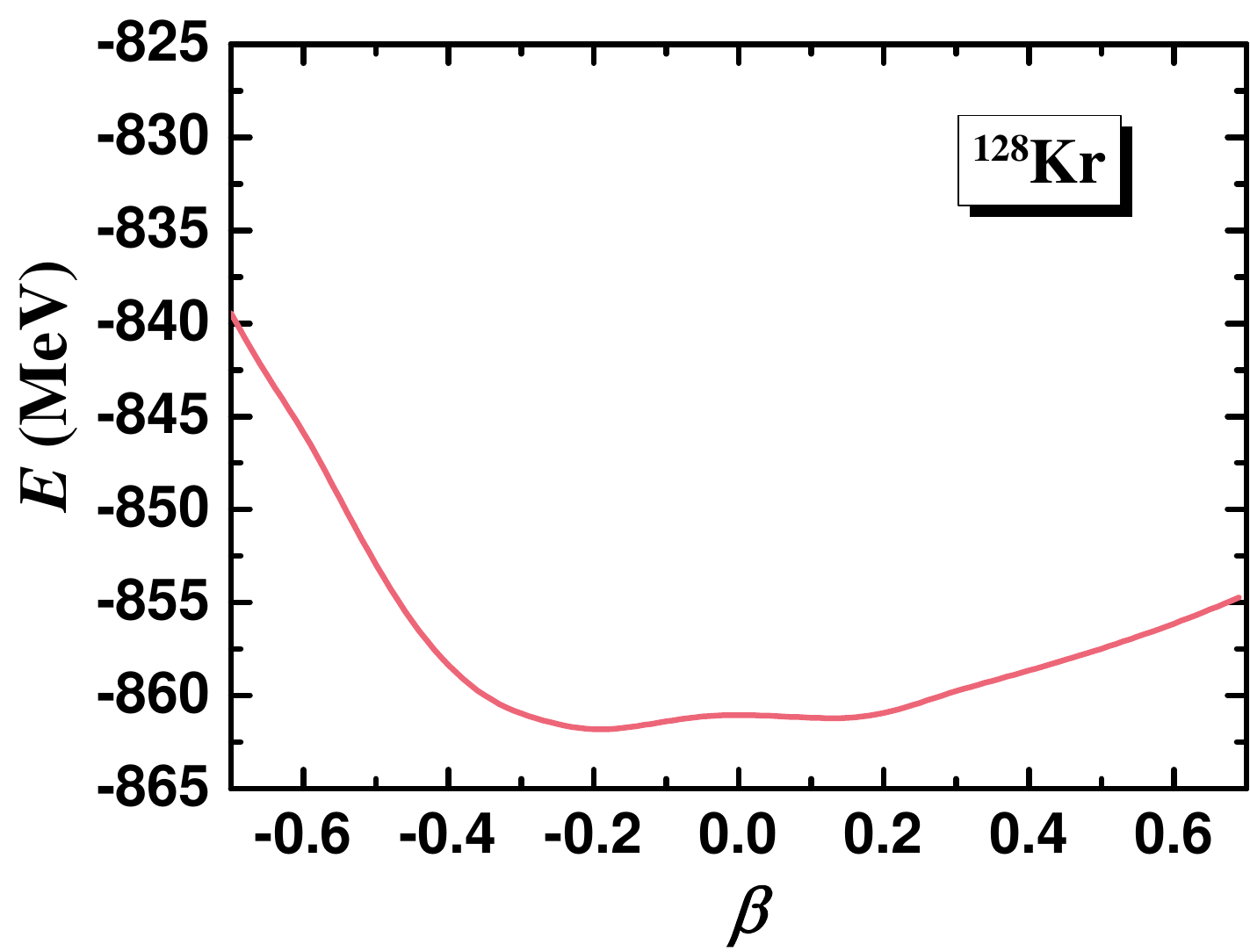}
\end{minipage}
\begin{minipage}{0.23\linewidth}
\includegraphics[width=\linewidth, angle=0]{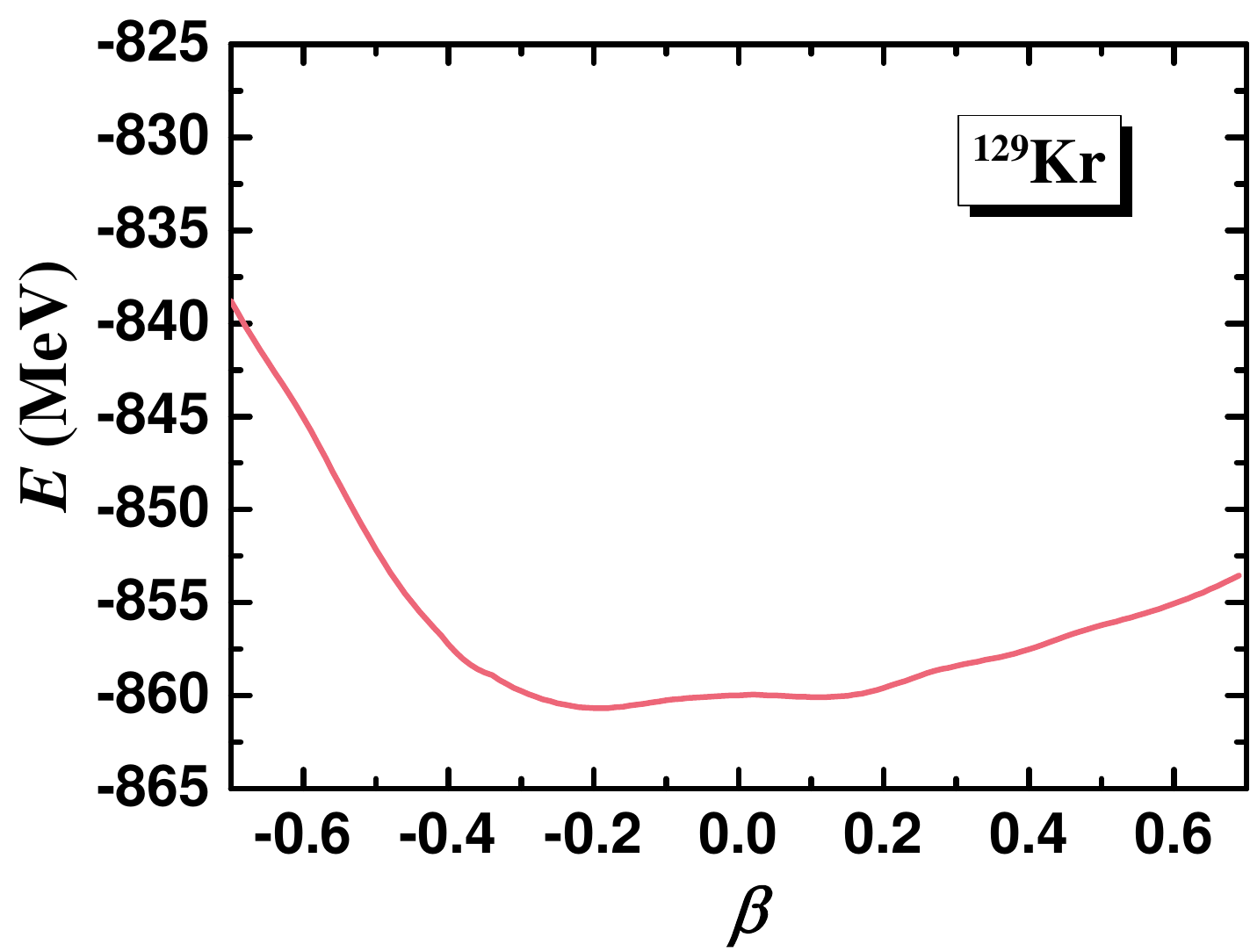}
\end{minipage}
\begin{minipage}{0.23\linewidth}
\includegraphics[width=\linewidth, angle=0]{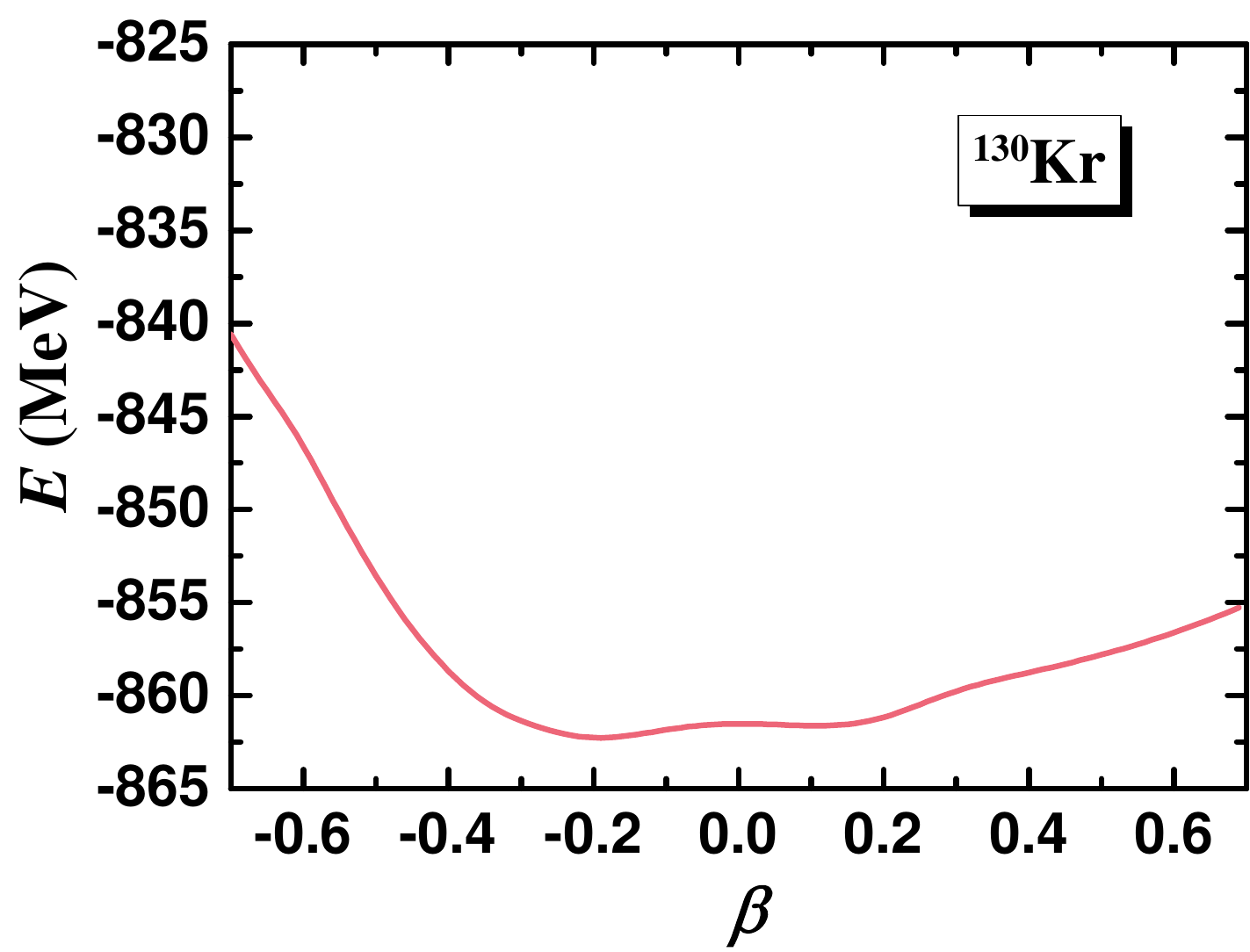}
\end{minipage}
\vspace{-1mm}
\begin{minipage}{0.23\linewidth}
\includegraphics[width=\linewidth, angle=0]{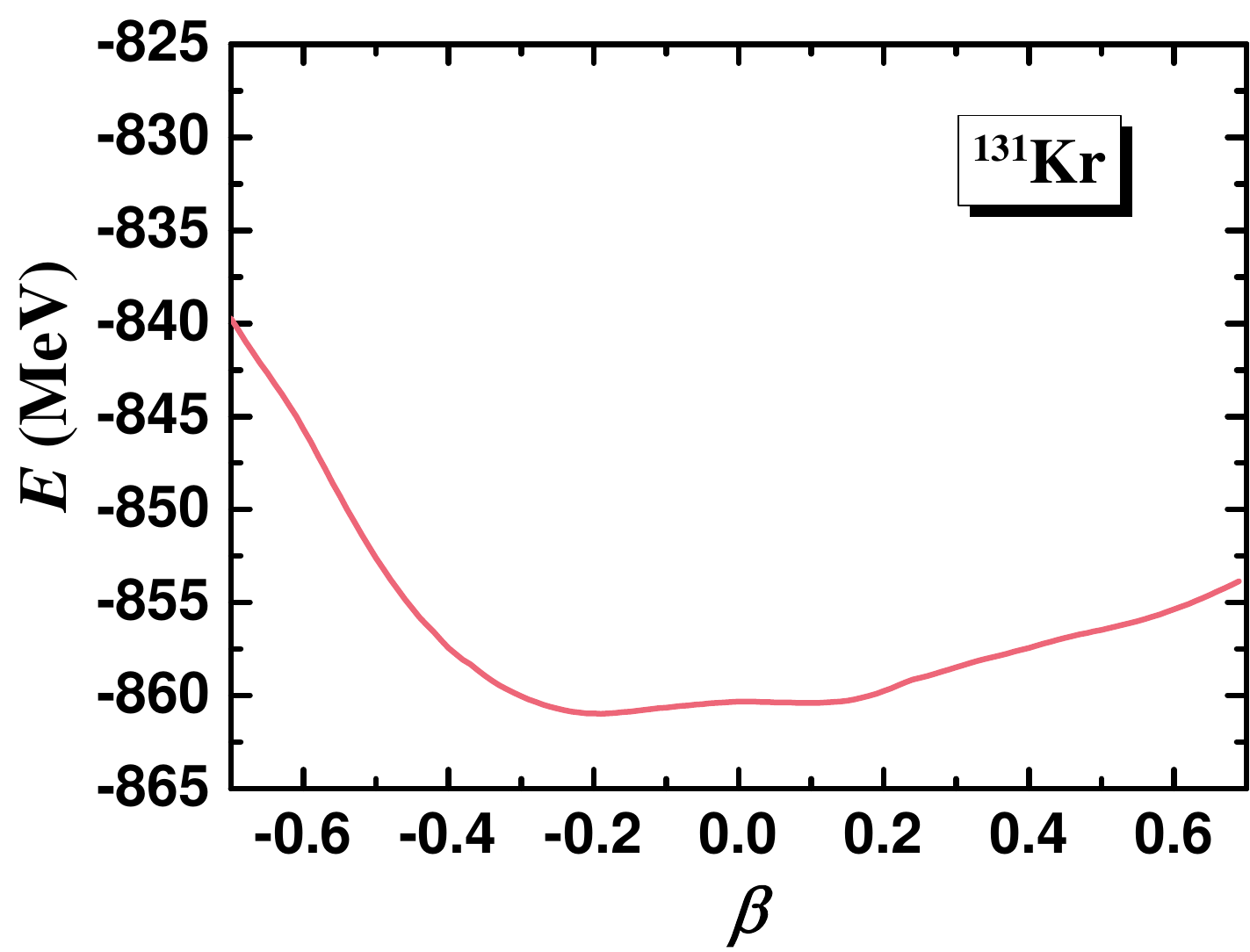}
\end{minipage}
\begin{minipage}{0.23\linewidth}
\includegraphics[width=\linewidth, angle=0]{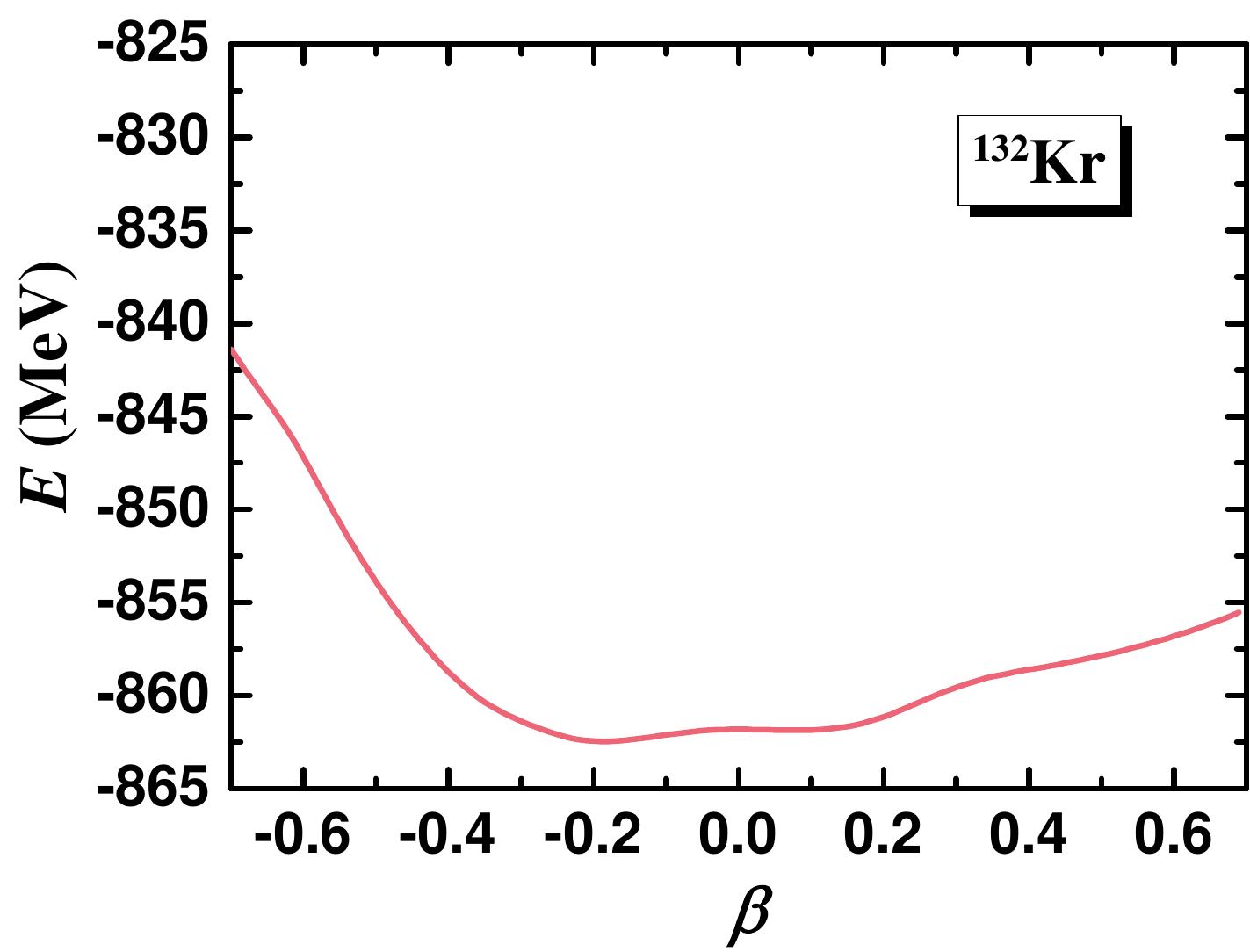}
\end{minipage}
\begin{minipage}{0.23\linewidth}
\hfill
\end{minipage}
\begin{minipage}{0.23\linewidth}
\hfill
\end{minipage}
\vspace{-1mm}
\caption{Potential energy curves as functions of the quadrupole $\beta_{2}$ deformation for the $^{123-132}$Kr isotopes computed with the relativistic Hartree-Bogoliubov approach with axially-symmetrical basis using the PC-L3R effective interaction and separable pairing force of finite range.}
\label{fig:Kr-PES3}
\vspace{-7mm}
\end{figure*}%

\subsection{Shape coexistence}

\begin{figure}[t!]%
\centering
\begin{minipage}{0.47\textwidth}
\centering
\includegraphics[width=\linewidth, angle=0]{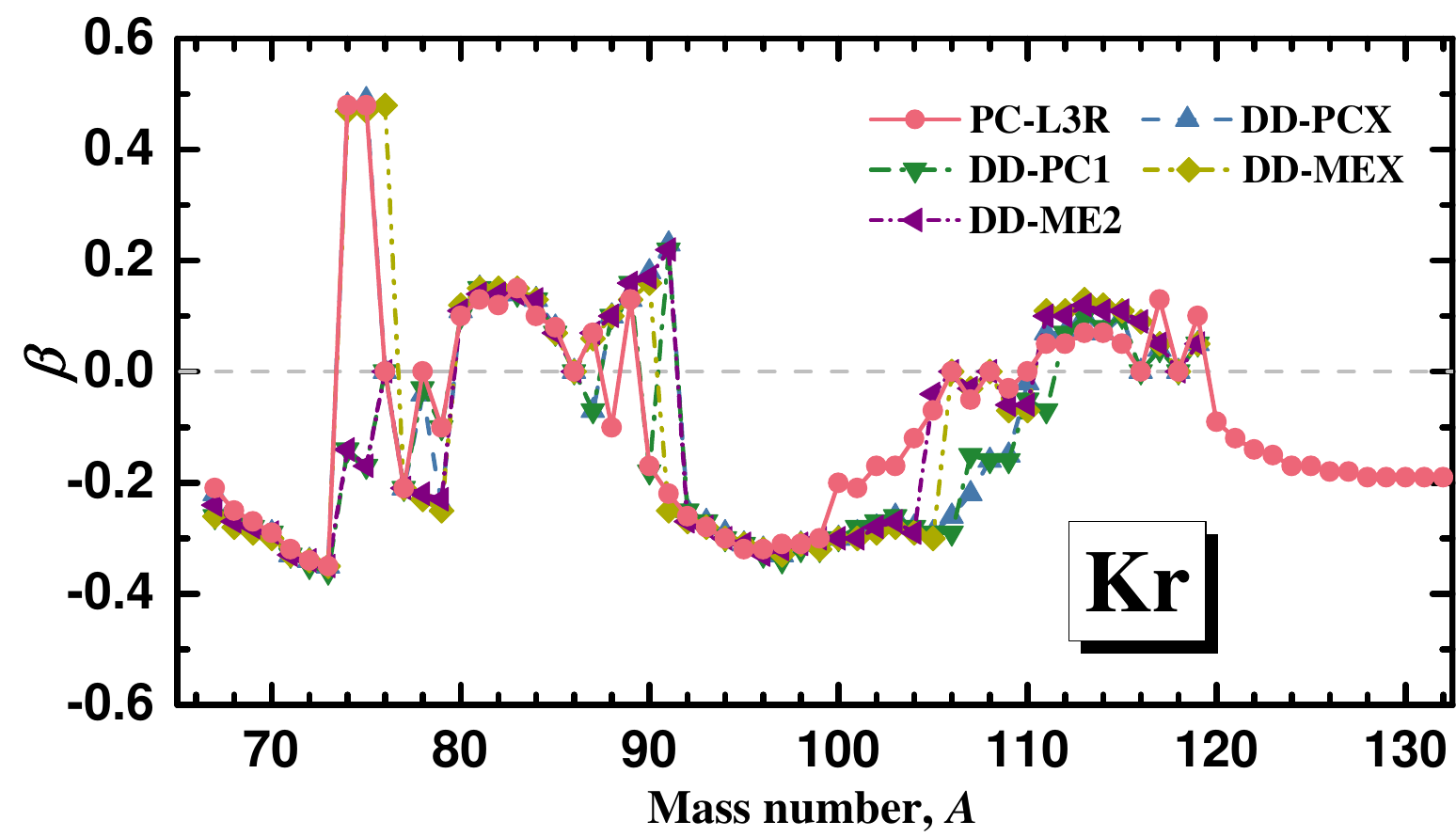}
\end{minipage}%
\caption{Quadrupole deformation $\beta_{2}$ of Kr isotopes is shown as a function of mass number $A$. The theoretical predictions are obtained from RHB calculations with the PC-L3R, DD-PCX, DD-PC1, DD-MEX, and DD-ME2 effective interactions.}
\label{fig:beta}
\vspace{-7mm}
\end{figure}

Figures~\ref{fig:Kr-PES1}, \ref{fig:Kr-PES2}, and \ref{fig:Kr-PES3} illustrate the potential energy curves (PECs) of the krypton isotopes, spanning from proton to neutron drip lines, which are produced from these effective interactions. The shape evolution along the krypton isotopic chain is clearly shown, manifesting a rich structure of the intriguing shape coexistence phenomena in the ground states. We systematically map the quadrupole deformation space defined by the quadrupole deformation, $\beta_{2}$. The $\beta_{2}$ of Kr isotopes is presented in Fig.~\ref{fig:beta}, showing clear evidence of the dependence of the global minima of PECs on the neutron number. Compared to the density-dependent covariant density functionals of DD-PCX, DD-PC1, DD-MEX, and DD-ME2, the nonlinear point-coupling effective interaction of PC-L3R predicts a more extensive Kr neutron drip line, beyond $^{118}$Kr ($N=82$) and up to $^{132}$Kr ($N=96$), which is in agreement with the drip line predicted by the axially-symmetric DRHBc approach with PC-PK1 interaction \cite{ADNDT2024Guo}. We discuss in detail the extensive drip line in the analysis of Fig.~\ref{fig:S2n}.

It is well understood that nuclei with magic numbers have predominantly spherical ground states. This perception is evidenced by the calculated PECs of $^{86}\text{Kr}$ ($N=50$) and $^{118}\text{Kr}$ ($N=82$) corresponding to the classical neutron magic numbers of $N=50$ and $N=82$, respectively (Figs.~\ref{fig:Kr-PES1} and \ref{fig:beta}). Calculations based on the PC-L3R, DD-PCX, or DD-MEX effective interactions predict a rapid shape transition from oblate shape ($^{73}$Kr) to prolate shape ($^{74}$Kr), whereas calculations using the DD-PC1 or DD-ME2 interactions produce a persistent oblate shape for these two isotopes. A similar abrupt shape transition was also reported by \citet{ADNDT2024Guo}, with $\beta_2$ changing from $-0.37$ to $0.48$, according to their DRHBc calculations using the PC-PK1 effective interaction. Microscopically, the divergent description of the shape transition between two groups of interactions, i.e., (i) PC-L3R, DD-PCX, and DD-MEX, and (ii) DD-PC1 and DD-ME2, is rooted in the arrangement of single-particle spectra around the Fermi surface. For the calculation of PC-L3R at $\beta_{2} = -0.15$, the occupation probabilities of nucleons in high-lying orbitals (e.g., $9/2^{+}$, $7/2^{+}$) are significantly increased by $\approx$200-300\% compared to the ones of DD-PC1 and DD-ME2. The Fermi energy obtained from PC-L3R exhibits an upward shift of $0.077$~MeV from the global minimum ($\beta_{2,\rm{1st}}=0.48$, $\lambda = -12.382$~MeV) to the secondary minimum ($\beta_{2,\rm{2nd}}=-0.15$, $\lambda = -12.305$~MeV). 
Conversely, for DD-PC1, the Fermi energy at the global minimum ($\beta_{2,\rm{1st}}=-0.15$, $\lambda = -12.490$~MeV) is 0.188~MeV lower than that at the secondary minimum ($\beta_{2,\rm{2nd}}=0.48$, $\lambda = -12.302$~MeV). 
With such an increment of the neutron Fermi surface, this upward shift in occupation reduces the total energy at the oblate deformation, thus establishing the prolate deformation as the global minimum. Nevertheless, DD-PC1 and DD-ME2 interactions do not produce such a pronounced effect, consequently preserving the oblate ground state.

Moreover, the PECs of $^{74}$Kr and $^{75}$Kr isotopes reveal a very $\beta$-soft structure (Figs.~\ref{fig:Kr-PES1} and \ref{fig:beta}). The PEC of $^{74}$Kr produced from DD-PCX features a global minimum of prolate shape ($\beta_2 = 0.48$, $E = -626.611$~MeV) and a secondary minimum of oblate shape ($\beta_2 = -0.15$, $E = -626.312$~MeV). The PEC around this secondary minimum exhibits a $\beta$-soft character, with the energy difference between these two minima being merely 0.298~MeV. On the contrary, the DD-ME2 calculation yields a global minimum of oblate shape ($\beta_2 = -0.14$, $E = -626.177$~MeV) and a secondary minimum of prolate shape ($\beta_2 = 0.48$, $E = -626.066$~MeV), separated by only 0.111~MeV. This behavior is also present in the $^{90}\text{Kr}$ to $^{92}\text{Kr}$ region. The respective $\beta$ and energy difference between two minima $\Delta E$ are given in Table~\ref{Tab1}. Therefore, $^{74,75}$Kr and $^{90-92}$Kr are potential candidates of shape coexistence with $\Delta E$ less than 1~MeV (criterion of $\Delta E \lesssim 1$~MeV as in Ref.~\cite{PRC2022Choi}). From $^{93}$Kr up to the neutron-rich region, Kr isotopes around the $N=82$ magic number exhibit (near) spherical minima in the PECs, while the remaining isotopes show distinct prolate or oblate deformations.

The present finding indicates the sensitivity of shape coexistence on the covariant density functionals for the Kr isotopic chain. Future precise experimental assessments on $^{73}$Kr, $^{74}$Kr, $^{90}$Kr, $^{91}$Kr, and $^{92}$Kr for verifying the oblate or prolate ground-state shapes would be greatly needed to constrain the density functionals.

\begin{table}[t]
\scriptsize
  \caption{\label{Tab1}Potential krypton isotopes with two shape coexistent configurations. Locations of the two ground state minima are indicated by $\beta_{2}$ for $^{74,75}$Kr and $^{90-92}$Kr isotopes based on PC-L3R, DD-PCX, DD-PC1, DD-MEX, and DD-ME2 effective interactions. $\Delta E$  represents the energy difference between the two minima, in units of MeV.}
  \begin{tabular*}{\linewidth}{@{\hspace{1.5mm}}r@{\hspace{1.5mm}}r@{\hspace{2.5mm}}r@{\hspace{2.5mm}}r@{\hspace{2.5mm}}r@{\hspace{2.5mm}}r@{\hspace{2.5mm}}r@{\hspace{1.5mm}}}
   \toprule
   \midrule[0.25pt]
            &                          & PC-L3R  & DD-PCX  & DD-PC1  & DD-MEX  & DD-ME2  \\
   \midrule[0.15pt]
   $^{74}$Kr&   $\beta_{2,{\rm 1st}}$  & $0.48 $ & $0.48 $ & $-0.15$ & $0.47 $ & $-0.14$ \\
            &   $\beta_{2,{\rm 2nd}}$  & $-0.15$ & $-0.15$ & $0.48$  & $-0.35$ & $0.47$  \\
            &   $\Delta E $            & $0.434$ & $0.298$ & $0.412$ & $0.579$ & $0.111$ \\
   \midrule[0.15pt]
  $^{75}$Kr&  $\beta_{2,{\rm 1st}}$    & $0.48 $ & $0.49 $ & $-0.18$ & $0.47 $ & $-0.17$ \\
           &  $\beta_{2,{\rm 2nd}}$    & $-0.17$ & $-0.17$ & $0.48 $ & $-0.37$ & $0.46 $ \\
           &   $\Delta E $             & $0.479$ & $0.154$ & $0.910$ & $1.179$ & $0.474$ \\           
  \midrule[0.15pt] 
  $^{90}$Kr&  $\beta_{2,{\rm 1st}}$    & $-0.17$ & $-0.17$ & $-0.19$ & $0.16 $ & $0.17 $ \\
           &  $\beta_{2,{\rm 2nd}}$    & $0.15 $ & $0.17 $ & $0.17 $ & $-0.19$ & $-0.18$ \\
           &   $\Delta E $             & $0.192$ & $0.017$ & $0.039$ & $0.213$ & $0.101$ \\
  \midrule[0.15pt]          
  $^{91}$Kr&  $\beta_{2,{\rm 1st}}$    & $-0.22$ & $0.23 $ & $0.21 $ & $-0.25$ & $0.22 $ \\
           &  $\beta_{2,{\rm 2nd}}$    & $0.19 $ & $-0.22$ & $-0.23$ & $0.23 $ & $-0.24$ \\
           &   $\Delta E $             & $0.345$ & $0.112$ & $0.121$ & $0.014$ & $0.006$ \\
  \midrule[0.15pt]       
  $^{92}$Kr&  $\beta_{2,{\rm 1st}}$    & $-0.26$ & $-0.24$ & $-0.26$ & $-0.27$ & $-0.27$ \\
           &  $\beta_{2,{\rm 2nd}}$    & $0.20 $ & $0.24$  & $0.24 $ & $0.24 $ & $0.25 $ \\
           &   $\Delta E $             & $0.650$ & $0.253$ & $0.174$ & $0.551$ & $0.430$ \\
  \bottomrule
  \end{tabular*}
\vspace{-7mm}
\end{table}

\begin{figure*}[ht]%
\centering
\includegraphics[width=0.5\textwidth, angle=0]{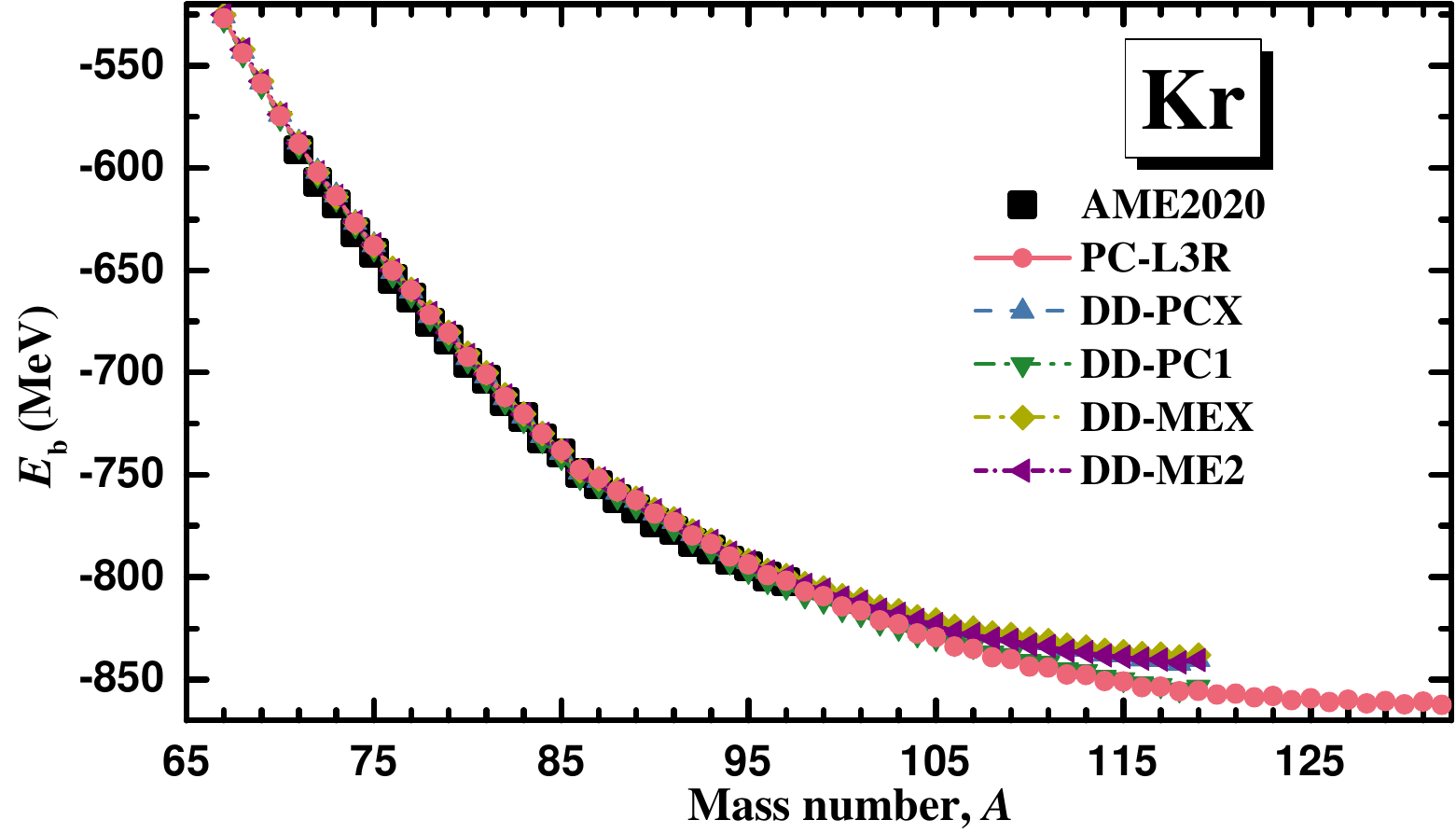}%
\includegraphics[width=0.48\textwidth, angle=0]{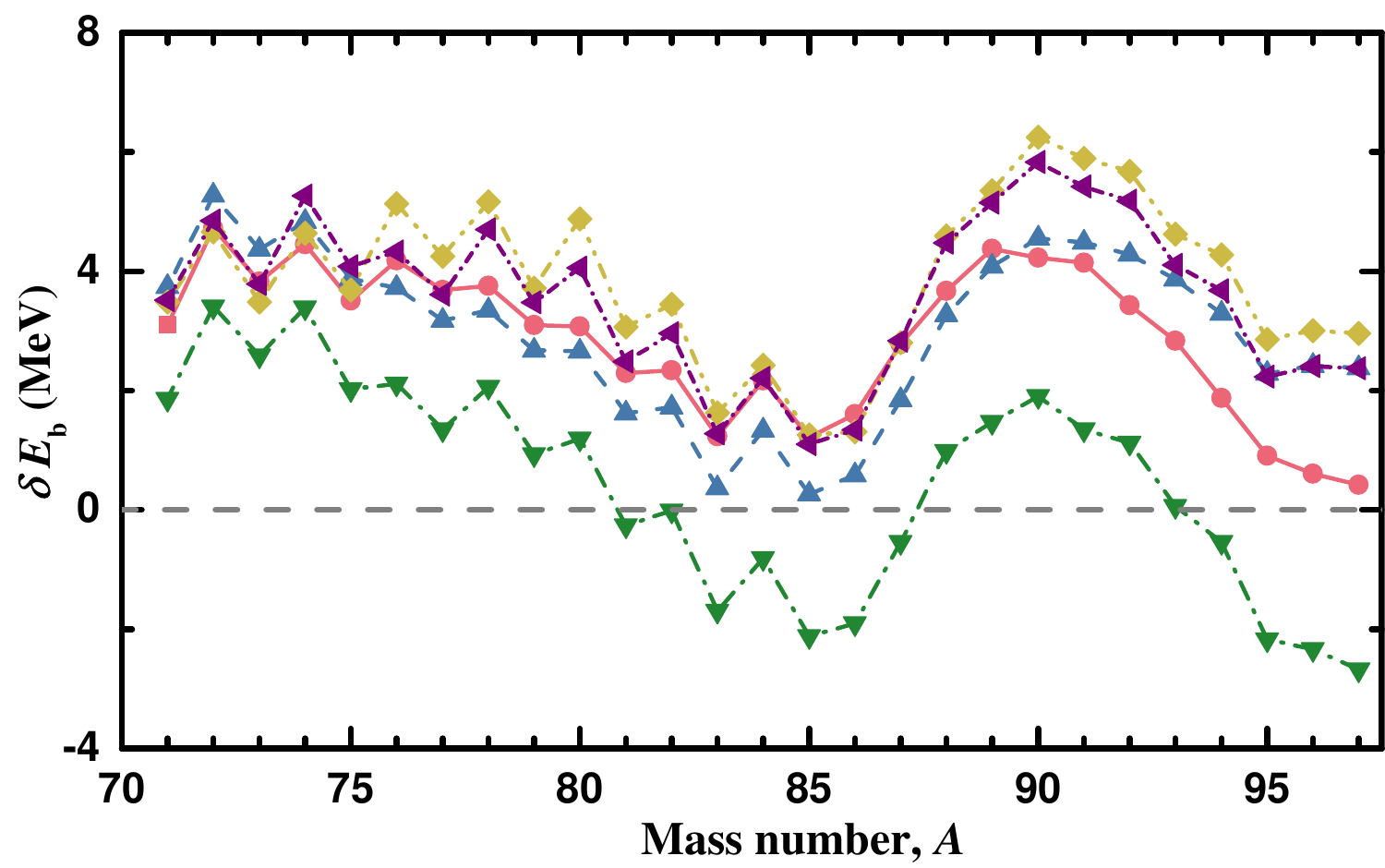}
\caption{Binding energies ($E_\mathrm{b}$) along the Kr isotopic chain. Left panel: experimental $E_\mathrm{b}$ obtained from AME2020~\cite{AME2020} (black squares) and theoretical binding energies derived from the RHB approach using the PC-L3R (red dots), DD-PCX (blue up triangles), DD-PC1 (green down triangles), DD-MEX (yellow diamonds), and DD-ME2 (left purple triangles) effective interactions. Right panel: deviations between theoretical and experimental $E_\mathrm{b}$, denoted as $\delta E_\mathrm{b} = E_\mathrm{b}^{\rm Theo} - E_\mathrm{b}^{\rm AME2020}$.}
\label{fig:rrs_BE}
\end{figure*}%

\subsection{Binding energies}

We calculate the binding energies of Kr isotopes using the self-consistent axially-symmetric RHB approach to assess the predictive capability of the PC-L3R, DD-PCX, DD-PC1, DD-MEX, and DD-ME2 effective interactions. Figure~\ref{fig:rrs_BE} illustrates the binding energies of isotopes along the Kr isotopic chain. The experimental values are taken from AME2020~\cite{AME2020}. The trend of experimental binding energies is successfully reproduced by the theoretical approaches presented here, although these theoretical frameworks systematically produce a set of somehow underbound binding energies (lower absolute magnitudes).

Among these effective interactions, DD-PC1 produces the lowest root-mean-square (rms) deviation of $1.822$~MeV, whereas other interactions of this work yield rms values at least $1.306$~MeV higher than that of DD-PC1. Meanwhile, the axially-symmetric DRHBc approach with PC-PK1 effective interaction used by \citet{ADNDT2024Guo} produces an rms deviation of $2.890$~MeV. This remarkable robustness of DD-PC1 comes directly from its optimization scheme, which is not considered in other effective interactions of this work. \citet{PRC2008Niksic} explicitly took into account the experimental binding energies of the deformed nuclei in optimizing the DD-PC1 interaction. Table~\ref{Tab2} presents the rms deviations between theoretical and experimental binding energies for both spherical and axially-deformed calculations. Compared with the spherical case, the outcome of the axially deformed calculation yields a substantial improvement in describing the binding energies, indicating the importance of degrees of freedom of deformation. This improvement is evident for all effective interactions, producing reductions of $1.548$~MeV, $2.171$~MeV, $1.046$~MeV, $3.214$~MeV, and $2.836$~MeV for PC-L3R, DD-PCX, DD-PC1, DD-MEX, and DD-ME2, respectively. The DD-MEX interaction exhibits the most pronounced enhancement. Using the self-consistent triaxially-symmetric RHB approach should shed light on the enhancement of reproducing experimental binding energies of the Kr isotopic chain and the information of single-particle structure. Such a systematic investigation of triaxial deformation is somehow inviting, but challenging. We leave these non-trivial works for the future.

\begin{table}[t]
\scriptsize
  \caption{\label{Tab2}The root-mean-square (rms) deviations between theoretical and experimental binding energies ($E_\mathrm{b}$), one-neutron separation energies ($S_{\rm n}$), and two-neutron separation energies ($S_{\rm 2n}$) for the ground states of $^{71-97}$Kr nuclei. Theoretical results are obtained from the RHB approach using the PC-L3R, DD-PCX, DD-PC1, DD-MEX, and DD-ME2 effective interactions based on axially-symmetrical or spherical basis.}
  \begin{tabular*}{\linewidth}{@{\hspace{1.0mm}}l@{\hspace{1.5mm}}c@{\hspace{1.5mm}}c@{\hspace{1.5mm}}c@{\hspace{1.5mm}}c@{\hspace{1.5mm}}c@{\hspace{1.0mm}}}
   \toprule
   \midrule[0.25pt]
                          & PC-L3R & DD-PCX  & DD-PC1   & DD-MEX & DD-ME2\\
   \midrule[0.15pt]
   rms$_{E_b}$ (Axial)      &$3.128 $&$3.272 $ &  $1.822$ &$4.094$ &$3.817$ \\
   rms$_{E_b}$ (Spherical)    &$4.676 $&$5.443 $ &  $2.868$ &$7.308$ &$6.653$ \\
   \midrule[0.15pt]
   rms$_{\rm Sn}$ (Axial)      &$0.764 $&$0.792 $ &  $0.977$ &$1.094$ &$1.024$ \\
   rms$_{\rm Sn}$ (Spherical)    &$0.724 $&$0.851 $ &  $0.872$ &$1.003$ &$0.988$ \\
   \midrule[0.15pt]
   rms$_{\rm S2n}$ (Axial)     &$0.988 $&$1.075 $ &  $1.265$ &$1.254$ &$1.211$ \\
   rms$_{\rm S2n}$ (Spherical)   &$1.248 $&$1.565 $ &  $1.537$ &$1.797$ &$1.765$ \\
  \bottomrule
  \end{tabular*}
\end{table}

\subsection{One- and two-neutron separation energies}

One- and two-neutron separation energies can be estimated from the ground state nuclear masses of three consecutive isotopes, i.e., $E_{\rm b}$$(Z,N)$, $E_{\rm b}$$(Z,N-1)$, and $E_{\rm b}$$(Z,N-2)$. One-neutron separation energy is expressed as 
\begin{equation}
    S_{\rm n} (Z,N) = E_{\rm b}~(Z,N) - E_{\rm b}~(Z,N-1),
\end{equation}
and two-neutron separation energy reads 
\begin{equation}
    S_{\rm 2n} (Z,N) = E_{\rm b}~(Z,N) - E_{\rm b}~(Z,N-2).
\end{equation}
These quantities dictate the stability of a nucleus against one- and two-nucleon emissions, thereby delineating the respective neutron drip lines. We use the self-consistent axially-symmetric RHB approach with the point-coupling (PC-L3R, DD-PCX, and DD-PC1) or the meson-exchange (DD-MEX and DD-ME2) effective interactions to compute the $E_{\rm b}$ of the $^{A}$Kr, $^{A-1}$Kr and $^{A-2}$Kr isotopes and the respective $S_{\rm n}$ and $S_{\rm 2n}$.

\begin{figure}[t!]%
\centering
\begin{minipage}{0.45\textwidth}
\centering
\includegraphics[width=\linewidth, angle=0]{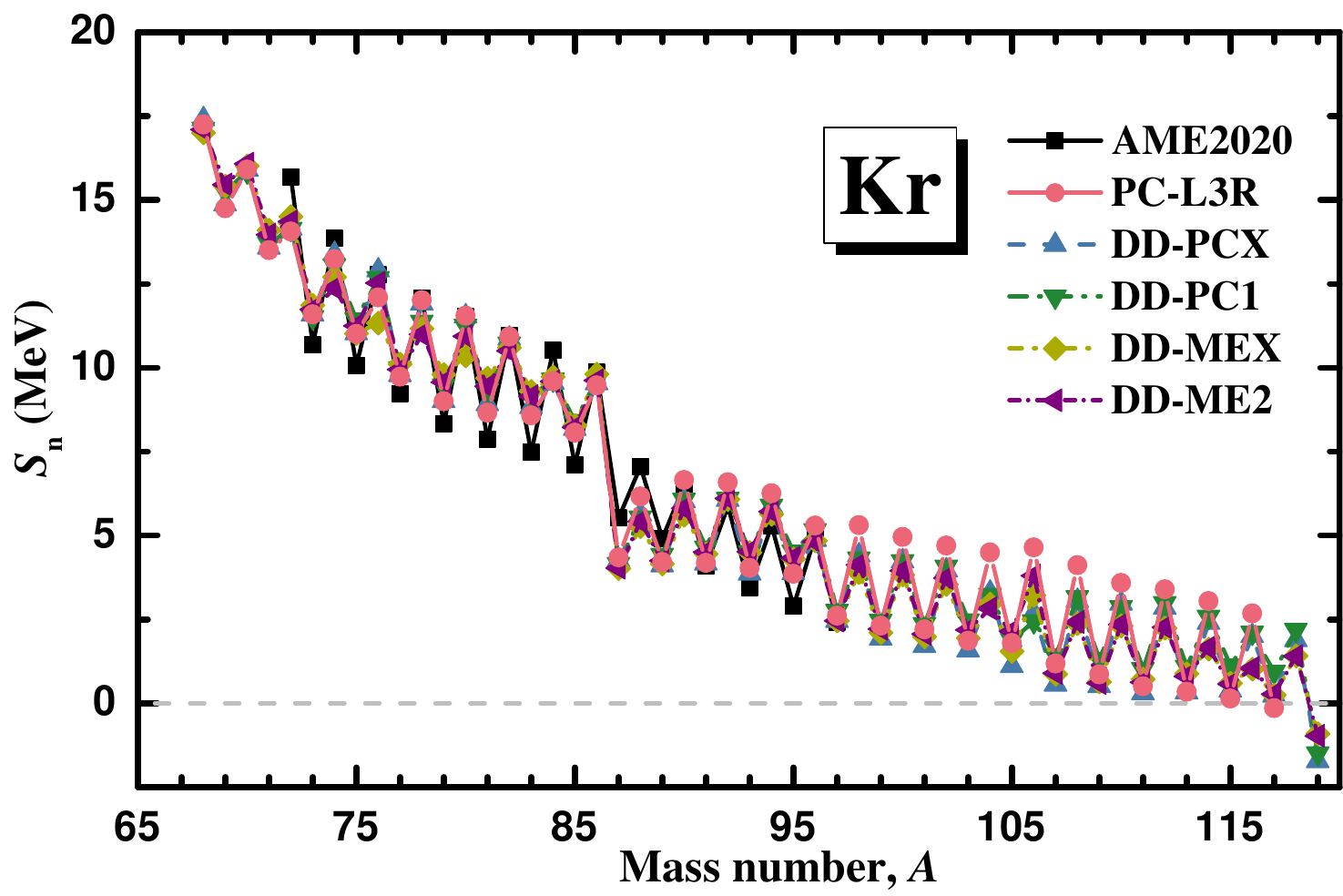}
\end{minipage}%
\centering
\vspace{-1mm}
\begin{minipage}{0.45\textwidth}
\includegraphics[width=\linewidth, angle=0]{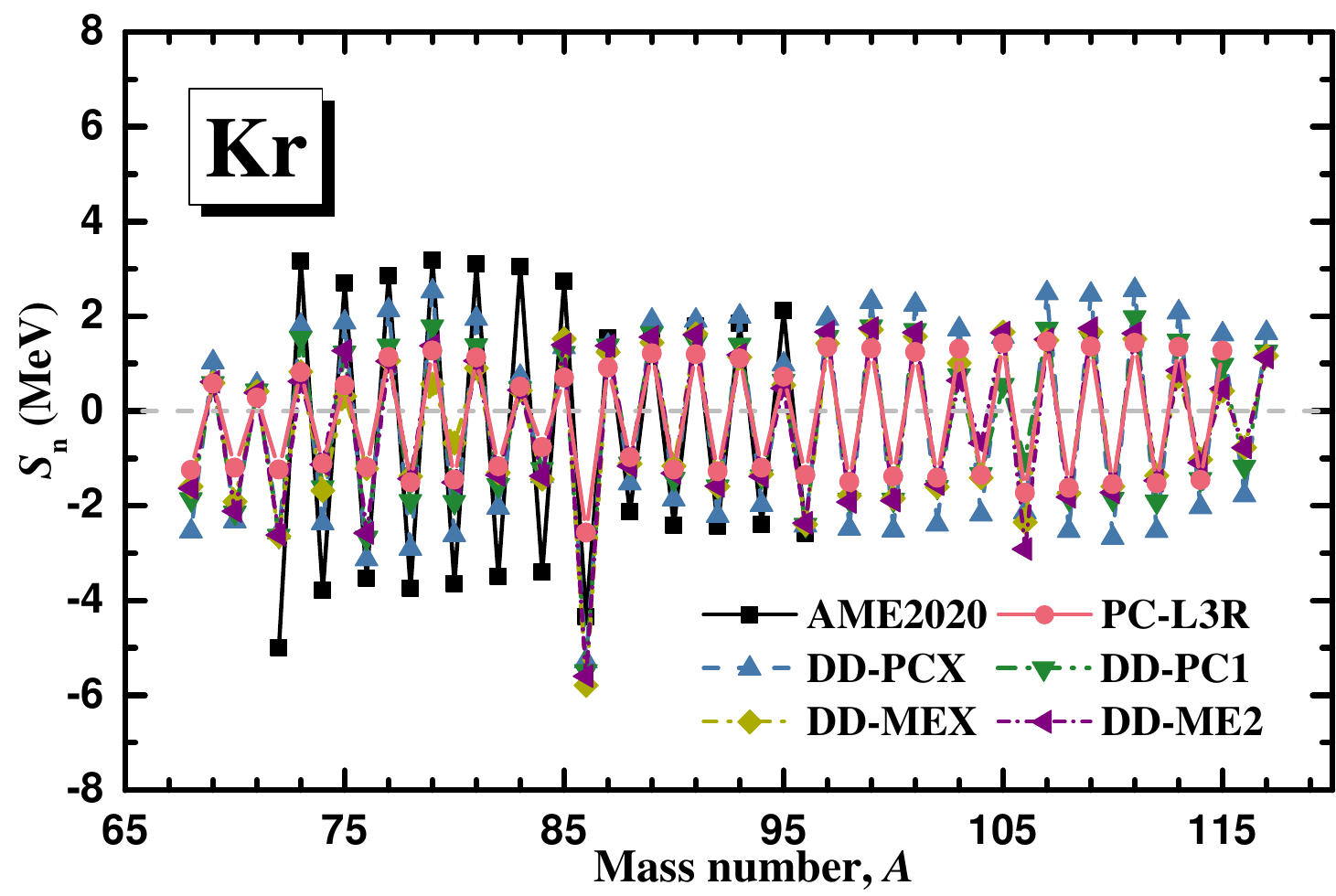}
\end{minipage}
\caption{One-neutron separation energy ($S_{\rm n}$) (top panel) and
the difference of the separation energy $\Delta S_{\rm n} = S_{\rm n} (Z,N) -  S_{\rm n} (Z,N-1)$ (bottom panel) of the Kr isotopic chain. Experimental $S_{\rm n}$ and $\Delta S_{\rm n}$ quoted from AME2020~\cite{AME2020} (black square) and theoretical $S_{\rm n}$ and $\Delta S_{\rm n}$ derived from the self-consistent axially-symmetric RHB approach with the PC-L3R (red dots), DD-PCX (blue up triangles), DD-PC1 (green down triangles), DD-MEX (yellow diamonds), and DD-ME2 (left purple triangles) effective interactions.}
\label{fig:Sn}
\vspace{-7mm}
\end{figure}%

Figure~\ref{fig:Sn} presents the $S_{\rm n}$ and the difference of one-neutron separation energy $\Delta S_{\rm n}$ of the bound Kr isotopes. The $S_{\rm n}$ generally decreases with increasing neutron number, exhibiting a pronounced odd-even staggering (OES) where odd-N isotopes have significantly lower $S_{\rm n}$ values than their adjacent even-even neighbors (top panel of Fig.~\ref{fig:Sn}).

\begin{figure}
\centering
\begin{minipage}{0.48\textwidth}
\centering
\includegraphics[width=\linewidth, angle=0]{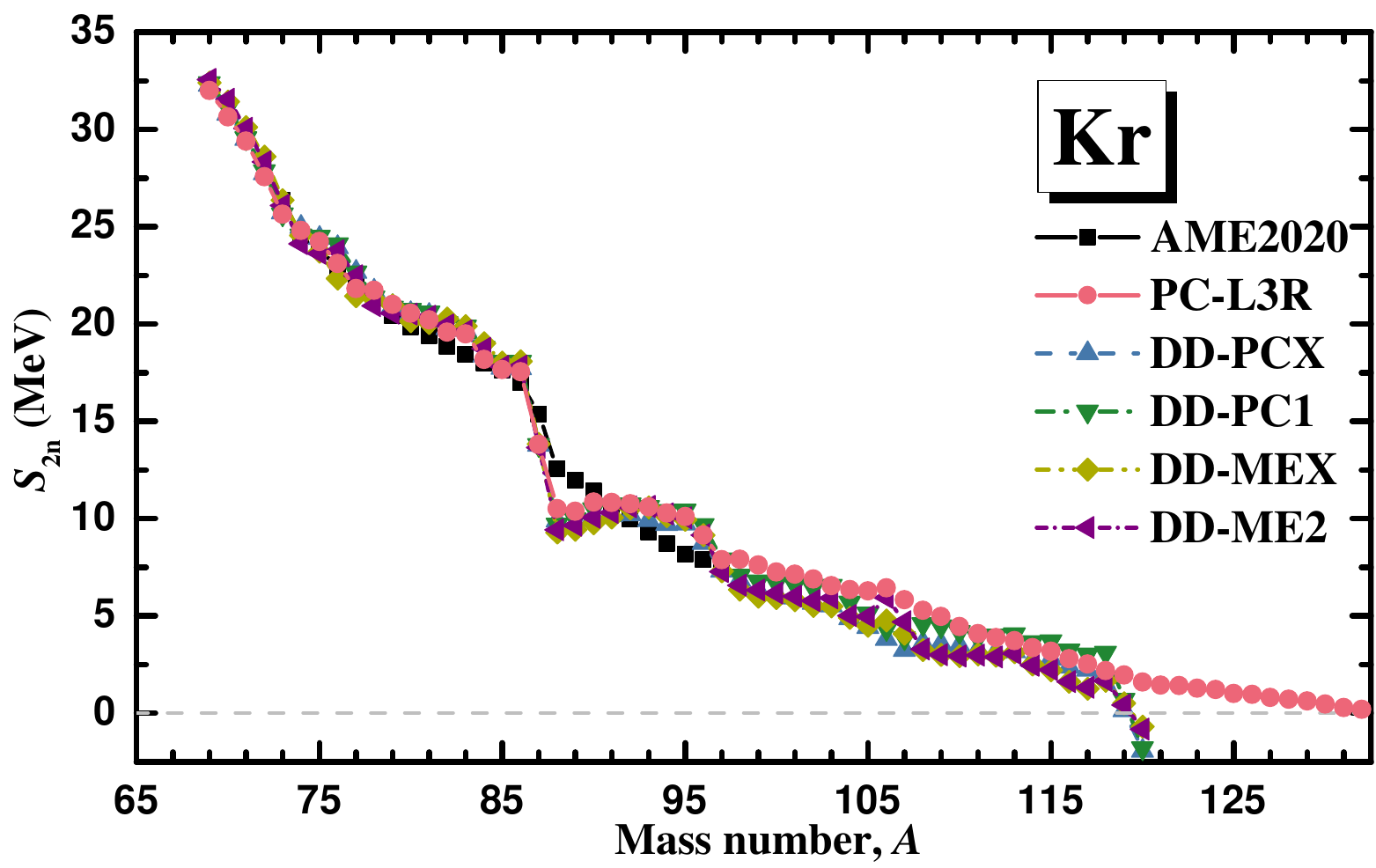}
\end{minipage}%
\centering
\vspace{-1mm}
\begin{minipage}{0.48\textwidth}
\includegraphics[width=\linewidth, angle=0]{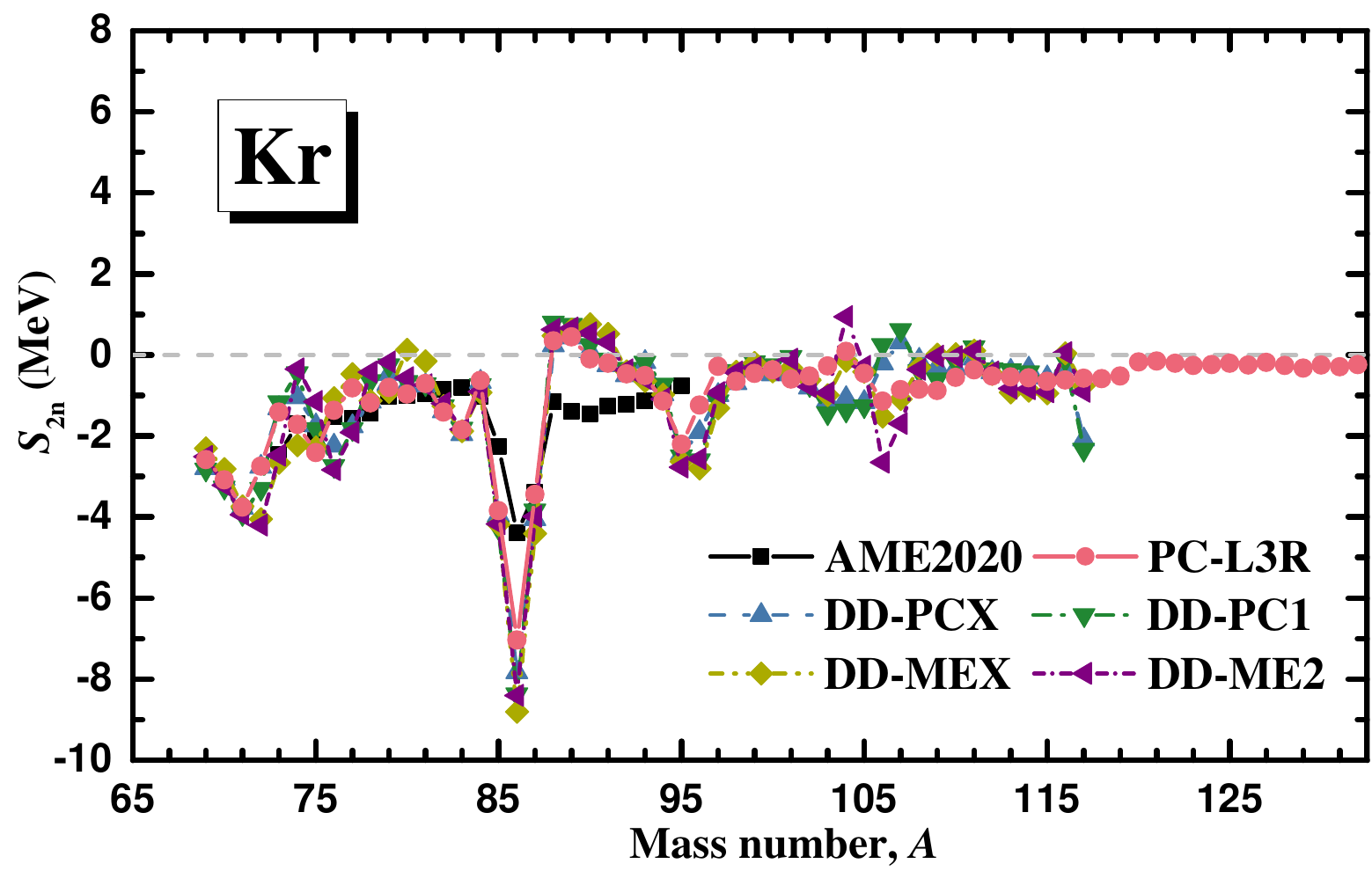}
\end{minipage}
\caption{Two-neutron separation energy $S_{\rm 2n}$ (top panel) and
the differential variation of the separation energy $\Delta S_{\rm 2n} =  S_{\rm 2n} (Z,N) -  S_{\rm 2n} (Z,N-2)$ (bottom panel) of the Kr isotopic chain.
Experimental $S_{\rm 2n}$ and $\Delta S_{\rm 2n}$ cited from AME2020~\cite{AME2020} (black square) and theoretical $S_{\rm 2n}$ and $\Delta S_{\rm 2n}$ derived from the self-consistent axially-symmetric RHB approach with the PC-L3R (red dots), DD-PCX (blue up triangles), DD-PC1 (green down triangles), DD-MEX (yellow diamonds), and DD-ME2 (left purple triangles) effective interactions.}
\label{fig:S2n}
\vspace{-7mm}
\end{figure}

Referring to the trend of $\Delta S_{\rm n}$ of isotopes from $A=67$ to $86$, the theoretical OES amplitude is noticeably smaller than that of the experimental counterpart. The discrepancy diminishes significantly beyond $^{86}$Kr. This deviation is attributed to the respective pairing interaction of each covariant density functional used for this work. With increasing neutron number, $S_{\rm n}$ exhibits abrupt drops near $^{86}$Kr ($N=50$) and $^{118}$Kr ($N=82$). These drops serve as clear evidence for the neutron shell closures. Comparing the trends of $S_{\rm n}$ and $S_{\rm 2n}$ (Figs.~\ref{fig:Sn} and \ref{fig:S2n}) indicates that the one-neutron drip line terminates at the lighter Kr isotope. Such a shortened neutron drip line is primarily due to the odd-even staggering effect, which is closely related to pairing correlations, causing the very neutron-rich odd-$N$ Kr isotope to be more unbound than the adjacent even-even Kr isotopes.

Table~\ref{Tab2} lists the rms deviations between experimental and theoretical $S_{\rm n}$ and $S_{\rm 2n}$, where the theoretical results include both spherical and axially-deformed calculations. Among the density functionals used for axially-deformed calculations for $S_{\rm n}$, PC-L3R and DD-PCX provide the two lowest rms values of $0.764$ and $0.792$~MeV, respectively. Compared to other effective interactions, this indicates an improvement in rms deviations by a reduction of $\approx200$~keV. Meanwhile, compared with the impact on theoretical binding energies, the consideration of deformation effects does not lead to an appreciable improvement in the description of $S_{\rm n}$. In regions where the deformation properties evolve smoothly along the isotopic chain, neighboring nuclei generally exhibit similar deformation effects, producing a negative shift in all binding energies and leaving only a limited net contribution to $S_{\rm n}$. This observation, however, does not imply that deformation effects are negligible. When the equilibrium deformation or the deformation-energy gain changes rapidly with neutron number, particularly in the vicinity of a shape transition, the spherical and deformed calculations may yield appreciably different values of $S_{\rm n}$, as found around $^{74}\mathrm{Kr}$ and $^{92}\mathrm{Kr}$.

Meanwhile, we observe that considering deformation effects does not significantly improve the description of $S_{\rm n}$. This is due to the tendency of neighboring nuclei to possess identical deformation structures, implying that the impact of deformation is suppressed in the corresponding energy differences (Figs.~\ref{fig:Kr-PES2} and \ref{fig:Kr-PES3}).

The $S_{\rm 2n}$ rms deviations generated from axially-symmetric calculations with PC-L3R, DD-PCX, DD-PC1, DD-MEX, and DD-ME2, are 0.988, 1.075, 1.265, 1.254, and 1.211 MeV, respectively. The axially-symmetric DRHBc approach with PC-PK1 yields an rms of 1.087~MeV \cite{ADNDT2024Guo}. It is worth noting that PC-L3R yields the smallest rms of $S_{\rm 2n}$, i.e., $0.988$ MeV, lower than those obtained with DD-PCX, DD-PC1, DD-MEX, and DD-ME2 by $0.087$, $0.277$, $0.266$, and $0.223$ MeV, respectively. Detailed numerical results are found in Table~\ref{Tab2}.

Figure~\ref{fig:S2n} presents the trend of $S_{\rm 2n}$ curve (top panel), indicating that $\Delta S_{\rm 2n}$ (bottom panel) manifests a steep drop towards a more bound $^{86}$Kr at the position of the $N=50$ neutron magic number. All density-dependent effective interactions used in this work consistently predict the two-neutron drip line at $^{119}$Kr. The two-neutron drip line predicted by PC-L3R is more extensive and up to $^{132}$Kr. A similar trend is predicted by the PC-PK1~\cite{ADNDT2024Guo} and NL3$^\ast$~\cite{PRC2014Agbemava} effective interactions, resulting the two-neutron drip lines at $^{133}$Kr and $^{130}$Kr, respectively. Such an extensive drip line arises from a more softened $N=82$ shell structure near the neutron drip lines, implying a quenched or even collapse of the traditional neutron shell closure in this region. Furthermore, the inclusion of axial deformation has no impact on the predictions of the nuclear drip lines for density-dependent energy functionals used in this work. For PC-L3R, considering axial deformation merely shortens the predicted two-neutron drip line by two isotopes. See the work of \citet{ADNDT2024Liu} for $S_{\rm n}$ and $S_{\rm 2n}$ of Kr isotopes generated from the spherically-symmetric RHB approach. This behavior is also observed in DRHBc calculations using the PC-PK1 interaction \cite{ADNDT2018Xia, ADNDT2024Guo}.

\section{SUMMARY}
\label{Sec:summary}
We perform a systematic study of the Kr isotopic chain, consisting of even-even and odd-$A$ isotopes using the axially deformed RHB approach with the PC-L3R, DD-PCX, DD-PC1, DD-MEX, and DD-ME2 effective interactions. A detailed analysis of the potential energy curves reveals the complex evolutionary features of the Kr isotopic chain. An abrupt shape transition from oblate to prolate emerges at $^{73}$Kr and $^{74}$Kr based on the PC-L3R, DD-PCX, and DD-MEX interactions, whereas an oblate ground state is maintained for these two isotopes based on the DD-PC1 and DD-ME2 interactions (Fig.~\ref{fig:beta}). Such a distinct outcome is due to the arrangement of the single-particle level structure near the Fermi surface described by the interaction. For PC-L3R, DD-PCX, and DD-MEX, the enhanced occupation of high-lying orbitals (e.g., $9/2^+$ and $7/2^+$) significantly alters the energy balance among competing minima. We observe that $^{74,75}$Kr and $^{90,91,92}$Kr can be identified as potentially representative shape coexistence candidates, as their potential energy curves exhibit competing prolate and oblate minima separated by less than 1~MeV (Table~\ref{Tab1}).

Overall, all trends in theoretical binding energies match the experimental counterpart (left panel of Fig.~\ref{fig:rrs_BE}) although theoretical estimates are systematically underbound (right panel of Fig.~\ref{fig:rrs_BE}). Notably, among these functionals, the DD-PC1 functional yields the lowest rms deviation, $1.822$~MeV, compared to the experimental data (Table~\ref{Tab2}). Other functionals produce rms deviations that of $1.3$~MeV higher than the one of DD-PC1, which was constructed by \citet{PRC2008Niksic}, focusing on reproducing experimental binding energies of deformed nuclei. In addition, the comparison of rms values obtained from spherically- and axially-symmetric RHB approaches demonstrate that degrees of freedom of deformation are essential for a reliable description of the Kr isotopic chain (Table~\ref{Tab2}). For instance, the rms deviation for DD-MEX is reduced by 3.214~MeV after taking into account the deformation, underscoring the importance of static deformation in this mass region. Future works based on the triaxial relativistic Hartree-Bogoliubov approach incorporating the odd-nucleon blocking effect may shed light on the description of binding energies, deformation, and shape coexistence for isotopic chains of this region.

The axially-symmetric RHB approach with the PC-L3R interaction produces the lowest two-neutron separation energy ($S_{\rm 2n}$) rms deviation, $0.988$~MeV, although the inclusion of deformation does not lead to a significant improvement in the description of the one-neutron separation energy ($S_{\rm n}$). A similar situation also happens on the axially-symmetric RHB approach with DD-PCX, DD-PC1, DD-MEX, and DD-ME2 interactions. Analysis of $S_{\rm 2n}$ and its derivative clearly reveals the $N=50$ and $N=82$ shell closures. The DD-PCX, DD-PC1, DD-MEX, and DD-ME2 interactions predict the one- and two-neutron drip lines at $^{118}$Kr and $^{119}$Kr, respectively. The PC-L3R interaction predicts a more extensive two-neutron drip line up to $^{132}$Kr. The NL3$^\ast$ and PC-PK1 interactions also predict an extensive two-neutron drip line, suggesting a possible softened $N=82$ shell near the drip line of the $Z=36$ region. This indicates a weakening or even vanishing traditional shell closure in neutron-rich nuclei. Our results highlight the need of future studies assessing the weakening shell closure based on the triaxial deformation and beyond-mean-field correlations for this isotopic region. 

\begin{acknowledgments}
\vspace{-5mm}
We greatly appreciate the anonymous Referee for carefully reviewing our manuscript with constructive suggestions on clarity. This work was supported by the National Natural Science Foundation of China (No. \textcolor{black}{11775277}), Dongjiang Laboratory S$\&$T Program (No. \textcolor{black}{DJ2025C008}), Science Foundation of Zhejiang Sci-Tech University (No. 25062123-Y). We greatly appreciate the computing resource provided by the Yukawa Institute Computer Facility, i.e., Yukawa-21 and Heian at the Yukawa Institute of Theoretical Physics (YITP) of Kyoto University, Japan. PR gratefully acknowledges financial support from the Deutsche Forschungsgemeinschaft (DFG, German Research Foundation) under Germany's Excellence Strategy EXC-2094-390783311, ORIGINS.
\end{acknowledgments}

\bigskip
\noindent
{\bf Data Availability Statement}
\smallskip\\
Data in Figs.~\ref{fig:Kr-PES1}, \ref{fig:Kr-PES2}, and \ref{fig:Kr-PES3} are publicly available at the Zenodo repository \cite{Zenodo}. These data can also be used to produce Figs.~\ref{fig:beta}, \ref{fig:rrs_BE}, \ref{fig:Sn}, \ref{fig:S2n}, and Tables~\ref{Tab1} and \ref{Tab2}. 

\bibliography{bibliography} 

\end{document}